\documentclass[final,3p,times]{elsarticle}
\usepackage{graphicx,color}
\usepackage{amssymb,amsmath,array}
\biboptions{sort,compress}
\journal{Annals of Physics}

\begin{document}

\begin{frontmatter}

\title{Fixed points of the SRG evolution and the on-shell limit of the nuclear force}

\author[1]{E. Ruiz Arriola}
\author[2]{S. Szpigel}
\author[3]{V. S. Tim\'oteo
\footnote{Corresponding author, tel.: +55 11 981 483 747, e-mail address: varese@ft.unicamp.br}}

\address[1]{Departamento de F\'isica At\'omica, Molecular y Nuclear and Instituto Carlos I de Fisica Te\'orica y Computacional \\
Universidad de Granada, E-18071 Granada, Spain}

\address[2]{Centro de R\'adio-Astronomia e Astrof\'\i sica Mackenzie, Escola de Engenharia,
Universidade Presbiteriana Mackenzie \\ 01302-907, S\~ao Paulo, SP, Brasil}

\address[3]{Grupo de \'Optica e Modelagem Num\'erica - GOMNI, Faculdade de Tecnologia - FT,
Universidade Estadual de Campinas - UNICAMP \\ 13484-332, Limeira, SP, Brasil}

\begin{abstract}
We study the infrared limit of the similarity renormalization group
(SRG) using a simple toy model for the nuclear force aiming to
investigate the fixed points of the SRG evolution with both the Wilson
and the Wegner generators. We show how a fully diagonal interaction at
the similarity cutoff $\lambda \rightarrow 0$ may be obtained from the
eigenvalues of the hamiltonian and quantify the diagonalness by means
of operator norms. While the fixed points for both generators are
equivalent when no bound-states are allowed by the interaction, the
differences arising from the presence of the Deuteron bound-state can
be disentangled very clearly by analyzing the evolved interactions in
the infrared limit $\lambda \to 0$ on a finite momentum grid. Another
issue we investigate is the location on the diagonal of the
hamiltonian in momentum-space where the SRG evolution places the
Deuteron bound-state eigenvalue once it reaches the fixed point.  This
finite momentum grid setup provides an alternative derivation of the
celebrated trace identities, as a by product. The different effects
due to either the Wilson or the Wegner generators on the binding
energies of $A=2,3,4$ systems are investigated and related to the
ocurrence of a Tjon-line which emerges as the minimum of an avoided crossing
between $E_\alpha= 4 E_t - 3 E_d$ and $E_\alpha= 2 E_t $.
All infrared features of the flow equations are illustrated using the
toy model for the two-nucleon $S$-waves.
\end{abstract}

\end{frontmatter}

%%%

\section{Introduction}

The similarity renormalization group (SRG) approach was proposed
independently by Glazek and Wilson and
Wegner~\cite{Glazek:1993rc,Glazek:1994qc,wegner1994flow,Kehrein:2006ti}
and was initially applied in solid-state physics to simplify
many-particle randomly disordered systems. Since the solution of a
many-particle problem requires diagonalization of the hamiltonian, a
transformation which makes it more diagonal would be of great
applicability, specially if the eigenvalues are preserved.  This idea
was then applied by Wegner in order to suppress the off-diagonal
matrix-elements of the hamiltonian by means of unitary transformations
\cite{wegner1994flow}. The unitarity of the transformations ensures
isospectrality and the generator of the transformations can be chosen
to be diagonal if one wants to drive the original hamiltonian towards
a band-diagonal form and to suppress its off-diagonal
matrix-elements. As a by-product one gets a framework where these
off-diagonal components may be better handled in perturbation theory.

The application of the SRG approach to nuclear physics was proposed by
Bogner, Furnstahl and Perry~\cite{Bogner:2006pc} with these
simplifications in mind and will be our concern here (for reviews see
e.g. \cite{Bogner:2009bt,Furnstahl:2012fn,Furnstahl:2013oba} and
references therein). The first applications of the SRG method
considered high-precision~\cite{Wiringa:1994wb,Stoks:1994wp} and
chiral effective field theory
(ChEFT)~\cite{Entem:2003ft,Epelbaum:2004fk} nucleon-nucleon ($NN$)
interactions as the input for the two-body flow equations.  Despite
different degrees of freedom and types of interaction, the nuclear
many-body problem has the same difficulty as the many-electron problem
in solid-state physics. More diagonal hamiltonians result in great
simplification and faster convergence in nuclear structure
calculations. This feature gave rise to a whole program of application
of the SRG methods in nuclear physics
\cite{Bogner:2006pc,Bogner:2009bt,Furnstahl:2012fn,Furnstahl:2013oba}. The
formalism was then extended to three-body forces in several schemes
\cite{Jurgenson:2009qs,Hebeler:2012pr,Wendt:2013bla}.

Yet in the two-nucleon system, the interplay between the SRG and a
subtractive renormalization approach
\cite{Frederico:1999ps,Timoteo:2005ia,Timoteo:2011mm,Szpigel:2012bc}
was investigated using the ChEFT $NN$ interaction at leading-order
\cite{Szpigel:2011bj}. Also, the role of long-distance symmetries in
effective interactions obtained via SRG flow equations have been
recently investigated by us, showing that there is a particular SRG
scale at which the $SU(4)$ spin-isospin Wigner symmetry is realized
almost exactly \cite{Arriola:2013nja,Timoteo:2012tt}. This is a
remarkable result which shows that despite the increasing popularity
of the SRG ideas, techniques and extensive computational applications
there is still much to be learned from dedicated analysis. The present
paper provides further insights along these lines.

The aim of this work is to explore the details underlying our recent
results scanning all values of the SRG cutoff and to extend them.  In
our previous letter \cite{Arriola:2014aia} a connection between the
infrared limit of the SRG evolution with the Wegner generator and
Levinson's theorem \cite{Ma:2006zzc} was established. Consequences of
the infrared interactions for few-nucleon systems and the nuclear
many-body problem were discussed in Ref. \cite{Arriola:2013gya} where
a theoretical and phenomenologically successful prediction for the
Tjon-line has been advanced. This is a well known existing linear
correlation between the binding energies of the triton and the
$\alpha$-particle which should be expected on the basis of scale
invariance~\cite{Delfino:2007zu} (see e.g.  \cite{Hammer:2012id} for a
review and references therein). Further consequences regarding unitary
neutron matter in the on-shell limit with a calculation of the Bertsch
parameter were addressed in \cite{Arriola:2014tva} and an SRG
discussion of the BCS pairing gap has also been undertaken in
Ref.~\cite{Arriola:2015hra}.

In this work we investigate the fixed points of the SRG evolution in
detail using a simple toy model for the nuclear force in the
two-nucleon $S$-waves as a particular illustration which simplifies
the computational effort considerably and allows for detailed
numerical studies in the infrared limit. However, in the infrared
region we will suggest that many features are fairly general and model
independent. Here we focus on the Wilson and the Wegner SRG
generators, since the evolution with a block-diagonal generator
\cite{Anderson:2008mu} in the infrared region has already been studied
in previous works \cite{Arriola:2013yca,Arriola:2013era}, where the
explicit renormalization of a simple $NN$ force and the implicit
renormalization of a pionless effective field theory (EFT) at
next-to-leading order were shown to be equivalent over a wide range of
the renormalization scale. The SRG equations are mostly solved
numerically on a finite momentum grid with sufficiently many points as
to approach the continuum and therefore the grid is viewed as an
auxiliary means. Here, we will analyze many effects which can only be
clearly disentangled with the aid of this momentum grid, which by
itself has some implications on its own and provides both an infrared
and ultraviolet cutoffs featuring two basic properties of finite
nuclei, namely, the long wavelength character of weak binding systems
as well as the finite size of atomic nuclei.  Kukulin and
collaborators in a series of recent and remarkable works have profited
from these finite grids in the few-body problem including both bound
and scattering states by analyzing Hamiltonian
eigenvalues~\cite{kukulin2009discrete,Rubtsova:2010zz,Pomerantsev:2014ija,Rubtsova:2015owa}. It
is conceivable that a judicious combination of SRG and this approach
may provide useful insights into the nuclear few-body problem.

This paper is organized as follows. In Section \ref{sec:Operator} we
present the SRG flow equations in operator form and introduce some
useful functional notation both in the continuum limit and in the
discretized form which allows for a discussion of fixed points and
their stability. In Section~\ref{sec:toy} we review the toy model
which provides a quite reasonable description of the $NN$ system in
the $S$-wave channels at low-momenta and will be used in practice to
carry out our infrared analysis. In Section~\ref{sec:SRG-grid} we deal
specifically with the SRG flow equations on a finite momentum
grid. The scattering problem as a Lippmann-Schwinger (LS) equation on
the finite momentum grid is analyzed in
Section~\ref{sec:scatt-grid}. There we provide a motivation to define
the phase-shift as an energy-shift which, unlike the conventional
phase-shifts determined from the solution of the LS equation, is
invariant under unitary transformations on the grid. We illustrate the
usefulness of the analysis by re-deriving a set of generalized trace
identities, first unveiled by Graham, Jaffe, Quandt and Weigel in
Ref.~\cite{Graham:2001iv}, but starting from a momentum grid. Our
numerical results on these issues are presented in
Section~\ref{sec:numerical}.  There we focus on the crucial issue of
the ordering of states along the SRG evolution trajectory and the
remarkable connection to Levinson's theorem through an energy-shift
formula. Our interpretation corrects the erroneous implementation of
Kukulin and collaborators~\cite{kukulin2009discrete}. Some of the
consequences of the correct identification and ordering of states for
the nuclear binding energies when approaching the infrared limit in
the $A=2,3,4$ systems are discussed in Section~\ref{sec:nucl}.  There
the SRG cutoff parameters triggers the avoided crossing pattern,
familiar from molecular physics, underlying the Tjon-line
correlation. Finally, in Section~\ref{sec:summary} we present a
summary of the results and our main conclusions.

%%%

\section{SRG flow equations in operator form}
\label{sec:Operator}

The similarity renormalization group (SRG) approach, developed by Glazek and Wilson~\cite{Glazek:1993rc,Glazek:1994qc} and independently by Wegner~\cite{wegner1994flow}, has been intensively
applied in the context of nuclear physics to handle multi-nucleon forces in order to soften the
short-distance core~\cite{Bogner:2006pc,Furnstahl:2013oba} with a
rather universal pattern for nuclear symmetries~\cite{Timoteo:2011tt,
  Arriola:2013nja} and interactions~\cite{Dainton:2013axa}. The basic
strategy underlying the application of the SRG methods to nuclear
forces is to evolve an initial (bare) interaction $H$, which has been
fitted to $NN$ scattering data, via a continuous
unitary transformation that runs a cutoff $\lambda$ on energy
differences. Such a transformation generates a family of unitarily
equivalent smooth interactions $H_{\lambda} =
U_{\lambda}~H~U_{\lambda}^{\dagger}$ with a band-diagonal structure of
a prescribed width roughly given by the SRG cutoff
$\lambda$.

We employ the formulation for the SRG developed by Wegner~\cite{wegner1994flow},
which is based on a non-perturbative flow equation that governs the
unitary evolution of the hamiltonian with a flow parameter $s$ that ranges from $0$ to $\infty$,
\begin{equation}
H_s=U_s H_0 U_s^\dagger\; ,
\label{eq:SRGflow}
\end{equation}
\noindent
where $H_0 \equiv H_{s=0}$ is the initial hamiltonian in the center-of-mass system (CM) and $U_s$ is the unitary transformation. The flow parameter $s$ has dimensions of $[{\rm energy}]^{-2}$ and in terms of the SRG cutoff $\lambda$ with dimension of momentum is given by the relation $s=\lambda^{-4}$. As usual, we split the hamiltonian as $H_s=T + V_s$, where $T \equiv T^{\rm cm}+T^{\rm rel}$ is the kinetic energy, which we assume to be independent of $s$, and $V_s$ is the evolved potential. For a translational invariant system, i.e. $V_s \equiv V_s^{\rm rel}$, we can separate the CM and consider only the SRG evolution of the hamiltonian for the relative motion, $H_s^{\rm rel}=T^{\rm rel} + V_s^{\rm rel}$, since the CM kinetic energy $T^{\rm cm}$ does not contribute~\cite{Jurgenson:2009}. For simplicity, in what follows we will drop the superscript $``{\rm rel}"$. The SRG flow equation in operator form can then be written as
\begin{equation}
\frac{d H_s}{ds}=\frac{d V_s}{ds}=[\eta_s,H_s]\; ,
\label{wegfloweq}
\end{equation}
\noindent
with
\begin{equation}
\eta_s=\frac{d U_s}{ds}U_s^{\dagger}=-\eta_s^\dagger\; ,
\end{equation}
\noindent
and is to be solved with the boundary condition $H_s|_{_{s \to 0}} \equiv H_0 = T + V_0$. The anti-hermitian operator $\eta_s$ which specifies the unitary transformation $U_s$ is usually taken as $\eta_s=[G_s,H_s]$, where $G_s$ is a hermitian operator which we will call the SRG generator since it defines $\eta_s$ and so the flow of the hamiltonian. The most popular choices for the generator are the relative kinetic energy $G_s=T$ (Wilson generator)~\cite{Bogner:2006pc}, the evolving diagonal part of the hamiltonian $G_s={\rm diag}(H_s)=H^{D}_s$ (Wegner generator)~\cite{Glazek:1994qc} and the so called block-diagonal generator $G_s=H^{BD}_s=PH_sP+QH_sQ$, where the operators $P$ and $Q=1-P$ are orthogonal projectors ($P^2 = P$, $Q^2 = Q$ , $QP=PQ=0$) for states below and above a given momentum scale $\Lambda_{BD}$~\cite{Anderson:2008mu}~\footnote{This is a unitary implementation
   to all energies of the previously proposed $V_{\rm low k}$ approach~\cite{Bogner:2001gq}.
   Novel generators have been proposed in \cite{Li:2011sr,Dicaire:2014fra}.}.

So far the SRG equations are quite general and can be used to evolve any hamiltonian. In what follows we will restrict to the case of $NN$ interactions in the center-of-mass (CM) system.

%%%

\subsection{Partial-wave equations}

After decomposition of the $NN$ interaction in partial waves the
structure of the SRG flow equations simplifies considerably in the
off-diagonal relative momentum-space basis (see
e.g.~\cite{Szpigel:2010bj} for an explicit derivation and details) for
which the following normalization in the completeness relation will be
assumed (here and in what follows we use units such that
$\hbar=c=M=1$, where $M$ is the nucleon mass):
\begin{eqnarray}
\frac{2}{\pi} \int p^2 dp | p \rangle \langle p |={\bf 1} \; .
\label{norm}
\end{eqnarray}
\noindent
Inserting this into Eq.~(\ref{wegfloweq}) with the Wilson generator, $G_s=T$, the flow equation for the SRG evolution of the $NN$ potential is given by (we drop the partial-wave quantum numbers for simplicity)
\begin{eqnarray}
\frac{dV_{s}(p,p')}{ds}=-(\epsilon_p-\epsilon_p')^2 \; V_{s}(p,p')+\frac{2}{\pi} \int_{0}^{\infty}dq \; q^2\;
(\epsilon_p+\epsilon_p'-2 \epsilon_q)\; V_{s}(p,q)\; V_{s}(q,p') \; ,
\label{flowWil}
\end{eqnarray}
\noindent
where $\epsilon_p =\langle p |T| p \rangle =p^2$. The flow equation for the SRG evolution with the Wegner generator, $G_s=H^{D}_s$, reads
\begin{eqnarray}
\frac{dV_{s}(p,p')}{ds} =-(\epsilon_p-\epsilon_p')\left[e_{p}(s) - e_{p'}(s)\right]V_{s}(p,p')+ \frac{2}{\pi} \int_{0}^{\infty}dq \; q^2\;
\left[e_{p}(s) + e_{p'}(s) - 2 e_q(s)\right]\; V_{s}(p,q)\; V_{s}(q,p') \; ,
\label{flowWeg}
\end{eqnarray}
\noindent
where $e_{p}(s) = \langle p |H^{D}_s| p \rangle=p^2 + V_s(p,p)$. As we
see, these are integro-differential equations which cannot generally be
solved analytically (see however~\cite{Szpigel:1999gf,Jones:2013cba})
and require a massive computational effort to be solved numerically. We will
tackle this problem below by the traditional discretization method of
the continuum~\footnote{However, the usefulness of discretizing the
  continuum equations goes beyond the practical need of numerically
  implementing the scattering problem; it provides a theoretical
  bridge between the energy-shift in the spectrum and the scattering
  phase-shifts. We will see that this becomes a crucial aspect when
  the number of grid points is reduced to a minimum.}.

%%%

\subsection{Operator space}

Most compact features of the SRG formalism can be best appreciated
within an operator space setup. We start with some remarks from the
operator theory for finite-dimensional operators in order to introduce
some notation (see e.g. the standard
textbook~\cite{kolmogorov1999elements}). For operators acting on a
Hilbert space endowed with a scalar product of states $\psi$ and
$\varphi$ such as $\langle \psi, \varphi \rangle$, we can also define
a further scalar product between operators $A$ and $B$, namely
\begin{eqnarray}
\langle A, B \rangle = {\rm Tr} \left[ A^\dagger B \right] \; .
\end{eqnarray}
\noindent
From here we define the Frobenius norm as usual
\begin{eqnarray}
|| A ||^2 = \langle A, A \rangle = {\rm Tr} \left[ A^\dagger A \right] \; ,
\end{eqnarray}
\noindent
and hence the induced distance between
operators as
\begin{eqnarray}
d[A,B]=||A-B||  \; .
\end{eqnarray}
\noindent
The norm can also be defined as
\begin{eqnarray}
|| A || = {\rm sup}_{||\psi||=1} || A \psi || \; ,
\end{eqnarray}
\noindent
where the standard scalar product induced norm, $|| \psi || =+ \sqrt{
  \langle \psi, \psi \rangle}$ has been introduced.  Of course, this
requires finite norm operators $||A|| < \infty$ which is unfortunately
not the case for the usual unbounded operators in quantum mechanics,
such as the kinetic energy. Therefore an ultraviolet cutoff $\Lambda$
is generally assumed to operate here. Using the partial-wave relative
momentum-space normalization metric given in Eq.~(\ref{norm}) we have
\begin{eqnarray}
\langle \psi , \psi \rangle = \frac{2}{\pi}\int_0^\infty dp ~ p^2  ~ |\psi(p)|^2 \; .
\end{eqnarray}
\noindent
Thus, the kinetic energy would have infinite norm , $ || T || \to
\Lambda^2/M \to \infty$ if everything is taken literally. However,
note that the continuum SRG flow equations do not need the kinetic energy
to be bound, but rather the potential energy. So, the implicit
assumption is that at very high energies the kinetic energy dominates
over the potential energy and hence SRG flow equations are well defined
provided the value of the potential does not boundlessly grow.  The
first problems we encounter with all these properties are: i) the fact
that the momentum-space basis spans a continuum set and ii) the
operators are unbounded. This difficulty is circumvented in practice
by using a finite grid in momentum-space and also introducing a
high-momentum cutoff $p_{{\rm max}} = \Lambda$ which will be assumed
below.

%%%

\subsection{Isospectral flow and fixed points}

In this section we briefly review the concepts of isospectral flow and fixed points incorporated in the
SRG evolution and the variational interpretation in operator space.

The isospectrality of the SRG flow equation becomes evident from the
trace invariance property of the evolved hamiltonian $H_s$. For any
choice of the SRG generator $G_s$ we have a unitary transformation and
hence
\begin{eqnarray}
{\rm Tr}~ H_s^n = {\rm Tr}~ H_0^n \; ,
\end{eqnarray}
for any integer $n$. This property follows directly from the
commutator structure of the SRG flow equation plus the regulator
assumption\footnote{Mathematically, such a property is ill defined in
  the continuum limit since even for $n=1$ one has $ {\rm Tr} (V_s)=
  \int_0^\infty p^2 V_s(p,p) = \int_0^\infty r^2 dr V(r,r)$ which for
  a local potential $V(r,r') = V(r) \delta (r-r')$ diverges as the
  momentum cutoff. Also, the trace of a commutator ${\rm Tr}[A,B]$
  only vanishes when both ${\rm Tr}(AB)$ and ${\rm Tr}(BA)$ are
  finite, as the choice $A=p$ and $B=x$ clearly illustrates, since
  $[p,x]= - i\hbar$ and hence ${\rm Tr}[p,x]= - i \hbar ~{\rm Tr}~
  {\bf I} = \infty$.}. Indeed, using the cyclic properties of the trace we get
\begin{eqnarray}
\frac{d}{ds}{\rm Tr}~H_s^n=n~{\rm Tr}\left(H_s^{n-1}\frac{d H_s}{ds}\right)=n~{\rm Tr}\left(H_s^{n-1}[\eta_s,H_s]\right)=0 \, .
\end{eqnarray}

Fixed points of the SRG evolution correspond to stationary solutions of the flow equation, Eq.~(\ref{eq:SRGflow}),
\begin{eqnarray}
\frac{d H_s}{ds} = [[G_s, H_s],H_s] = 0 \; .
\label{eq:fixed}
\end{eqnarray}
\noindent
This condition implies that there is a basis in which both $[G_s, H_s]$ and $H_s$ become simultaneously diagonal at a fixed point. The question is what choices of the generator $G_s$ actually drive the hamiltonian $H_s$ to the diagonal form. For generators $G_s$ which satisfy $d/ds({\rm Tr}~ G_s^2)=0$ and using the cyclic properties of the trace and the invariance of ${\rm Tr}~ H_s^n$, we get that
\begin{eqnarray}
\frac{d}{ds}{\rm Tr}~(H_s-G_s)^2=2~{\rm Tr}~[G_s,H_s]^2=-2~{\rm Tr}\left[\left(i~[G_s,H_s]\right)^\dagger \left(i~[G_s,H_s]\right)\right]\le 0  \; ,
\end{eqnarray}
\noindent
because $ A \equiv i~[G_s,H_s] = A^\dagger$ is a self-adjoint operator and therefore $A^\dagger A $ is a semi-definite positive operator. Since ${\rm Tr}~ (H_s-G_s)^2$ is positive but its derivative is negative, the limit $s \to \infty$ exists and corresponds to the infrared fixed point of the SRG evolution ($\lambda \to 0$), at which the hamiltonian $H_s$ becomes diagonal. Thus, the SRG flow equation just provides a continuous procedure to diagonalize the initial hamiltonian $H_0 = T + V_0$.

In the case of the SRG evolution with the Wilson generator, $G_s=T$, the flow equation is given by
\begin{eqnarray}
\frac{d H_s}{ds}=\frac{d V_s}{ds} = [[T,H_s],H_s]=[[T,V_s],T+V_s] \; .
\end{eqnarray}
\noindent
We then get for $||V_s|| \equiv {\rm Tr}~V_s^2$ that
\begin{eqnarray}
\frac{d}{ds}{\rm Tr}~V_s^2 = 2~{\rm Tr}~[T,V_s]^2 = -2~{\rm Tr}\left[\left(i~[T,V_s]\right)^\dagger \left(i~[T,V_s]\right)\right]
\le 0  \; .
\end{eqnarray}
\noindent
As a consequence of this and using the unitary equivalence, $H_s = T+V_s = U_sH_0 U_s^\dagger$, we get that
\begin{eqnarray}
0 < {\rm Tr}~ V_s^2 \le {\rm Tr}~ V^2_0  \; .
\end{eqnarray}
\noindent
Therefore there must be a minimum value obtained at the limit $s \to
\infty$ which also implies in $[T,V_{s \to \infty}]=0$ due to the stationary
condition on the derivative for the infrared fixed point. Hence
\begin{eqnarray}
\lim_{s \to \infty} {\rm Tr}~V_s^2 = \min_{V_s} {\rm Tr}~ V_s^2 \Big|_{H_s= T+V_s = U_s H_0 U_s^\dagger}  \; .
\end{eqnarray}
\noindent
Thus, in the infrared limit $s \to \infty$ ($\lambda \to 0$) the SRG
evolution with the Wilson generator yields asymptotically to the
smallest potential, in the Frobenius norm sense, giving the same
spectrum as the initial potential $V_0$ and commuting with the kinetic
energy $T$, i.e.  being diagonal in momentum-space. This is a rather
interesting result as it provides a working definition on the ``size''
of the potential, and moreover different potentials can actually be
compared using the distance between the operators induced by the
Frobenius norm.  Furthermore, this can be interpreted as a
quantitative measure of the off-shellness of the interaction. In the
partial-wave relative momentum-space basis the orbital degeneracy
induced by the $(2/\pi) p^2 dp$ integration measure in the Frobenius
norm has two complementary effects. While it suppresses the
contribution to the norm from low-energy states it also enhances the
contribution from high-energy components. Thus, minimizing the
potential along the SRG evolution trajectory transfers very
efficiently high-energy components into low-energy components. This
provides a working scheme where any short-distance, or equivalently
high-momentum core, becomes softer. It is fair to say that this is the
main reason why SRG methods have become popular in realistic nuclear
applications. For completeness let us mention that there is an
alternative interpretation of softness of the interaction not based on
the Frobenius norm, and based on the insightful work of
Weinberg~\cite{Weinberg:1963zza} and taken up by recent studies from
several viewpoints~\cite{Bogner:2006tw,Perez:2014bua} where the
repulsive character of the interaction at short-distance plays a key
role.  At present the connection between these two alternatives, while
suggesting different measures of the softness, is somewhat vague and
we will not dwell into it here. In our case, we will deal with a
potential toy model where the repulsive piece is absent from the start
(see Section~\ref{sec:toy}).

In the case of the SRG evolution with the Wegner generator, $G_s=H^{D}_s$, the flow equation is given by
\begin{eqnarray}
\frac{d H_s}{ds} = \frac{d V_s}{ds}=[[H^{D}_s,H_s],H_s] \; .
\end{eqnarray}
\noindent
Then, in this case we get for $|| H_s-H^{D}_s ||^2 \equiv {\rm Tr}~ (H_s-H^{D}_s)^2$ that
\begin{eqnarray}
\frac{d}{ds}{\rm Tr}~(H_s-H^{D}_s)^2=2~{\rm Tr}~[H^{D}_s,H_s]^2=-2~{\rm Tr}~\left[\left(i~[H^{D}_s,H_s]\right)^\dagger \left(i~[H^{D}_s,H_s]\right)\right]
\le 0  \; ,
\end{eqnarray}
\noindent
such that $|| H_s-H^{D}_s || \to 0$ and so
\begin{eqnarray}
\lim_{s \to \infty} H_s = H^{D}_s = \min_{H_s} || H_s-H^{D}_s ||  \; ,
\end{eqnarray}
\noindent
which just shows that the Wegner generator minimizes the distance to
its diagonal matrix-elements keeping the eigenvalues of the original
Hamiltonian. Of course, if the hamiltonian becomes diagonal the
eigenvectors cannot be free momentum eigenstates.

Finally, for the block-diagonal generator we define two orthogonal projection operators $P+Q=1$ which split the states
below or above a given momentum scale $\Lambda_{BD}$. In this case the flow equation is given by
\begin{eqnarray}
\frac{d H_s}{ds} = [[P H_s P + Q H_s Q ,H_s],H_s]  \; .
\end{eqnarray}
\noindent
We then get that the evolution makes the asymptotic hamiltonian block-diagonal hence minimizing the off-diagonal matrix-elements,
\begin{eqnarray}
\lim_{s \to \infty} H_s = P H_s P+Q H_s Q = \min_{H_s} || H_s-P H_s P-Q H_s Q||  \; .
\end{eqnarray}

%%%

\subsection{Discrete equations}
\label{sec:discrete}

Only in few cases the SRG operator equations can be handled in the
continuum~\cite{Szpigel:1999gf,Jones:2013cba}. In this section we
analyze the details of the implementation of the SRG when a finite
dimensional reduction of the model space is imposed. The particular
case of a momentum grid discretization of the continuum will be
discussed specifically in a later section. For simplicity we will consider a basis of eigenstates $| n \rangle$ of the kinetic energy operator $T$ on a finite $N$-dimensional Hilbert space ${\cal H}_N$, namely $T | n \rangle = \epsilon_n | n \rangle~(n=1,2,\dots,N)$, and assume that the corresponding spectrum of eigenvalues $\epsilon_n$ is non-degenerate (similar to what happens in the partial-wave relative momentum-space basis for which $\epsilon_n=p_n^2$). Thus, the matrix-elements of the hamiltonian $H= T + V$ in this basis read $H_{nm}\equiv \langle n |H| m \rangle=\delta_{nm}\epsilon_n + V_{nm}$.

The discrete SRG flow equations for the matrix-elements of the hamiltonian in the case of the Wilson generator can be written in the form
\begin{eqnarray}
\frac{d H_{nm}(s)}{ds}&=&\sum_{k} (\epsilon_n+\epsilon_m-2 \epsilon_k) H_{nk}(s) H_{km}(s)\nonumber\\
 &=&-(\epsilon_n-\epsilon_m)\left[e_{n}(s)-e_{m}(s)\right] H_{nm}(s) +
\sum_{k \neq n,m} (\epsilon_n+\epsilon_m-2 \epsilon_k) H_{nk}(s) H_{km}(s)  \; ,
\label{eq:discWil}
\end{eqnarray}
\noindent
where $H_{nm}(s)=\delta_{nm}\epsilon_n + V_{nm}(s)$ and $e_{n}(s)\equiv H_{nn}(s)=\epsilon_n + V_n(s)$. In the case of the Wegner generator the discrete SRG flow equations read
\begin{eqnarray}
\frac{d H_{nm}(s)}{ds} &=& \sum_k \left[e_{n}(s)+ e_{m}(s)- 2 e_{k}(s) \right] H_{nk}(s) H_{km}(s) \nonumber\\
&=&-\left[e_{n}(s)-e_{m}(s)\right]^2 H_{nm}(s) +
\sum_{k \neq n,m} \left[e_{n}(s)+e_{m}(s)-2 e_{k}(s)\right] H_{nk}(s) H_{km}(s) \; .
\label{eq:discWeg}
\end{eqnarray}
\noindent
These equations are to be solved with the boundary conditions $H_{nm}(s)|_{_{s \to 0}}\equiv H_{nm}(0)=\delta_{nm}\epsilon_n + V_{nm}(0)$.

The fixed points of the SRG evolution with a given generator $G_s$ correspond to the stationary solutions of the SRG flow equations for the matrix-elements of the hamiltonian,
\begin{equation}
\frac{d H_{nm}(s)}{ds} = \langle n| [[G_s,H_s],H_s]|m \rangle=0 \; ,
\end{equation}
\noindent
which implies, for both the Wilson ($G_s=T$) and the Wegner ($G_s=H^{D}_s$) generators, that in the infrared limit $s \to \infty$ ($\lambda \to 0$) the hamiltonian $H_s$ becomes diagonal \footnote{One should note that the stationary condition in operator form, Eq.~(\ref{eq:fixed}), in principle just requires that at a fixed point both $[G_s,H_s]$ and $H_s$ become diagonal in the same basis, not necessarily the one in which the generator $G_s$ is diagonal. However, we can show that the condition $[[G_s,H_s],H_s]=0$ also implies that $[G_s,H_s]=0$. Taking a discrete basis in which $H_s$ is diagonal, i.e. $H_{\alpha \beta}(s)= \delta_{\alpha \beta}H_{\alpha}(s)$, we have
\begin{eqnarray}
\langle \alpha |[G_s, H_s] | \beta \rangle=\sum_\gamma \left[G_{\alpha \gamma}(s) H_{ \gamma \beta}(s) - H_{\alpha \gamma}(s)
G_{\gamma \beta}(s)\right]=  G_{\alpha \beta}(s)\left[H_{\beta}(s)- H_{\alpha}(s)\right] \; . \nonumber
\end{eqnarray}
\noindent
Thus, for $\alpha \neq \beta$ we get, in the absence of degeneracies, that $\langle \alpha |[G_s, H_s] | \beta \rangle = 0$ only if $G_{\alpha \beta}(s)= 0$ and so both the hamiltonian $H_s$ and the generator $G_s$ are diagonal in the same basis. Of course, this becomes a trivial result for generators $G_s$ which by definition are diagonal in the basis of eigenstates of the kinetic energy operator $T$, such as the Wilson and the Wegner generators.}. Thus, we have that
\begin{eqnarray}
\lim_{s\to \infty} H_{nm}(s)=\delta_{nm} E_n\; ,
\end{eqnarray}
\noindent
where $\{E_n\}_{n=1}^N$ denotes the spectrum of discrete eigenvalues of the hamiltonian $H_s$ obtained in the infrared limit $s \to \infty$, which are given by
\begin{eqnarray}
E_n \equiv \lim_{s\to \infty}e_n(s)=\epsilon_n + V_n(s \to \infty)\; .
\end{eqnarray}
%%

%%%

\subsubsection{Ordering of the spectrum induced by the SRG evolution}

As we have shown, the SRG evolution with both the Wilson and the Wegner generators on a finite $N$-dimensional discrete space have infrared fixed points at which the hamiltonian becomes diagonal. Thus, we have two interpolating SRG trajectories between the initial bare hamiltonian, $H_0$, and the final one, $H_{s \to \infty}$. From this point of view, there seems to be no conceptual difference between the SRG evolution with both generators, since the isospectrality of the SRG flow equation guarantees the invariance of the spectrum of eigenvalues of the hamiltonian $H_{s}$. This naive argument overlooks an important detail: the fact that through the SRG evolution the basis of eigenstates $|\psi_{\alpha} (s)\rangle$ of the hamiltonian $H_s$ is actually changing, i.e.
\begin{equation}
U_{s}|\psi_{\alpha} (s=0)\rangle = |\psi_{\alpha} (s)\rangle \;,
\end{equation}
\noindent
and thus the isospectrality does not necessarily fix the final
ordering of the eigenvalues which is obtained in the infrared limit $s
\to \infty$ ($\lambda \to 0$). This is not a peculiar feature of the
SRG evolution; it is shared by {\it any} diagonalization procedure. In
the Gauss elimination method~\cite{demidovich1973computational}, for
instance, one makes an arbitrary choice on how to reduce the original
matrix to a diagonal form in a finite number of steps; the re-ordering
of eigenstates has to be over-imposed at the end by hand, arbitrarily
choosing one of the possible permutations of the eigenvalues. Such a
re-ordering of eigenstates is equivalent to a unitary
transformation. However, unlike the conventional diagonalization
methods, the SRG evolution can be interpreted as a {\it continuum}
diagonalization through the flow parameter $s$, and thus an infinite
number of steps is involved. As a consequence, the SRG evolution
induces a very specific ordering of the eigenstates. This, of course,
does not provide {\it exact} results since in practice the infrared
limit $s \to \infty$ is never reached, but the errors scale
exponentially with the flow parameter $s$ and are of order ${\cal
  O}[e^{-{\rm min}~(s~\epsilon_n^2)}]$. The crucial aspect is that
since on a finite dimensional discrete space the SRG flow equation
becomes a set of non-linear first-order coupled differential equations
for the matrix-elements of the evolved hamiltonian $H_{nm}(s)$ with
the boundary conditions $H_{nm}(s)|_{_{s \to 0}}\equiv H_{nm}(0)$, the
uniqueness of the solution implies that just one particular ordering
of the eigenvalues takes place asymptotically in the infrared limit $s
\to \infty$, which may depend on the choice of the SRG generator
$G_s$.

Let us consider the initial bare hamiltonian $H_0$ in the basis of eigenstates $|n \rangle$ of the kinetic energy operator $T$ on a finite $N$-dimensional Hilbert space ${\cal H}_N$. The spectrum of eigenvalues $\{\epsilon_n\}_{n=1}^N$ of the operator $T$ is assumed to be non-degenerate and arranged in ascending order, i.e. $\epsilon_1 < \epsilon_2 < \dots < \epsilon_N$. If we denote by $\{E_{\alpha}^0\}_{\alpha=1}^N$ the spectrum of $N$ discrete eigenvalues of $H_0$ obtained by any conventional matrix diagonalization method and arranged (by hand) in ascending order similarly to the spectrum of eigenvalues of $T$, i.e. $E_1^0 < E_2^0 < \dots < E_N^0$, then we have that the final ordering of the spectrum of eigenvalues $\{E_n\}_{n=1}^N$ of the SRG evolved hamiltonian $H_{s}$, obtained asymptotically in the infrared limit $s \to \infty$, is specifically given by
\begin{eqnarray}
\{E_n\}_{n=1}^N \equiv \{E_{\pi(\alpha)}^0\}_{\alpha=1}^N \; ,
\end{eqnarray}
\noindent
where $\pi(\alpha)$ is one of the $N!$ possible permutations of the spectrum of $H_0$.

It is important to note that in the {\it continuum} diagonalization of
the hamiltonian through the SRG evolution, the correspondence between
the kinetic energies $\epsilon_n$ and the diagonal matrix-elements of
the potential $V_n(s)$ is maintained all the way along the SRG
trajectory, as one can clearly see from the expression for the
diagonal matrix-elements of the hamiltonian, $H_{nn}(s)=\epsilon_n +
V_n(s)$. Thus, we have a unique well-defined pairing of the kinetic
energies $\epsilon_n$ with the eigenvalues $E_n$ obtained
asymptotically in the infrared limit $s \to \infty$, namely
$E_n=H_{nn}(s \to \infty)=\epsilon_n + V_n(s \to \infty)$, which is
indeed what determines the specific ordering of the spectrum induced
by the SRG evolution. One should also note that depending on which
particular final ordering of the eigenvalues takes place in the
infrared limit $s \to \infty$ there may be crossing amongst diagonal
matrix-elements of the hamiltonian $H_{nn}(s)$ along the SRG
trajectory. On the other hand, in the conventional diagonalization
methods there are $N!$ different ways of pairing the kinetic energies
$\epsilon_n$ with the eigenvalues $E_{\alpha}^0$, corresponding to the
possible orderings of the spectrum. As we have discussed in
Ref.~\cite{Arriola:2014aia}, this is a crucial issue to establish an
isospectral definition of the phase-shift based on an energy-shift
approach, which necessarily involves a prescription to order the
eigenvalues $E_{\alpha}^0$ and set their pairing with the kinetic
energies $\epsilon_n$.

In section \ref{sec:numerical} we will illustrate through numerical
calculations that when bound-states are allowed by the interaction the
phase-shifts evaluated using the energy-shift approach do not comply
to Levinson's theorem \cite{Ma:2006zzc} at low-energies if a naive
pairing is set just by ordering the spectrum of eigenvalues
$E_{\alpha}^0$ in ascending order as the kinetic energies
$\epsilon_n$; we further show that the specific ordering of the
spectrum induced by the SRG evolution with the Wegner generator in the
infrared limit $s \to \infty$ remarkably provides a prescription to
evaluate the phase-shifts using the energy-shift approach which allows
to obtain results that fulfill Levinson's theorem in the presence of
bound-states. In section \ref{sec:nucl} we dicuss the inequivalent
behaviour of both the Wilson and the Wegner generators in simple
variational calculations beyond some critical SRG cutoff approaching
the infrared limit.

%%%

\subsubsection{Stability analysis of the infrared fixed points}

As we have pointed out, the uniqueness of the solution of the SRG flow
equations on a finite $N$-dimensional discrete space implies that a
very specific final ordering of the eigenvalues of the hamiltonian is
obtained in the infrared limit $s \to \infty$ ($\lambda \to 0$). Of
course, the uniqueness of the solution further implies that the
infrared fixed point to which the SRG evolved hamiltonian is steadily
driven must be asymptotically stable. In this section we will carry
out a perturbative stability analysis of the infrared fixed points for
both the Wilson and the Wegner generators, which is based on a
linearization of the SRG flow equations similar to that described in
Ref.~\cite{Brockett1991}, as an attempt to determine {\it a priory}
the final ordering of the spectrum induced by the SRG evolution in the
infrared limit. As we will see, the perturbative analysis is
well-succeeded in predicting the ordering of the spectrum only in the
case of the Wilson generator, although the results for both generators
are consistent with the analytical proof of diagonalization of the SRG
evolved hamiltonian presented in
Refs~\cite{Furnstahl:2012fn,Jurgenson:2009}.

Let us consider a perturbation of the matrix-elements of the SRG
evolved hamiltonian $H_{nm}(s)$ near an infrared fixed point $H_{nm}
(s \to \infty) = \delta_{nm} E_n$, namely
\begin{eqnarray}
H_{nm} (s) = \delta_{nm} E_n + \Delta H_{nm}(s) \; ,
\end{eqnarray}
\noindent
with the matrix-elements of the perturbation $\Delta H_{nm}(s)$ required to satisfy the condition $\Delta H_{nm}(s \to \infty) = 0$.

By inserting the perturbed hamiltonian into the SRG flow equation and taking only the terms to first-order in the perturbation we can obtain a set of linearized flow equations for the matrix-elements $\Delta H_{nm}(s)$. In the case of the Wilson generator $G_s=T$, we get from Eq.~(\ref{eq:discWil})
\begin{eqnarray}
\frac{d \Delta H_{nm}(s)}{ds}  = - (\epsilon_n - \epsilon_m) (E_n - E_m) \Delta H_{nm}(s) \; .
\end{eqnarray}
\noindent
The solutions of these equations for the diagonal matrix-elements ($n = m$) are just constants which actually vanish due to the condition $\Delta H_{nn}(s \to \infty)=0$, namely $\Delta H_{nn}(s) \equiv C_{nn}=0$, while for the non-diagonal matrix-elements ($n \neq m$) the solutions are given by
\begin{eqnarray}
\Delta H_{nm}(s) = C_{nm} e^{-s(\epsilon_n - \epsilon_m) (E_n - E_m)} \; ,
\end{eqnarray}
\noindent
where the integration constants $C_{nm}$ will depend on the initial conditions set for the matrix-elements $H_{nm}(s)$ of the perturbed hamiltonian. Thus, we have
\begin{eqnarray}
H_{nm} (s)  = E_n \delta_{nm}+ C_{nm} e^{-s(\epsilon_n-\epsilon_m)(E_n-E_m)} + \dots \; .
\end{eqnarray}
\noindent
Clearly, in the absence of degeneracies the off-diagonal matrix-elements will be ensured to monotonically decrease with $s$ provided $(\epsilon_n - \epsilon_m) (E_n-E_m) > 0$. This implies that from all $N!$ possible final orderings of the spectrum only the one in which the eigenvalues $E_n$ are arranged according to the kinetic energies $\epsilon_n$, i.e. in ascending order, corresponds to an asymptotically stable infrared fixed point. Thus, in the Wilson generator case the perturbative stability analysis to first-order in the perturbation allows to predict beforehand the specific final ordering of the spectrum induced by the SRG evolution in the infrared limit.

In the case of the Wegner generator $G_s=H^{D}_s$, we get from Eq.~(\ref{eq:discWeg})
\begin{eqnarray}
\frac{d \Delta H_{nm}(s)}{ds} = -(E_n-E_m)^2 \Delta H_{nm}(s) \; .
\end{eqnarray}
\noindent
which yields the solution
\begin{eqnarray}
\Delta H_{nm}(s) = C_{nm} e^{-s(E_n-E_m)^2} \; ,
\end{eqnarray}
\noindent
such that
\begin{eqnarray}
H_{nm} (s)  = E_n \delta_{nm}+ C_{nm} e^{-s(E_n-E_m)^2} + \dots \; .
\end{eqnarray}
\noindent
As one can see, in this case the off-diagonal matrix-elements will monotonically decrease with $s$ (in the absence of degeneracies) regardless the ordering of the eigenvalues and so in principle all $N!$ possible final orderings of the spectrum correspond to asymptotically stable infrared fixed points. Thus, in the Wegner generator case the specific final ordering of the spectrum induced by the SRG evolution in the infrared limit cannot be determined {\it a priory} through the perturbative stability analysis to first-order in the perturbation and we have to rely on numerical analysis.

%%%

\section{A simple toy model for the $NN$ interaction}
\label{sec:toy}

We intend to explore the infrared limit of the SRG evolution $(\lambda
\to 0)$. Quite generally, the equations to be discussed involve heavy
numerical calculations with a discretized continuum spectrum in the
case of nuclear physics. The problem is that most of the
high-precision potentials, which fit $NN$ scattering data up to the
pion-production threshold ($\sqrt{m_\pi M_N} \sim 350~{\rm MeV}$),
have a very long tail in momentum space which requires many points and
large momentum cutoffs not to miss important contributions. As a
consequence the flow equation gets extremely stiff as the SRG
cutoff $\lambda$ approaches zero, such that the computational effort
becomes unduly expensive. Actually, there is currently a gap in SRG
calculations below $\lambda \sim 1~ {\rm fm}^{-1}$ for high-precision
\cite{Wiringa:1994wb,Stoks:1994wp} and
ChEFT~\cite{Entem:2003ft,Epelbaum:2004fk} $NN$ potentials.

Therefore, we will illustrate most of our points by using a simple toy
model for the $NN$ interaction which reduces the computational time
and allow us to push the SRG evolution towards the infrared limit in a
way which is not practical with realistic interactions. The simplicity
of the toy model does not affect the main features of the $NN$
interaction in the $S$-wave channels and gives us the opportunity to
investigate the infrared fixed point of the SRG flow equations with
any generator. Here we concentrate on the Wilson and the Wegner
generators.

Our framework is defined by a toy model for the $NN$ force in the $^1S_0$ and the $^3S_1$ channels which consists of a separable gaussian potential, given by
\begin{equation}
V(p, p') = C~ g_L(p) g_L(p') = C~ \exp \left[ -\left( p^{2}+p'^2 \right)/L^2 \right] \; .
\label{gaupot}
\end{equation}
\noindent
The parameters $C$ and $L$ are determined from the solution of the LS equation for the on-shell transition matrix $T$ by fitting the experimental values of the parameters of the Effective Range Expansion (ERE) to second order in the on-shell momentum, i.e. the scattering length $a_0$ and the effective range $r_e$. Namely, we solve the partial-wave LS equation for the $T$-matrix with the toy model potential,
\begin{equation}
T(p,p';E)=V(p,p')+\frac{2}{\pi}\; \int_{0}^{\infty} \; dq \; q^2 \;
\frac{V(p,q)}{E-q^2+i \; \epsilon} \; T(q,p';E) \; ,
\end{equation}
\noindent
where $E$ is the scattering energy, and match the resulting on-shell $T$-matrix to the ERE expansion,
\begin{eqnarray}
T^{-1}(k,k;k^2)=-\left[-\frac{1}{a_0}
+\frac{1}{2}~r_e~k^2 + {\cal O}(k^4)-i~k \right] =-\left[k~{\rm cot}~\delta(k)-i~k \right]\; ,
\label{eq:ERE}
\end{eqnarray}
\noindent
where $k=\sqrt{E}$ is the on-shell momentum in the CM frame and $\delta(k)$ stands for the phase-shifts. In order to avoid the numerical integration on a contour in the complex plane, we switch to the LS equation for the partial-wave reactance matrix $K$ with standing-wave boundary conditions,
\begin{equation}
K(p,p';k^2)=V(p,p')+\frac{2}{\pi}\; {\cal P}\int_{0}^{\infty} \; dq \; q^2 \;
\frac{V(p,q)}{k^2-q^2} \; K(q,p';k^2) \; ,
\label{LSKNN}
\end{equation}
\noindent
where ${\cal P}$ denotes the Cauchy principal value. The relation between the $K$-matrix and the $T$-matrix on-shell is given by
\begin{equation}
K^{-1}(k,k;k^2) = T^{-1}(k,k;k^2) \; - \; i \, k = - \; k \cot \delta(k)\; .
\label{KTrel}
\end{equation}
\noindent
Following the method introduced by Steele and Furnstahl~\cite{Steele:1998un}, we fit the difference between the inverse on-shell $K$-matrices corresponding to the toy model potential and the ERE expansion to an interpolating polynomial of degree $k^2$ for a spread of very small on-shell momenta ($k \le 0.1~{\rm fm^{-1}}$), namely
\begin{eqnarray}
\Delta K^{-1}=K^{-1}(k,k;k^2) - K^{-1}_{\rm ERE}(k,k;k^2) = A_0 + A_2~k^2 \; .
\end{eqnarray}
\noindent
and then minimize the coefficients $A_0$ and $A_2$ with respect to the variations of the parameters $C$ and $L$.

In the case of the separable gaussian potential toy model, given by
Eq.~(\ref{gaupot}), it is straightforward to determine the
phase-shifts $\delta(k)$ from the solution of the LS equation for the
$T$-matrix using the {\it ansatz}
\begin{eqnarray}
T(p,p';k^2) = g_L(p) ~t(k)~ g_L(p') \; ,
\end{eqnarray}
\noindent
where $t(k)$ is called the reduced on-shell $T$-matrix. This leads to
the simple relation (valid for separable potenials only)
\begin{eqnarray}
 k \cot \delta(k)= - \frac{1}{V(k,k)} \left[1- \frac{2}{\pi}\;
{\cal P}\int_0^\infty dq~q^2 \frac{1}{k^2-q^2} V(q,q) \right] \; .
\label{phasetoy}
\end{eqnarray}

In Table \ref{tab:toy} we display the values of the parameters for the toy model potential in the $^1S_0$ and the $^3S_1$ channels used in our numerical calculations, which are adjusted to reproduce the corresponding experimental values of $a_0$ and $r_e$ by the method described above.

\begin{table}[h]
\caption{Parameters for the toy model potential in the $^1S_0$ and the $^3S_1$ channels used in the numerical calculations.}
\vskip 0.3cm
\centering
\renewcommand\arraystretch{1.3}
\begin{tabular}{c c c c c c c }
\hline\hline
 - & $a_0~({\rm fm})$  &  $r_e~({\rm fm})$  & $ C~({\rm fm})$ & $1/L^{2}~({\rm fm}^{2})$  \\
\hline\hline
$^1S_0$   & -23.74  & 2.77 & -1.915884  &  0.6913 \\
$^3S_1$   & 5.42    & 1.75 & -2.300641  &  0.4151 \\
\hline
\end{tabular}
\renewcommand\arraystretch{1.0}
\label{tab:toy}
\end{table}

\begin{figure}[t]
\begin{center}
\includegraphics[width=7.5cm]{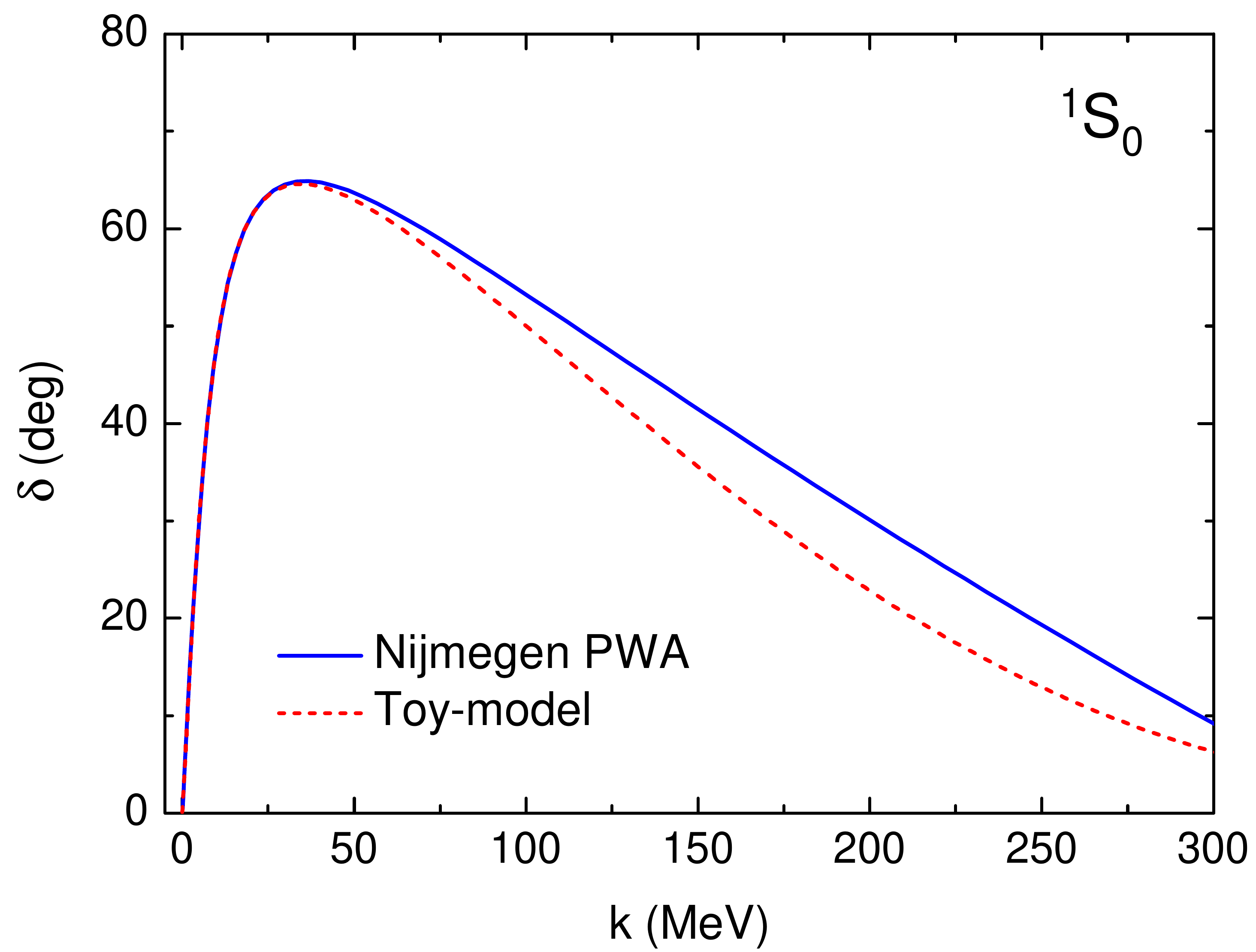}\hspace{0.5cm}
\includegraphics[width=7.65cm]{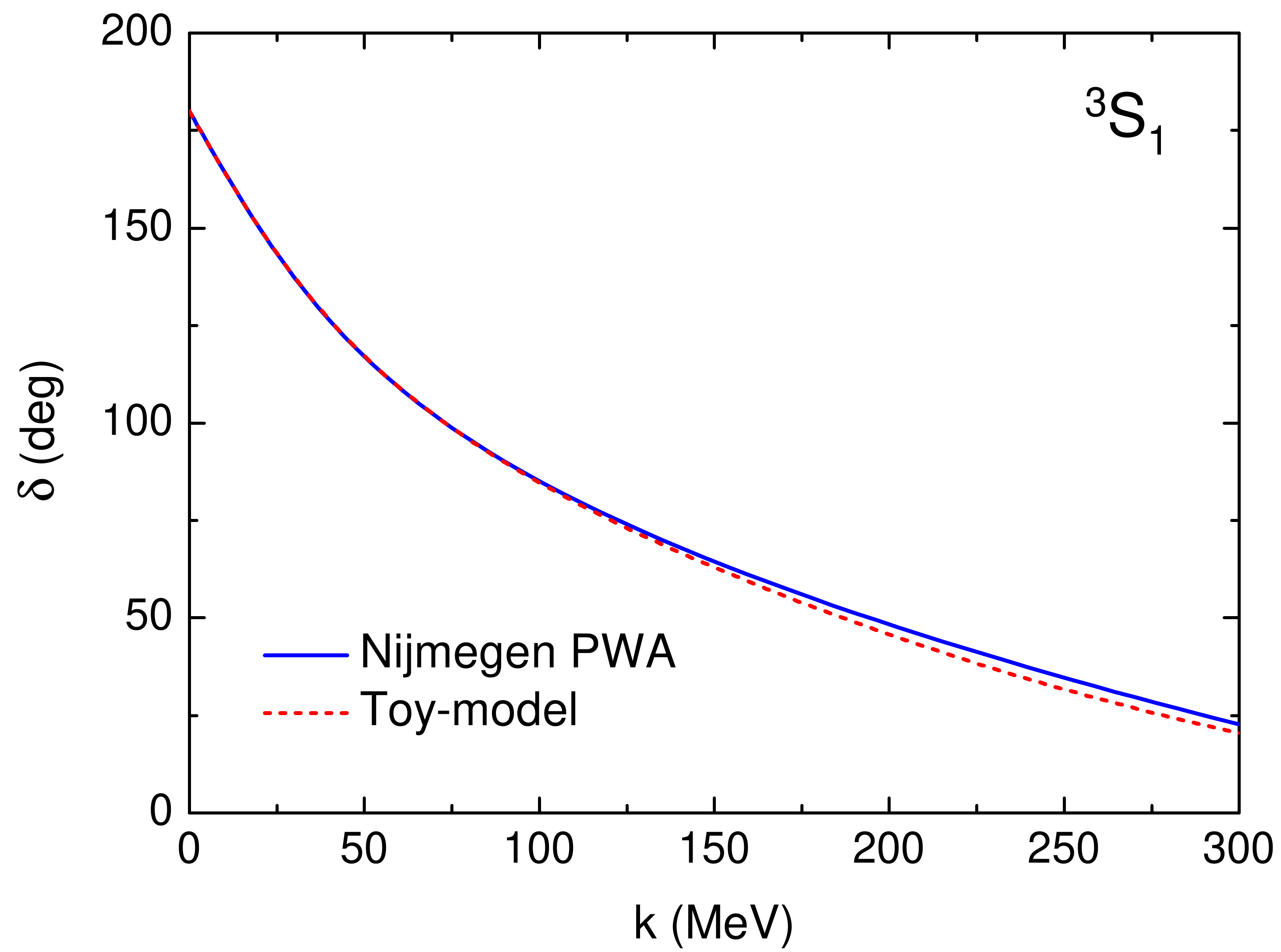}
\end{center}
\caption{Phase-shifts for the toy model potential in the $^1S_0$ and the $^3S_1$ channels compared to the results obtained from the Nijmegen PWA \cite{Stoks:1993tb}.}
\label{fig:1}
\end{figure}

The phase-shifts for the toy model potential in the $^1S_0$ and the
$^3S_1$ channels evaluated from Eq.~(\ref{phasetoy}) are shown in
Fig.~\ref{fig:1}, together with the results obtained from the 1993
Nijmegen partial-wave analysis (PWA) \cite{Stoks:1993tb} or the more
recent 2013
upgrades~\cite{Perez:2013mwa,Perez:2013jpa,Perez:2013oba,Perez:2014yla}. As
one can see, despite the simplicity of the potential and the fact that
the $^3S_1$ channel is not treated as a coupled channel, our toy model
for the $NN$ interaction provides a reasonable qualitative description
of the $S$-wave phase-shifts. Moreover, the on-shell $T$-matrix for
the $^3S_1$ channel toy model potential has a pole located at an
imaginary momentum $k= i~\gamma=i~0.2314~{\rm fm^{-1}}$, corresponding
to a satisfactory Deuteron binding-energy $B_d \simeq 2 ~ {\rm MeV}$.

%% Toy phases versus Nijmegen

%%

%%%

\section{SRG on a finite momentum grid}
\label{sec:SRG-grid}

\subsection{Momentum-space grid}

For most cases of interest the SRG flow equations are solved numerically on a finite momentum grid with $N$ integration points
$p_n$ and weights $w_n$ ($n=1, \dots N$) by implementing a high-momentum ultraviolet (UV) cutoff, $p_{\rm max}=\Lambda$, and an
infrared (IR) momentum cutoff, $p_{\rm min} = \Delta p$. The integration rule becomes
\begin{eqnarray}
\int_{\Delta p}^\Lambda dp f(p) \to \sum_{n=1}^N w_n f(p_n) \; .
\end{eqnarray}

Taking Chebychev-Gauss points~\cite{demidovich1973computational}, for
example, we get after re-scaling to the interval $[0,\Lambda]$,
\begin{eqnarray}
p_n = \frac{\Lambda}{2}\left\{ 1- \cos \left[ \frac{\pi}N (n-1/2) \right] \right\}
\;  , \qquad
w_n =  \frac{\Lambda}{2} \frac{\pi}{N} \sin \left[\frac{\pi}N (n-1/2) \right] \; ,
\end{eqnarray}
\noindent
and thus
\begin{eqnarray}
p_{\rm min} = p_1 =  \Lambda \sin^2 \left( \frac{\pi}{4N}\right)
\;  , \qquad
p_{\rm max} = p_N =  \Lambda \sin^2 \left[ \frac{\pi}{2N} (N-1/2) \right]  \; .
\end{eqnarray}
\noindent
For a large grid and for $n << N$ we have $p_n= \Lambda (\pi n/2N)^2/2$ which differs from the spherical box quantization. As it is well known, this grid choice guarantees an exact result for polynomials in p to order $M \le N$, i.e.
\begin{eqnarray}
\int_{\Delta p}^\Lambda dp \frac{P_M (p)}{\sqrt{\Lambda^2-p^2}}  = \sum_{n=1}^N w_n \frac{P_M(p_n)}{\sqrt{\Lambda^2-p_n^2}} \; .
\end{eqnarray}
\noindent
The completeness relation in discretized partial-wave relative momentum-space basis is given by
\begin{eqnarray}
\frac{2}{\pi}\sum_{n=1}^N w_n ~p_n^2 ~| p_n \rangle \langle p_n |={\bf 1} \; .
\label{completeness}
\end{eqnarray}

Once we have the finite momentum grid we may seek to diagonalize the hamiltonian, whose matrix-elements read
\begin{eqnarray}
H(p_n,p_m)=p_n^2 ~ \delta_{nm} + \frac{2}{\pi}w_n~p_n^2 ~V_{nm}  \; ,
\label{gridev}
\end{eqnarray}
\noindent
where $V_{nm}$ denotes the matrix-elements of the potential $V(p_n,p_m)$. The discrete eigenvalue equations on the finite momentum grid are given by
\begin{eqnarray}
p_n^2 ~\psi_\alpha (p_n) + \frac{2}{\pi}\sum_k w_k ~p_k^2 ~V_{nk}
~\psi_\alpha(p_k) = P_\alpha^2 ~ \psi_\alpha (p_n) \; ,
\label{gridev}
\end{eqnarray}
where $P_\alpha^2$ and $\psi_\alpha (p_n)$ stand respectively for the eigenvalues and the eigenfunctions of the hamiltonian.

The weight factors $w_n$ complicate the discrete eigenvalue equations and hide the hermiticity of the hamiltonian. Making a change of variables
\begin{eqnarray}
\psi_\alpha(p_n) = \frac{\varphi_\alpha(p_n)}{p_n \sqrt{2 w_n/\pi}} \; ,
\end{eqnarray}
\noindent
we have
\begin{eqnarray}
p_n^2 ~\varphi_\alpha (p_n) + \frac{2}{\pi}\sum_k \sqrt{w_n} ~p_n
~V_{nk} ~\sqrt{w_k}~ p_k ~ \varphi_\alpha(p_k) = P_\alpha^2 ~
\varphi_\alpha (p_n) \; .
\end{eqnarray}
\noindent
Thus the square-integrability condition reads
\begin{eqnarray}
\frac2{\pi}\sum_{n=1}^N w_n p_n^2 | \psi(p_n) |^2 = \sum_{n=1}^N |
\varphi(p_n) |^2 \; ,
\end{eqnarray}
\noindent
and the multiplication of operators corresponds to the plain matrix multiplication
\begin{equation}
\langle A, B \rangle = \left(\frac{2}{\pi}\right)^2 \int_0^\infty
p^2 dp \int_0^\infty k^2 dk A(k,p)^* B(p,k) \to
\left(\frac{2}{\pi}\right)^2 \sum_{n,k=1}^N w_n w_k p_n^2 p_k^2
A_{nk}^* B_{kn} \; \equiv \sum_{n,m=1}^N \bar{A}^*_{nk} ~
\bar{B}_{kn} \; ,
\end{equation}
\noindent
where
\begin{eqnarray}
\bar{A}_{nm} = \frac{2}{\pi} ~ p_n \sqrt{w_n} ~ A_{nm} ~p_m \sqrt{w_m} \; .
\end{eqnarray}
\noindent
In terms of this new basis the discrete eigenvalue equations become very simple,
without the disturbing extra factors, as it was shown in Section~\ref{sec:discrete}. Note that in the free case, $V_{nm}=0$, we have $\psi_\alpha(p_n)\sim \delta_{\alpha,n}$, so that labeling states as distorted states
stemming from a given free state, $p_n \to P_n $, means we have an {\it interacting momentum}. Moreover, in the particular case of a diagonal potential, i.e. $~V_{nm}=\delta_{nm}~V_n~$, Eq.~(\ref{gridev}) yields a relation which resembles a self-energy equation,
\begin{eqnarray}
p_n^2  ~ + ~  \frac{2}{\pi} ~ w_n ~ p_n^2 ~  V_n = P_n^2  \; ,
\label{self}
\end{eqnarray}
\noindent
where $P_n^2$ denotes the eigenvalues of the hamiltonian.

%%%

\subsection{SRG flow equations}

The SRG flow equations on the finite momentum grid for the matrix-elements of the potential in a given partial wave follow from inserting the completeness relation, Eq.~(\ref{completeness}), into the SRG flow equation in operator form, Eq.~(\ref{wegfloweq}). In the case of the Wilson generator we get
\begin{equation}
\frac{d V_{nm}(s)}{ds} = -(p_n^2-p_m^2)^2 V_{nm}(s) +
\frac{2}{\pi}\sum_{k}w_k~ p_k^2 (p_n^2+p_m^2-2 p_k^2) V_{nk}(s) V_{km}(s)  \; ,
\end{equation}
\noindent
and in the case of the Wegner generator,
\begin{equation}
\frac{d V_{nm}(s)}{ds} = -(p_n^2-p_m^2)\left[e_n(s) - e_m(s)\right] V_{nm}(s)+\frac{2}{\pi}\sum_{k}w_k~ p_k^2 \left[e_n(s)+e_m(s)-2 e_k(s) \right] V_{nk}(s) V_{km}(s)  \; ,
\end{equation}
\noindent
where $e_n(s)=p_n^2 + \frac{2}{\pi} w_n ~p_n^2 ~V_s(p_n,p_n)$

The eigenvalue problem on the finite momentum grid for the SRG-evolved hamiltonian $H_{\lambda} = U_{\lambda}~H~U_{\lambda}^{\dagger}$ may be formulated as
\begin{eqnarray}
H_{\lambda}~ | \alpha,\lambda \rangle = P_\alpha^2 ~| \alpha,\lambda \rangle   \; ,
\end{eqnarray}
\noindent
where $| \alpha,\lambda \rangle$ are the eigenstates of $H_{\lambda}$. The matrix-elements of the hamiltonian $H_{\lambda}$ and the corresponding eigenfunctions $\psi_{\alpha,\lambda}$ in momentum-space representation are given respectively by
\begin{equation}
H_{\lambda}(p_n,p_m) = p_n^2 ~ \delta_{nm} + \frac{2}{\pi} w_n ~p_n^2 ~V_{\lambda}(p_n,p_m)  \;
\end{equation}
\noindent
and
\begin{equation}
\psi_{\alpha,\lambda} (p_n) = \langle n | \alpha,\lambda \rangle  \, .
\end{equation}
\noindent
A bound-state with (negative) eigenvalue $P_\alpha^2 =- B_\alpha$
corresponds to a pole in the scattering amplitude at imaginary
momentum $P_\alpha= i~ \gamma$. Because of the commutator structure of the SRG flow equation the isospectrality property still holds on the grid and therefore the eigenvalues $P^{2}_{\alpha}$ of $H_{\lambda}$ are $\lambda$-independent, i.e.
\begin{eqnarray}
\frac{d P_\alpha}{d \lambda}=0  \; .
\end{eqnarray}

As we have discussed in section \ref{sec:discrete}, although the
isospectrality property ensures that the eigenvalues of the
hamiltonian remain invariant along the SRG trajectory, the ordering of
the states may depend on the specific choice of the SRG generator
$G_s$. This yields to an interesting feature observed systematically
in SRG calculations when bound-states are allowed by the
interaction. In the presence of bound-states (real or spurious),
hamiltonians evolved using the Wilson and the Wegner generators begin
to flow differently when the SRG cutoff $\lambda$ approaches some
critical momentum $\Lambda_c$, which corresponds to the threshold
scale where the bound-state emerges \cite{Glazek:2008pg,Wendt:2011qj}
(one should note that the critical momentum $\Lambda_c$ is distinct
from the characteristic bound-state momentum scale $\gamma$). This is
explicitly verified by a block-diagonal generator
analysis~\cite{Arriola:2014fqa} (see also an analysis based just on
scattering information~\cite{Arriola:2014aia} using Effective Field
Theory ideas). In the Wilson generator case the bound-state remains
coupled to the low-momentum scales as $\lambda$ approaches
$\Lambda_c$, such that the bound-state eigenvalue is pushed towards
the lowest momentum available on the grid, $p_1$, which corresponds to
the IR momentum cutoff $\Delta p$. Moreover, matrix-elements of the
potential at low-momentum diverge when $\Delta p \to 0$ in order to
force the bound-state eigenvalue to smaller momenta, such that the SRG
evolution may become numerically unstable. In the Wegner generator
case the bound-state decouples from the low-momentum scales as
$\lambda$ approaches $\Lambda_c$, being placed on the diagonal of the
hamiltonian as an isolated negative eigenvalue at a momentum
$p_{n_{\rm BS}}$ between the lowest momentum on the grid $p_1$ and the
critical momentum $\Lambda_c$. As pointed out in
Ref.~\cite{Wendt:2011qj}, the {\it a priori} determination of the
position at which the bound-state is placed on the diagonal when using
the Wegner generator is still an open problem. It is important to note
that when the SRG cutoff $\lambda$ is kept well above the critical
momentum $\Lambda_c$ or in the absence of bound-states the SRG
evolutions using the Wilson and the Wegner generators are nearly
identical, a behavior that can be traced to the dominance of the
kinetic energy.

%%%

\section{The scattering problem on a finite momentum grid}
\label{sec:scatt-grid}

So far we have dealt with the evolution of the hamiltonian according
to the SRG flow equations, which take a manageable form in the discrete
momentum basis. In this section we exploit the same basis to analyze
the scattering problem and point out several specific features
unveiled in our previous work~\cite{Arriola:2014aia} and extending them.

Of course, the momentum grid should in principle be arbitrary. In the
next section we consider a physically motivated procedure based on
considering a spherical box which introduces a quantization condition
for the momentum and also allows a straightforward determination of
the phase-shifts at the quantized momentum values. However, this
finite box scheme makes the choice of the grid to depend on the
angular momentum. Afterwards we will consider the standard and fixed
gaussian integration grid which is valid for all partial waves,
although the identification of the phase-shifts is less obvious.

%%%

\subsection{LS equation and phase-shifts}

For the case of nuclear hamiltonians and more specifically for the
$NN$ situation, the scattering problem requires defining phase-shifts.
Unfortunately, the standard methods to handle the scattering problem
in momentum-space do it through the numerical solution of the LS
equation, which introduce a momentum grid but do not allow to compute
the phase-shifts on the {\it same} grid points. In fact there appear
two different grids where a difference between the so-called
observation points in discrete energy values and the momenta on
discrete values are not in one-to-one correspondence. For our
purposes, it becomes necessary to use a method where the phase-shifts can
be defined on the momentum grid {\it without} extra observation
points.

In operator form, the LS equation for the $T$-matrix of a two-body system reads
\begin{eqnarray}
T(E) &=& V + V~G_{0}^{+}(E)~T(E) \; ,
\label{LS}
\end{eqnarray}
\noindent
where $E$ is the scattering energy and $G_{0}^{+}(E) = (E - h_0 + i
\epsilon)^{-1}$ is the free Green's function with outgoing-wave
boundary conditions given in terms of the free hamiltonian $h_0$
(which corresponds to the kinetic energy operator $T$).

The LS equation for the $T$-matrix on the finite momentum grid is
obtained by taking the matrix-elements in the discretized partial-wave
relative momentum-space basis and reads
\begin{eqnarray}
T_{nm}(p) = V_{nm} + \frac{2}{\pi} \sum_{k=1}^N w_k \frac{p_k^2}{p^2-p_k^2+ i \epsilon} V_{nk} T_{km}(p) \; ,
\end{eqnarray}
\noindent
where $p^2$ is the scattering energy $E$. Note that the momentum $p$
corresponding to the observation point can in principle be chosen
independently of the momentum grid. The on-shell limit is obtained by
taking $p=p_n$ on the grid. We switch to the $R$-matrix, which on the
grid yields the LS equation for the half-on-shell amplitude
\begin{eqnarray}
R_{nm}(p_n) = V_{nm} + \frac{2}{\pi} \sum_{k \neq n} w_k \frac{p_k^2}{p_n^2-p_k^2} V_{nk} R_{km}(p_n) \; ,
\label{LSreaction}
\end{eqnarray}
\noindent
where the excluded sum embodies the principal value prescription of the continuum version, Eq.~(\ref{LSKNN}). This equation can be solved by inversion for any grid point $p_n$ and thus we obtain the LS phase-shift~\footnote{The explicit, and apparently pedantic, LS label in $\delta^{\rm LS}(p_n)$ is not redundant; it does reflect the dependence on the choice of the LS equation, the grid and the number of grid points. This will become clear below when we show the inequivalence of these phase-shifts on a finite momentum grid along the SRG trajectories.}
\begin{eqnarray}
- ~ \frac{\tan \delta^{\rm LS}(p_n)}{p_n} = R_{nn} (p_n) \; .
\label{LSphase}
\end{eqnarray}
%%

%% Toy phases on the grid

\begin{figure}[t]
\begin{center}
\includegraphics[width=7.5cm]{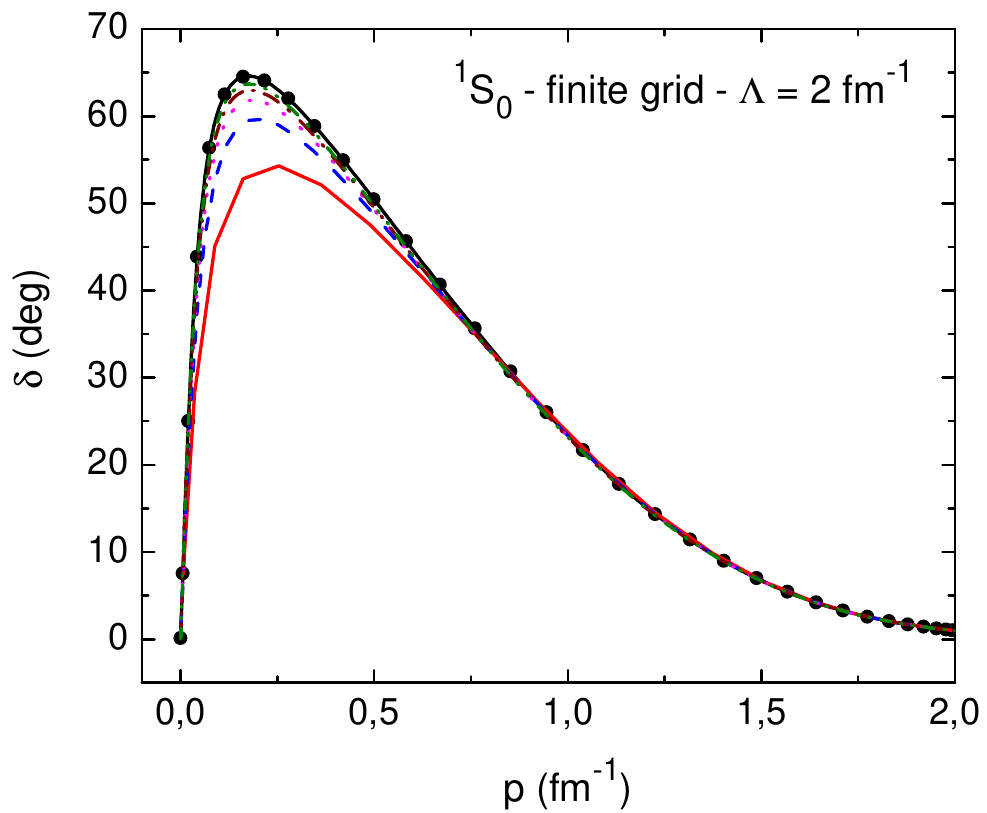} \hspace{0.5cm}
\includegraphics[width=7.65cm]{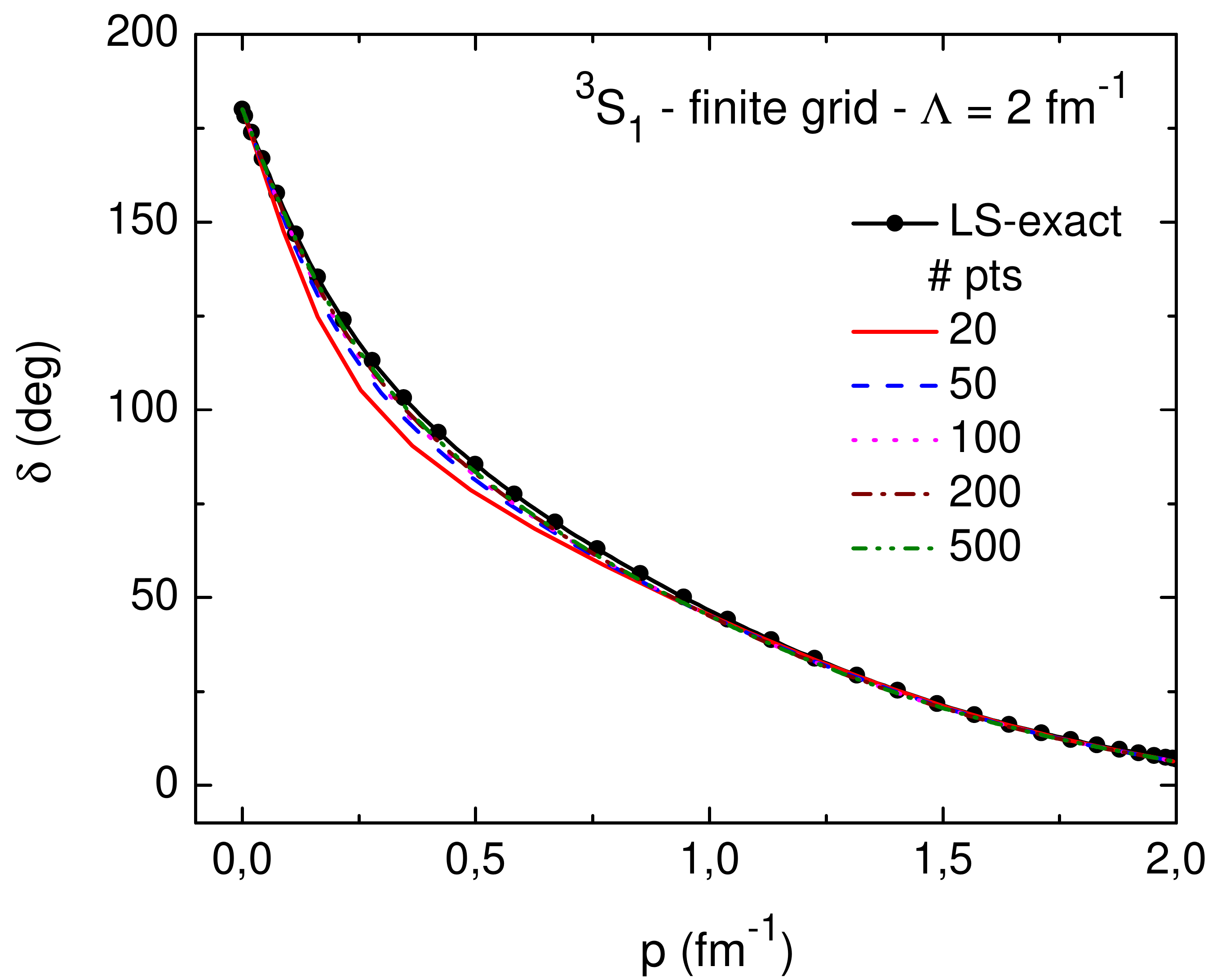}
\end{center}
\caption{Phase-shifts for the toy model separable gaussian potential in the $^1S_0$ and the $^3S_1$ channels evaluated from the solution of the LS equation on the grid {\it at the grid points} for a high-momentum UV cutoff $\Lambda=2~{\rm fm}^{-1}$ and different number of grid points $N$. For comparison, we also show the exact phase-shifts obtained from the solution of the standard LS equation in the continuum limit ($N \to \infty$).}
\label{fig:2}
\end{figure}

In Fig.~\ref{fig:2} we show the phase-shifts for the toy model separable gaussian
potential in the $^1S_0$ and the $^3S_1$ channels evaluated from the solution of the LS equation on the grid {\it at the observation grid points} for a high-momentum UV cutoff $\Lambda=2~{\rm fm}^{-1}$ and different number of grid points $N$, compared to the exact results obtained from the solution of the standard LS equation in the continuum limit ($N \to \infty$).

%%%

\subsection{Energy-shift in a box: coordinate space}

While the momentum grid provides a viable alternative to
solve both the SRG and the scattering equations, since it
corresponds to a discretization of the integration rule, it is useful
to consider a physically motivated situation where this momentum
discretization naturally appears. Thus, we analyze the effects of
having a large spherical box in three dimensions for a central
potential. The case of a cubic box, which would be relevant for finite
volume lattice calculations, has the disadvantage of breaking the
symmetries of the problem and will be studied elsewhere.

Let us consider for simplicity a local and finite range potential
$V(r)$ with a range larger than a certain distance $a$. For the
$S$-wave at large distances, i.e. outside of the range of the
potential, we have
\begin{eqnarray}
u(r) = \sin\left[pr+\delta(p)\right] \; .
\end{eqnarray}
\noindent
If the size of the box $L$ is larger than the range $a$ of the potential then
the condition of the particle confined to be inside the finite box is
\begin{eqnarray}
u(L)=0 \to p_n L + \delta(p_n) = n \pi  \;  ,
\end{eqnarray}
\noindent
which is a quantization condition for the allowed momentum $p$.  Thus,
the effect of the finite box is to discretize the momenta.  In the
free case we have $p_{n,0} L = n \pi $, and then we may write the
shift in momentum due to the interaction as follows
\begin{eqnarray}
\Delta p_n= p_n -p_{n,0} = - \frac1{L} \delta(p_n) \to - \frac1{L} \delta(p) +{\cal O}(L^{-2}) \; ,
\label{eq:p-box}
\end{eqnarray}
\noindent
where in the last line we assume a large box and a smooth behavior of
the phase-shift. This means that the spectrum gets a distortion which
scales inverse proportional to the size of the box. Note that this
result is strictly valid for large box sizes. In the case where the
box size is not much larger than the range of the potential, but still
$L \gtrsim a $ the quantization condition is actually a non-linear
eigenvalue equation.

The interesting thing is that integrals over momentum get discretized and one has
\begin{eqnarray}
\int \frac{dp}{2\pi} f(p) \to \frac1{L} \sum_n f( \frac{n \pi}{L}) \; .
\end{eqnarray}
\noindent
For instance if we sum over functions of eigenvalues of the hamiltonian
\begin{eqnarray}
\sum_n F(E_n) = \sum_n F(E_n^{(0)}) - \frac{1}{L} \sum_n F'(E_n^{(0)}) \delta(p_{n,0})  \; ,
\end{eqnarray}
\noindent
which can be re-written as
\begin{eqnarray}
{\rm Tr} F(H) = {\rm Tr} F(H_0) - \int \frac{dp}{2\pi}  F'(E) \delta(p)  \; .
\label{eq:trace}
\end{eqnarray}

The procedure above can be generalized to all partial waves, since far from the potential we have
\begin{eqnarray}
u_l(r)/r = \sin \delta_l(p) ~ j_l( pr) -  \sin \delta_l(p) ~ y_l(pr) \; ,
\end{eqnarray}
\noindent
so that the box boundary condition yields
\begin{eqnarray}
\sin \delta_l(p) ~ j_l( p L) -  \sin \delta_l(p) ~ y_l(p L)=0  \; ,
\end{eqnarray}
\noindent
and obviously the momenta are quantized but also depend on angular momentum.

%%%

\subsection{From energy-shift to phase-shift}

The previous analysis motivated Lifshitz many years ago, when studying
the impurities in a crystal lattice, to propose a generalization of
the quantization condition based on an arbitrary finite momentum
grid. This approach, also pursued independently by
DeWitt~\cite{DeWitt:1956be} and Fukuda and
Newton~\cite{Fukuda:1956zz}, has been taken over in nuclear physics
for the analysis of the few-body scattering problem by Kukulin and
collaborators in a series of recent and remarkable
works~\cite{kukulin2009discrete,Rubtsova:2010zz,Pomerantsev:2014ija,Rubtsova:2015owa}. The
basic idea is to solve the scattering problem without solving the
scattering equations but just the eigenvalue equations for the
discretized hamiltonian in the continuum limit. One should note that
while the box quantization condition corresponds to a specific
momentum grid choice, $p_n = n \pi/L $ and $w_n = \pi/L$ in the free
case, a general momentum grid does not have an obvious meaning.

One important feature of the previous form of the LS equation for the
$R$-matrix on a finite momentum grid, Eq.~(\ref{LSreaction}), is that
the Cauchy principal value for the integral becomes an excluded
summation which is not symmetric since the distance above and below
the pole differs, i.e. $p_{n+1}^2-p_n^2 \neq p_n^2-p_{n-1}^2$. This
induces some finite asymmetry which is ultimately pointwise removed in
the continuum since it scales as $w_n$. However, when summing up the
spectrum with weight $w_n$ the net effect is finite. Thus, if we
consider the matrix $R_{nm}(p)$ as a function of the external
momentum $p=p_n$ on the grid corresponding to the scattering energy
$p_n^2$, we get trivially from Eq.~(\ref{LSreaction}) that
$R_{nn}(p_n)=V_n$ for a diagonal potential $V_{nm}=\delta_{nm}~V_n$,
such that Eq.~(\ref{self}) can be written in the form
\begin{eqnarray}
p_n^2+ \frac2\pi w_n p_n^2 ~ R_{nn}(p_n) = P_n^2  \;  .
\end{eqnarray}
\noindent
A comparison of this relation in the continuum limit, $w_n \to 0$, with the treatment presented in the previous section for the large box quantization in coordinate space suggests the identification
\begin{equation}
R_{nn}(p_n)  = \frac{\pi}{2} ~ \frac{\left( P^2_n - p^2_n \right)}{ w_n~p^2_n}  \to
-\frac{\delta^{\rm ES}(p_n)}{p_n} \; ,
\end{equation}
\noindent
which provides an energy-shift formula to define the phase-shift,
\begin{equation}
\delta^{\rm ES}(p_n) =  - \frac{\pi}{2} ~ \frac{\left( P^2_n - p^2_n \right)}{ w_n~p_n} \; .
\label{ES-formula}
\end{equation}
\noindent
Of course, both the energy-shift (ES) and the LS definitions for the
phase-shift lead to the same results in the continuum limit, $w_n \to
0$. However, for a finite momentum grid there are important
differences which are relevant for the SRG evolution, namely, for
unitary transformations on the grid the energy-shift definition of the
phase-shift provides invariant results along the SRG trajectory unlike
the LS phase-shift (see below). One should also note that the
energy-shift formula for the phase-shift is not unique. We could
instead use a momentum-shift formula very much in agreement with the
finite box treatment. Quite generally, all these possible formulas
become equivalent in the continuum limit, but for a finite number $N$
of grid points we should expect differences. It would be highly
interesting to design formulas where improved accuracies for finite
$N$ are displayed. We leave this interesting topic for future
research. Finally, it is important to note that for a finite momentum
grid with dimension $N$ there are $N!$ possible permutations for the
eigenvalues $P^2_n$ obtained from the diagonalization of the
hamiltonian and so the evaluation of phase-shifts using the
energy-shift approach necessarily involves an ordering prescription.

%%%

\subsection{Phase-equivalence and inequivalence}
\label{phaseineq}

As pointed out before the original motivation for the application of the SRG in the context of nuclear physics was to
soften the $NN$ interaction while keeping the phase-shifts invariant. In our previous work \cite{Arriola:2014aia} we have shown that on a finite momentum grid phase-equivalence does not hold along the SRG trajectory for the LS definition, while for the isospectral definition based on the energy-shift formula phase-equivalence is preserved. In this section we elaborate more on this issue.

First, let us consider the eigenstate decomposition of the LS equation for the $T$-matrix on a finite momentum grid with the SRG-evolved potential, which can be written in operator form as
\begin{eqnarray}
T(E)= V_{\lambda} + V_{\lambda} (p^2 - H_{\lambda}+ i \epsilon)^ {-1} V_{\lambda} \; .
\end{eqnarray}
\noindent
Taking the matrix-elements in the discretized partial-wave relative momentum-space basis and inserting a complete set of eigenstates of the SRG-evolved hamiltonian $H_{\lambda}=U_{\lambda}HU_{\lambda}^\dagger$, namely
\begin{eqnarray}
\sum_{\alpha} ~| \alpha,\lambda \rangle \langle \alpha,\lambda|={\bf 1} \; ,
\label{completeness}
\end{eqnarray}
\noindent
we obtain
\begin{eqnarray}
T_{nm}(p) = V_{nm}(\lambda) +\left(\frac{2}{\pi}\right)^2 \sum_{k=1}^N \sum_{l=1}^N \left(w_k ~p_k^2 \right) \left(w_l~p_l^2\right)~\sum_{\alpha} V_{nk}(\lambda)\frac{\psi_{\alpha,\lambda} (p_k)~
\psi_{\alpha,\lambda} (p_l)}{p^2-P_\alpha^2 + i~\epsilon}V_{lm}(\lambda) \; ,
\label{LSsrgT}
\end{eqnarray}
\noindent
Switching to the $R$-matrix at the grid points $p=p_n$, we get
\begin{eqnarray}
R_{nm}(p_n) = V_{nm}(\lambda) +\left(\frac{2}{\pi}\right)^2 \sum_{k\neq n}\sum_{l} \left(w_k ~p_k^2 \right) \left(w_l~p_l^2\right)~\sum_{\alpha} V_{nk}(\lambda)\frac{\psi_{\alpha,\lambda} (p_k)~
\psi_{\alpha,\lambda} (p_l)}{p_n^2-P_\alpha^2}V_{lm}(\lambda) \; .
\label{LSsrgR}
\end{eqnarray}
\noindent
A surprising result is that for unitarily equivalent operators on the
grid, such as those generated by the SRG evolution, the phase-shifts
computed from the solution of the LS equation are {\it not} the same;
i.e. $\delta_{N}^{\rm LS}(p_n,H) \neq \delta_{N}^{\rm LS}(p_n,
U_{\lambda}HU_{\lambda}^\dagger) $.  This is evident from
Eqs.~(\ref{LSsrgT}) and (\ref{LSsrgR}), where we can see explicitly
that the LS phase-shifts depend {\it both} on the eigenvalues
$P_\alpha^2$, which are isospectral, and the eigenfunctions
$\psi_{\alpha,\lambda} (p_{k})$, which in turn change along the SRG
evolution trajectory and so are {\it not} independent of the SRG
cutoff $\lambda$. Of course, in the continuum limit one has $\lim_{N
  \to \infty} \delta_{N}^{\rm LS}(p_n, H) = \lim_{N \to
  \infty}\delta_{N}^{\rm LS}(p_n, U_{\lambda}H~U_{\lambda}^\dagger)
$.

On the other hand, the SRG-evolved potential becomes diagonal in the infrared limit $\lambda \to 0$,
\begin{eqnarray}
\lim_{\lambda\to 0} V_{nm}(\lambda) = \delta_{nm}~V_{n}(\lambda \to 0)   \;  ,
\end{eqnarray}
\noindent
and hence we get
\begin{eqnarray}
V_{n}(\lambda \to 0) = R_{nn}(p_n) \; ,
\end{eqnarray}
\noindent
such that the infrared fixed-point of the SRG evolution can be related to the energy-shift formula Eq.~(\ref{ES-formula}),
\begin{eqnarray}
V_{n}(\lambda \to 0) = \frac{\left(P^2_n - p^2_n \right)}{\frac{2}{\pi} ~ w_n~p^2_n}= -\frac{\delta^{\rm ES}(p_n)}{p_n} \; ,
\label{eigenpot}
\end{eqnarray}
\noindent
thus providing an isospectral definition for the phase-shifts which therefore preserves phase-equivalence along the SRG trajectory. We refer to $V_n(\lambda \to 0)$ and $\delta^{\rm ES}(p_n)$ as the "eigenpotential" and the "eigenphases" since they are obtained directly from the eigenvalues of the hamiltonian, $P^2_n$. It should be mentioned that in the coupled channel case (which actually occurs for the $^3S_1 - ^3D_1$ Deuteron channel) one needs not only the eigenvalues but also the eigenvectors
to determine both the eigenphases and the mixing angles, as discussed in Refs.~\cite{kukulin2009discrete,Rubtsova:2010zz}.

%%%

%%% Needs to be revised

%%%

\subsection{Trace Identities and Levinson's theorem}

According to Parisi~\cite{parisi1988statistical} the trace identities
are one of the most beautiful results in quantum mechanics. One of the
advantages of using the momentum grid basis is the almost
straightforward derivation of the finite energy sum rules in potential
scattering found in Ref.~\cite{Graham:2001iv} which are a
generalization of the trace
identities~\cite{parisi1988statistical}. While the analyticity of
interactions was extensively used there, our derivation is almost
completely trivial, and we will just show here some particular examples
grasping the basic essence of the approach.

We can compute the following expressions involving the trace of the
hamiltonian $H=T+V$ on a finite momentum grid with $N$ integration points
$p_n \in [0,\Lambda]$ and weights $w_n$ ($n=1, \dots N$),
\begin{eqnarray}
I_k = {\rm Tr}~ (T+V)^k - {\rm Tr}~ T^k \; ,
\end{eqnarray}
\noindent
for any integer $k$. On one hand we can saturate with the spectrum including possible
bound-states and on the other hand with the momentum states, which yields
\begin{eqnarray}
I_k
&=& \sum_\alpha (-\gamma_\alpha^2)^k + \sum_{n} \frac{2}{\pi} w_n p_n^2 p_n^{2k-2}~
k\left[-\frac{\delta^{\rm ES}(p_n)}{p_n}\right]= \sum_{n,m,\dots} \left[p_n^2 \delta_{nm}+\frac{2}{\pi}w_n p_n^2 V_{nm}\right]^k - \sum_{n} \left[p_n^2\right]^k\; .
\label{tracek}
\end{eqnarray}

For $k=1$ we obtain
\begin{eqnarray}
I_1 = \sum_\alpha (-\gamma_\alpha^2) + \sum_{n} \frac{2}{\pi} w_n
p_n^2 \left[-\frac{\delta^{\rm ES}(p_n)}{p_n}\right] = \sum_{n} \frac{2}{\pi}
w_n p_n^2 V_n \; ,
\end{eqnarray}
\noindent
which in the continuum yields
\begin{eqnarray}
I_1  \to \sum_\alpha (-\gamma_\alpha^2) + \frac{2}{\pi} \int_0^\Lambda dp~p^2 \left[-\frac{\delta(p)}{p}\right] =
\frac{2}{\pi} \int_0^\Lambda dp~p^2 V(p,p) \; .
\end{eqnarray}
\noindent
In terms of the first Born approximation for the phase-shifts, $\delta^B(p)$, this can be written as
\begin{eqnarray}
\sum_\alpha (-\gamma_\alpha^2) + \frac{2}{\pi} \int_0^\Lambda dp~p^2 \left[\frac{-\delta(p)+ \delta^B (p)}{p}\right] =0  \; .
\end{eqnarray}
\noindent
This still holds for the finite $\Lambda$. Taking the infinite cutoff
limit $\Lambda \to \infty$ we get the form of the finite energy sum
rule of Ref.~\cite{Graham:2001iv}. Note that our result does not
require taking the $\Lambda \to \infty$ limit and hence no assumption
on analyticity is required for finite $\Lambda$.

For $k=2$ we obtain
\begin{eqnarray}
I_2 = \sum_\alpha (-\gamma_\alpha^2)^2 + 2 \sum_{n} \frac{2}{\pi}w_n p_n^4 \left[-\frac{\delta^{\rm ES}(p_n)}{p_n}\right] = 2 \sum_{n}
\frac{2}{\pi} w_n p_n^4 V_n + \sum_{n,m} \left[\frac{2}{\pi}w_n p_n^2 V_{nm}\right]\left[ \frac{2}{\pi} w_m p_m^2 V_{mn}\right] \; .
\end{eqnarray}
\noindent
Similarly to the case $k=1$, in the continuum this can be written as of Ref.~\cite{Graham:2001iv} in terms of the second Born approximation for the phase-shifts, $\delta^{B2} (p)$,
\begin{eqnarray}
\sum_\alpha (-\gamma_\alpha^2)^2 + \frac{4}{\pi} \int_0^\Lambda dp~p^4 \left[\frac{-\delta(p)+ \delta^{B2} (p)}{p}\right] =0 \; .
\end{eqnarray}

The case $k=0$ is a bit more tricky and obtained by taking the limit $k \to 0$ in Eq.~(\ref{tracek}),
\begin{eqnarray}
I_0= \sum_\alpha (-\gamma_\alpha^2)^0 + \lim_{k \to 0} \sum_{n} \frac{2}{\pi} w_n p_n^2 p_n^{2k-2} k
\left[-\frac{\delta^{\rm ES}(p_n)}{p_n}\right]=\sum_{n} (1)_{T+V}-\sum_{n} (1)_{T}= 0 \; .
\label{trace0}
\end{eqnarray}
\noindent
The last equality just expresses the conservation of dimensions
regardless of the interaction. The first term ater the first equality
is simply the number of bound-states. The term with the phase-shift
can be transformed by using the identity $ 2 k p_n^{2k-2} = d p_n^{2k}
/dp_n^2$ going to the continuum and integrating by parts. Thus in the
continuum limit
\begin{eqnarray}
N_B  + \frac{\delta(\Lambda)-\delta(0)}{\pi}=0  \; ,
\end{eqnarray}
\noindent
which becomes the usual Levinson's theorem for $\Lambda \to \infty$.
Reinstating the finite momentum grid we get
\begin{eqnarray}
\delta(p_1)-\delta(p_N)=N_B ~\pi  \; ,
\end{eqnarray}
\noindent
where $N_B$ is the number of bound-states allowed by the
interaction.
%Note that the last equality in Eq.~(\ref{trace0})
%expresses the conservation of the number of states regardless of the
%interaction.
%%%

\section{Numerical results}
\label{sec:numerical}
%%%

\subsection{SRG evolution}

We solve the SRG flow equations for the toy model potential numerically on a finite momentum grid with $N=50$ gaussian points and a high-momentum UV cutoff $\Lambda=2~{\rm fm}^{-1}$. The discretization of the momentum-space leads to a system of $N^2$ non-linear first-order coupled differential equations which is solved numerically by using an adaptative variable-step fifth-order Runge-Kutta algorithm. The SRG cutoff $\lambda$ is varied in a range from $0.05$ to $2.0~{\rm fm}^{-1}$.

\begin{figure}[t]
\begin{center}
\includegraphics[width=7.5cm]{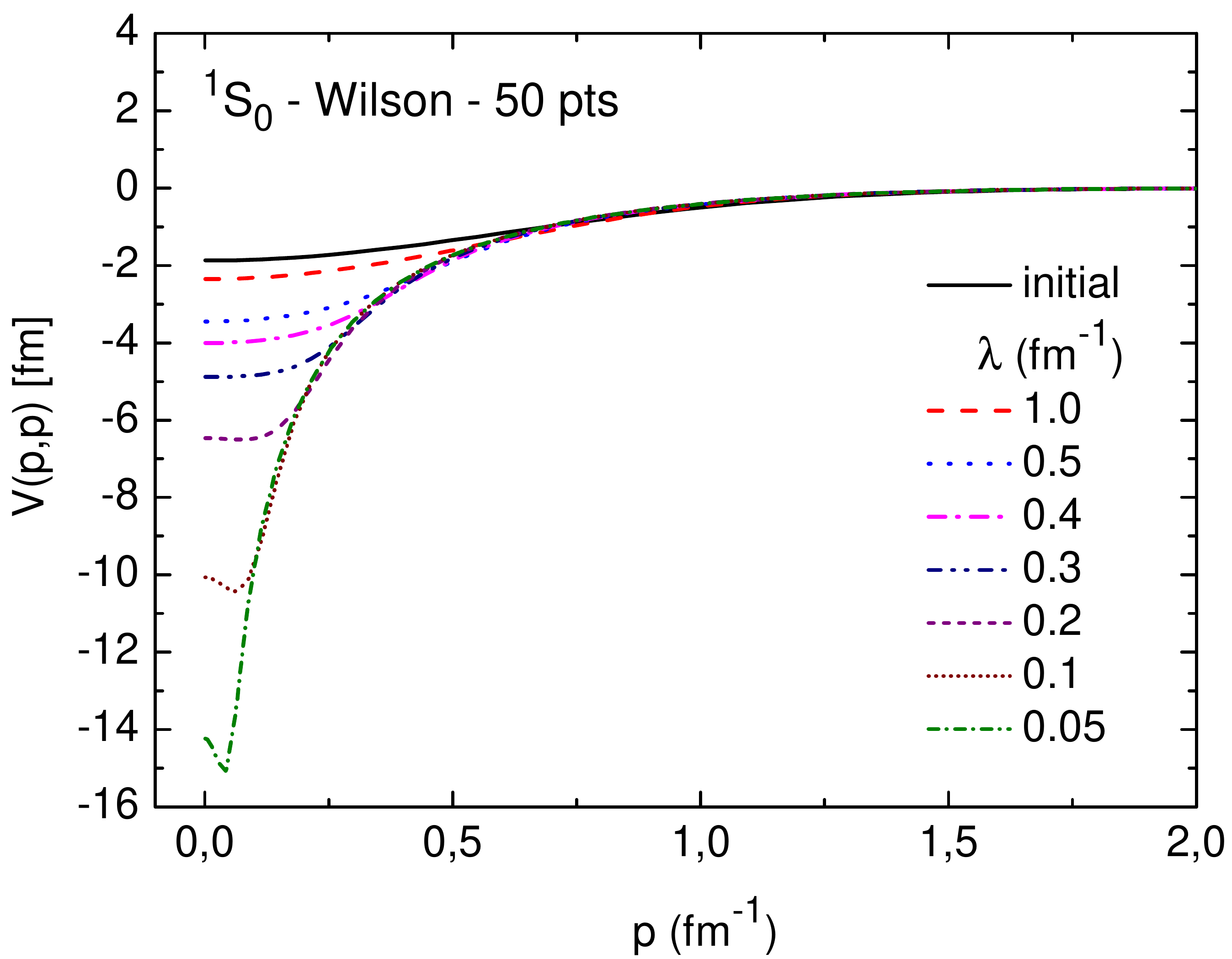}\hspace{0.5cm}
\includegraphics[width=7.5cm]{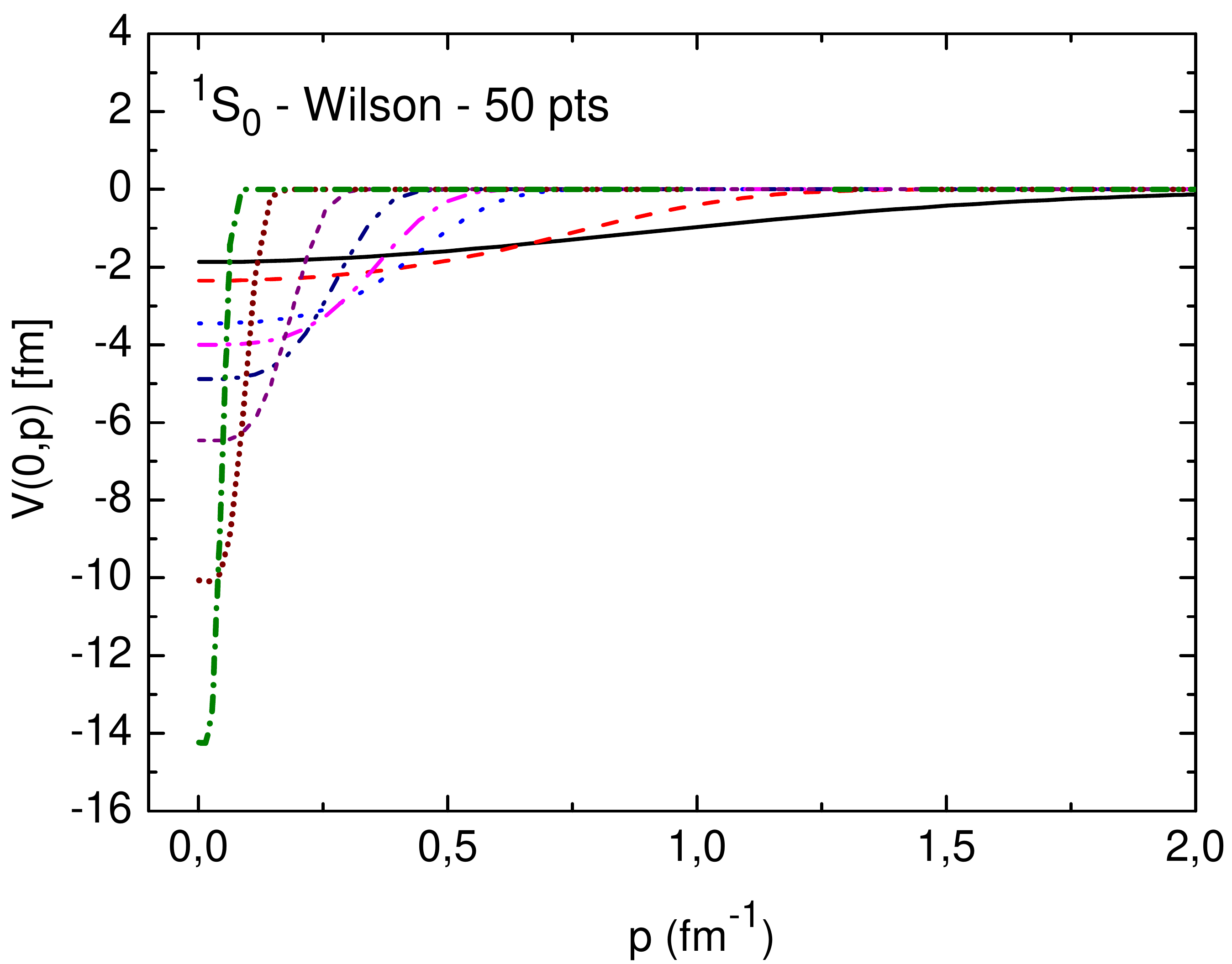} \\ \vspace{0.6cm}
\includegraphics[width=7.5cm]{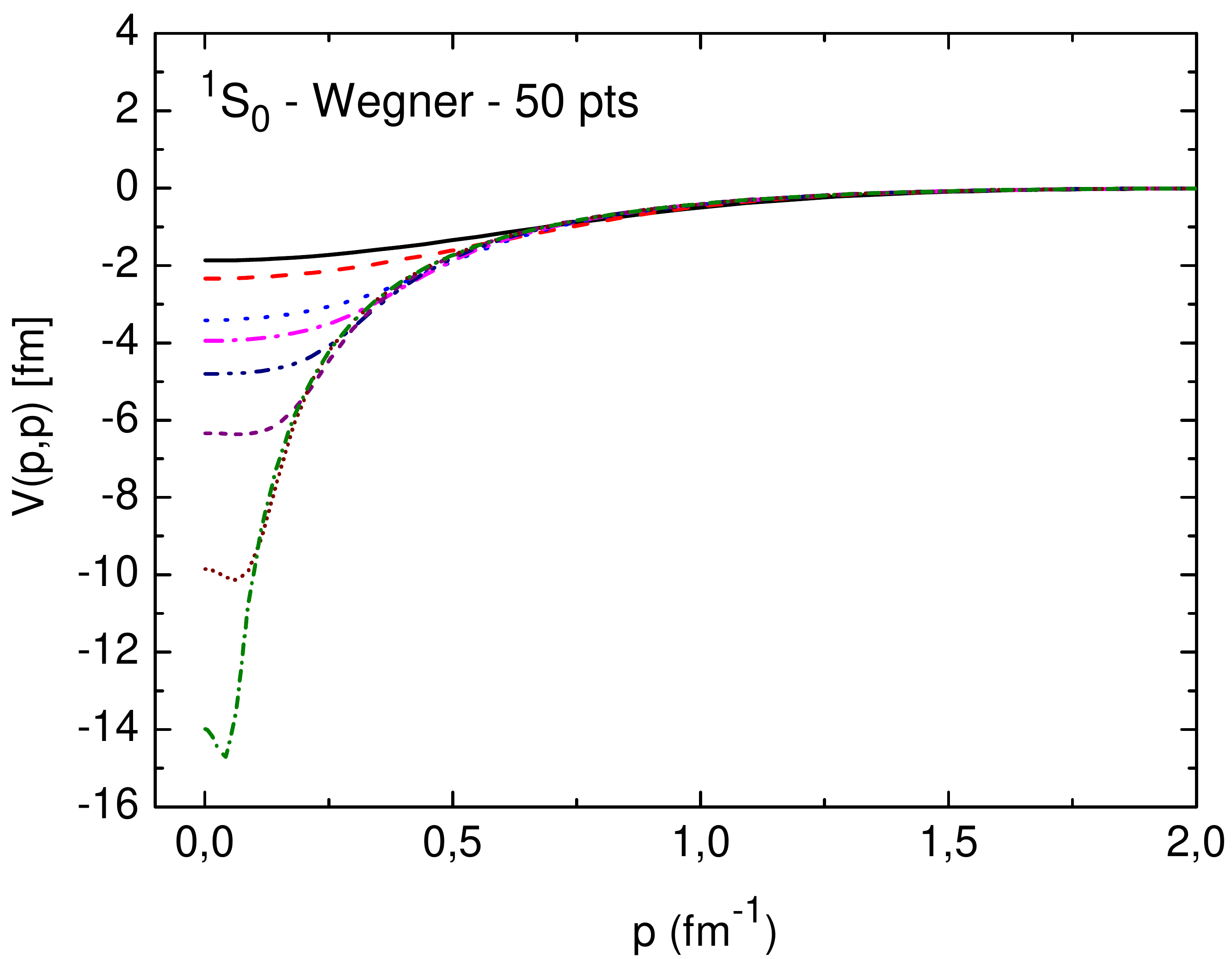}\hspace{0.5cm}
\includegraphics[width=7.5cm]{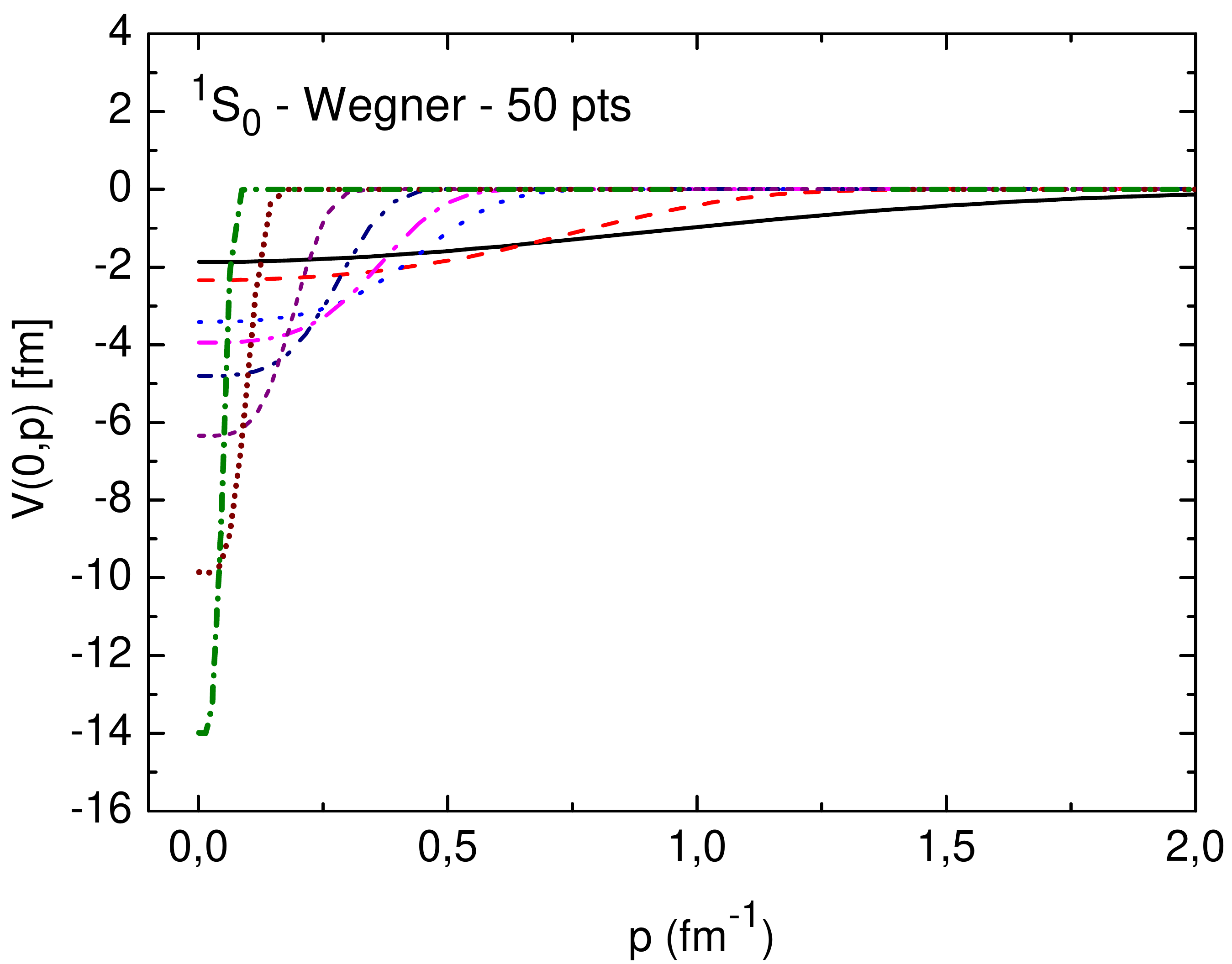}
\end{center}
\caption{SRG evolution of the toy model separable gaussian potential in the $^1S_0$ channel ($N=50$ and $\Lambda=2~{\rm fm}^{-1}$) using the Wilson and the Wegner generators. Left panels: diagonal matrix-elements; Right panels: fully off-diagonal matrix-elements.
}
\label{fig:3}
\end{figure}

\begin{figure}[t]
\begin{center}
\includegraphics[width=14cm]{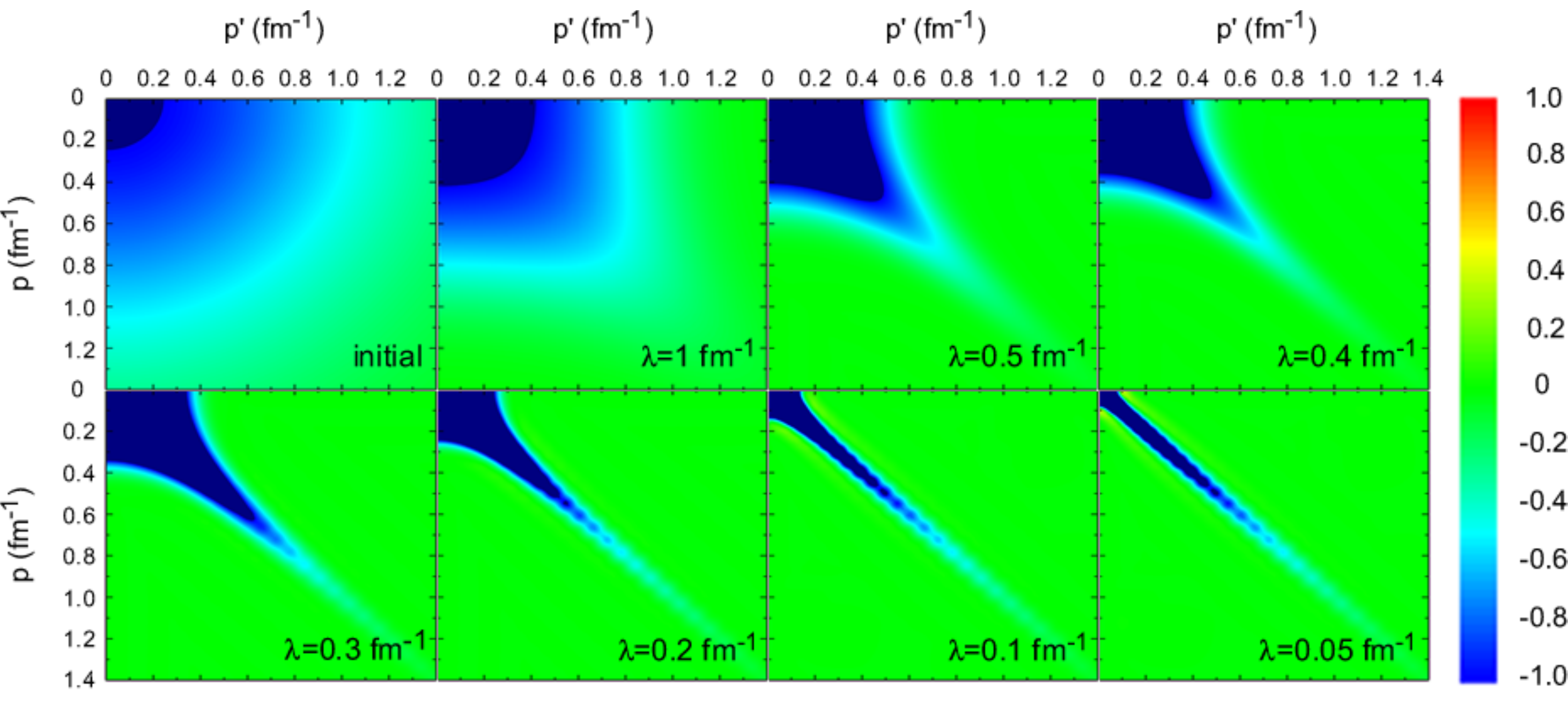} \\ \vspace{0.7cm}
\includegraphics[width=14cm]{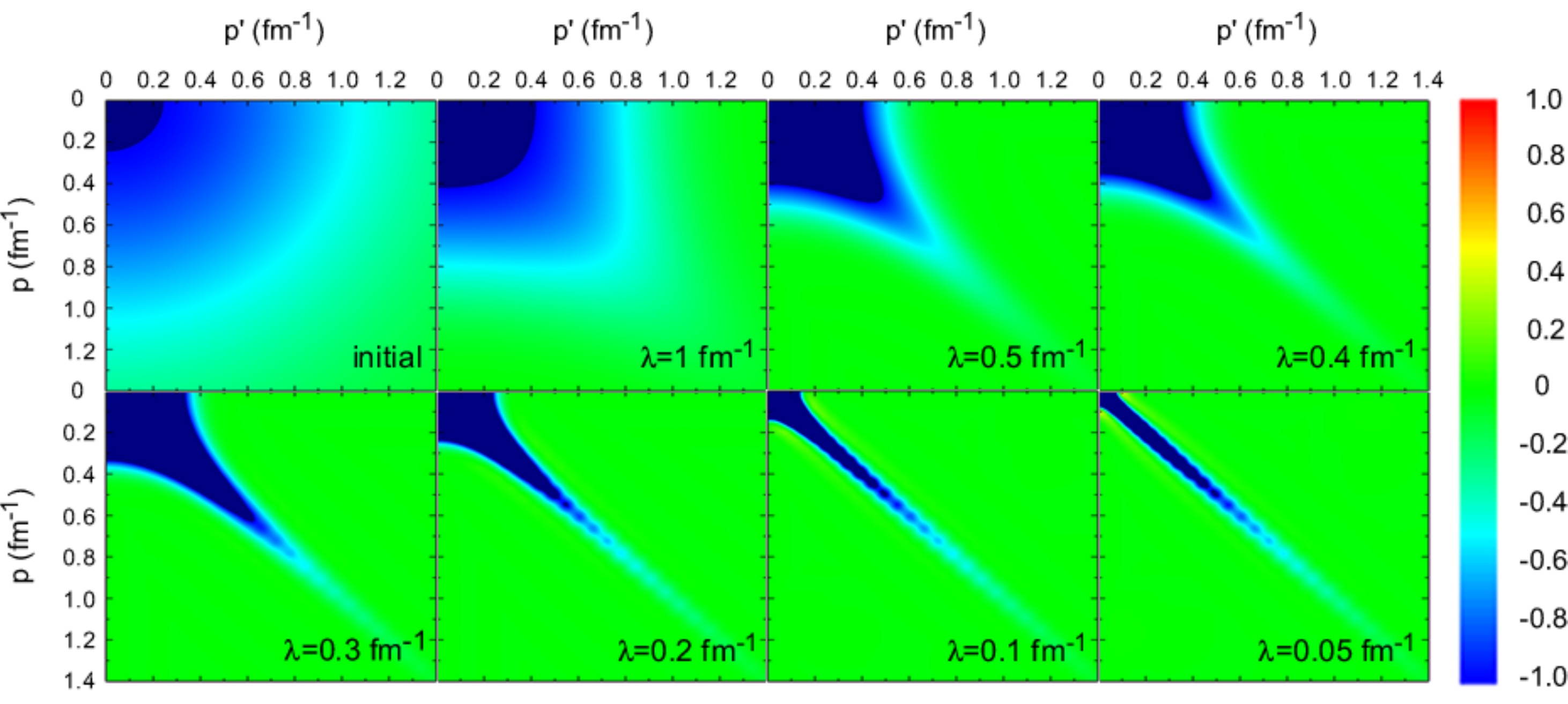}
\end{center}
\caption{Density plots for the SRG evolution of the toy model separable gaussian potential in the $^1S_0$ channel ($N=50$ and $\Lambda=2~{\rm fm}^{-1}$) using the Wilson (top panel) and the Wegner (bottom panel) generators.}
\label{fig:4}
\end{figure}

In Fig.~\ref{fig:3} we show the SRG evolution of the diagonal and the
fully off-diagonal matrix-elements of the toy model potential in the
$^1S_0$ channel using the Wilson and the Wegner generators. As
expected, since there are no bound-states in this channel, the SRG
evolutions using both generators are nearly identical. As the
potential evolves with the SRG cutoff $\lambda$ the off-diagonal
matrix-elements are gradually suppressed such that the potential is
driven towards the diagonal form. This can be seen very clearly from
the density plots displayed in Fig.~\ref{fig:4}.

%%%%   1S0 Wilson & Wegner

%%%%   1S0 Density Wilson & Wegner

\begin{figure}[t]
\begin{center}
\includegraphics[width=7.5cm]{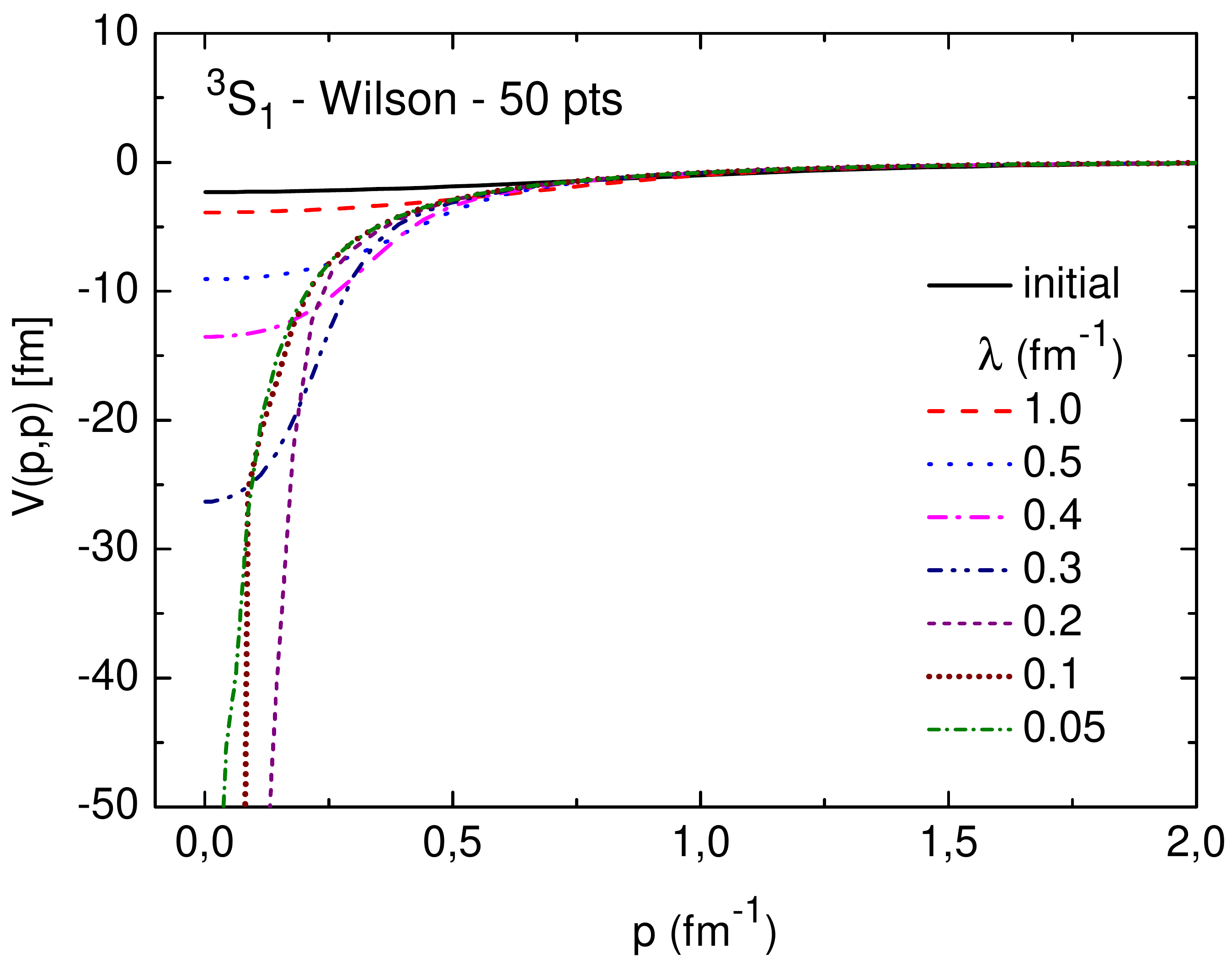}\hspace{0.5cm}
\includegraphics[width=7.5cm]{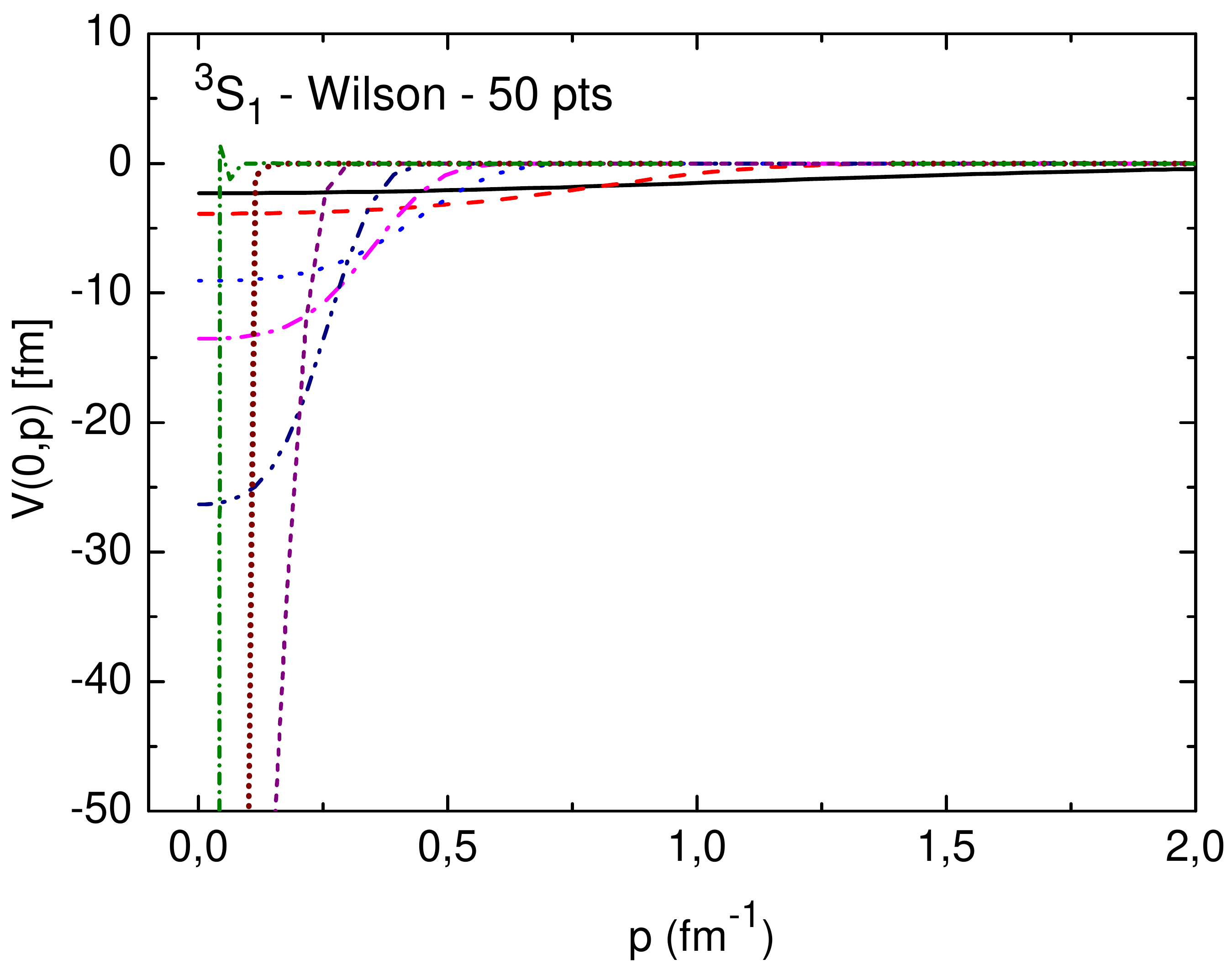} \\ \vspace{0.6cm}
\includegraphics[width=7.6cm]{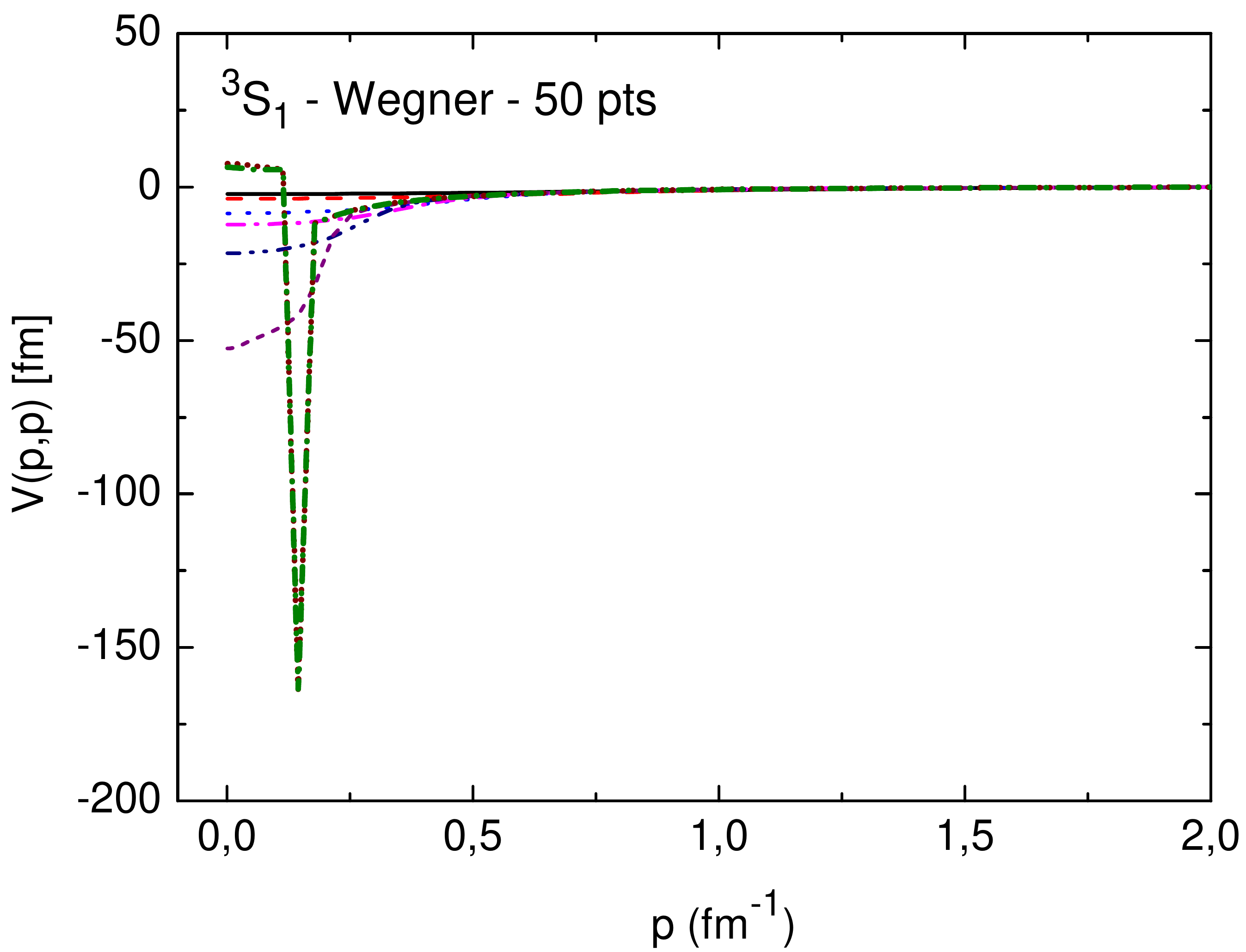}\hspace{0.5cm}
\includegraphics[width=7.5cm]{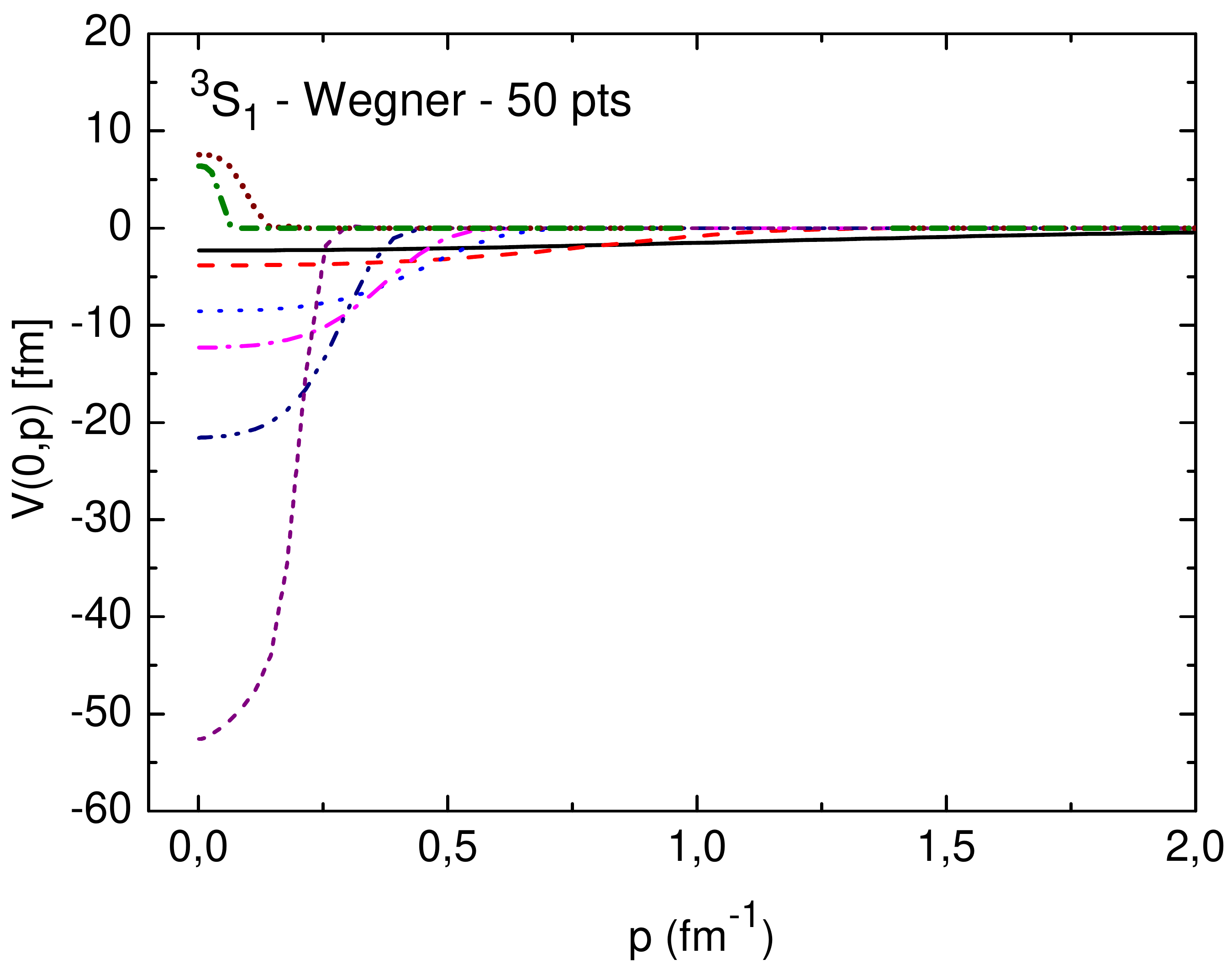}
\end{center}
\caption{SRG evolution of the toy model separable gaussian potential in the $^3S_1$ channel ($N=50$ and $\Lambda=2~{\rm fm}^{-1}$) using the Wilson and the Wegner generators. Left panels: diagonal matrix-elements; Right panels: fully off-diagonal matrix-elements.}
\label{fig:5}
\end{figure}

\begin{figure}[t]
\begin{center}
\includegraphics[width=14cm]{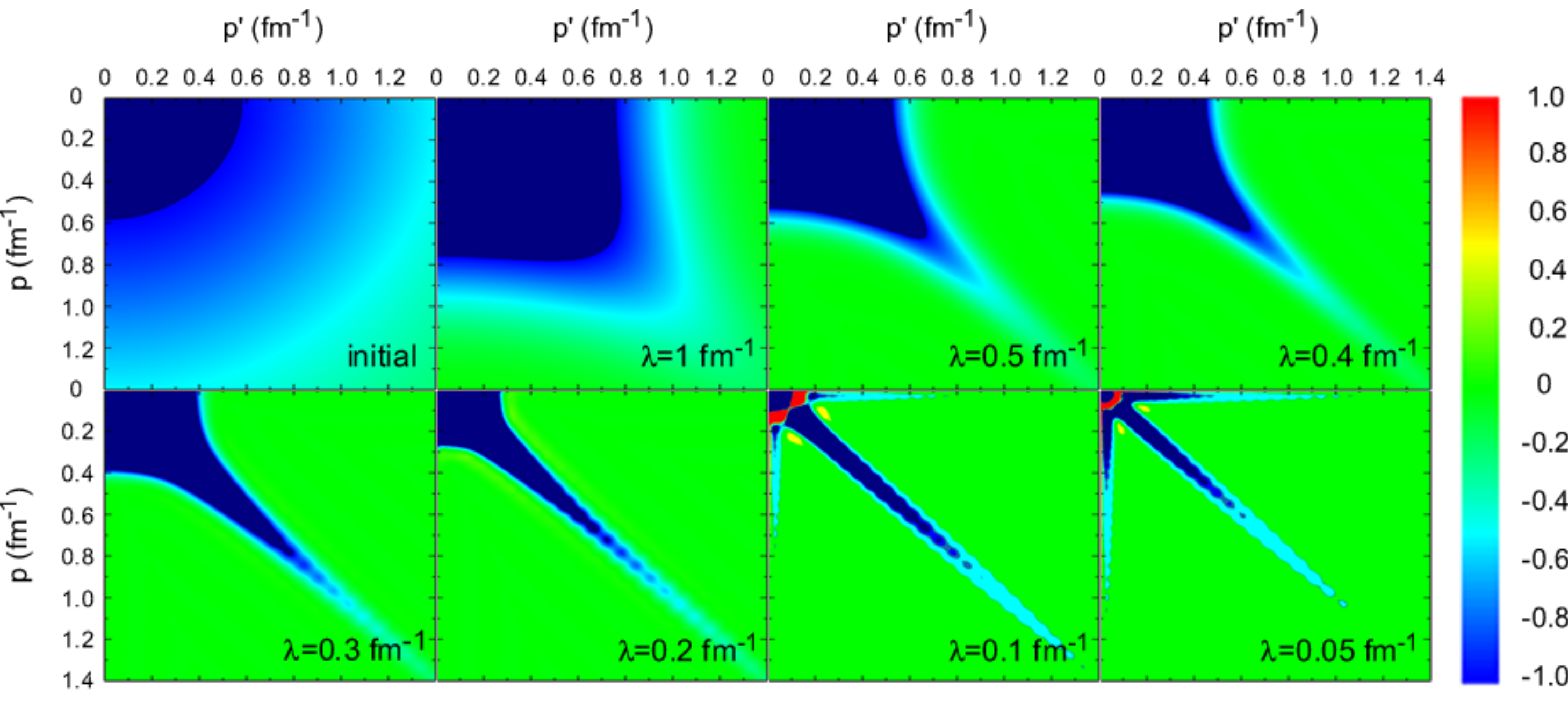} \\ \vspace{0.7cm}
\includegraphics[width=14cm]{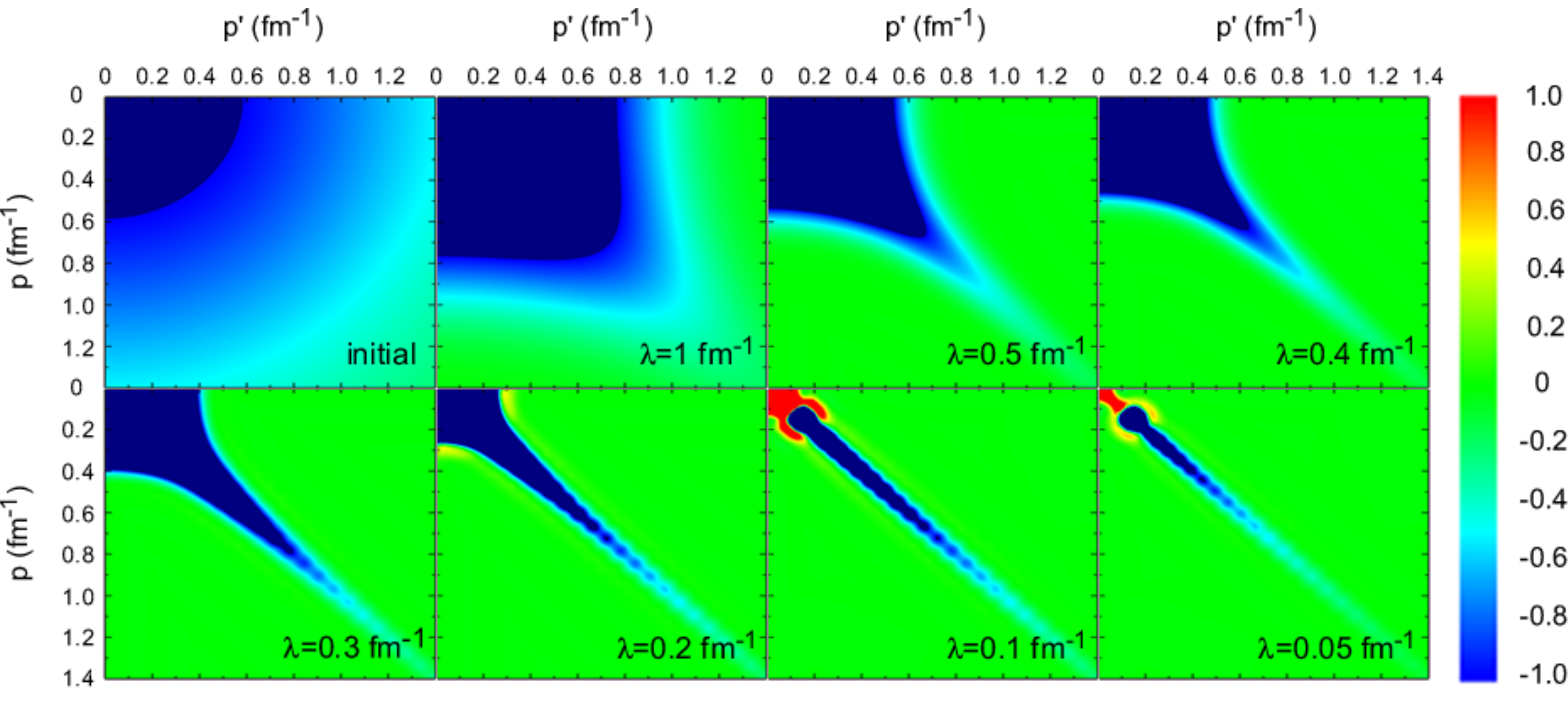}
\end{center}
\caption{Density plots for the SRG evolution of the toy model separable gaussian potential in the $^3S_1$ channel ($N=50$ and $\Lambda=2~{\rm fm}^{-1}$) using the Wilson (top panel) and the Wegner (bottom panel) generators.}
\label{fig:6}
\end{figure}

In Fig.~\ref{fig:5} we show the SRG evolution of the diagonal and the fully off-diagonal matrix-elements of the toy model potential in the $^3S_1$ channel using the Wilson and the Wegner generators. For both generators the potential is driven towards a diagonal form though, as expected for the $^3S_1$ channel, the SRG evolutions become very different when the SRG cutoff $\lambda$ goes below the critical momentum scale $\Lambda_c \sim 0.3 ~{\rm fm}^{-1}$ where the Deuteron bound-state emerges. In the case of the Wilson generator, the low-momentum matrix-elements of the SRG-evolved potential are driven to large negative values, since the Deuteron bound-state is pushed towards the lowest momentum $p_1$ available on the grid. In the case of the Wegner generator, the Deuteron bound-state decouples from the low-momentum scales when the SRG cutoff $\lambda$ goes below the critical momentum $\Lambda_c$ and is placed at a higher momentum $p_{n_{\rm BS}}$. We observe in our SRG calculations for the toy model potential on a finite grid that the position at which the bound-state is placed changes when using different values for the number of grid points $N$ and/or the UV cutoff $\Lambda$, similar to what is observed in Ref.~\cite{Wendt:2011qj} for the SRG evolution of LO ChEFT interactions with large momentum cutoffs $\Lambda_{\rm EFT}$, in which the (spurious) bound-state position also changes with the cutoff. Here, for a grid with $N=50$ points and $\Lambda=2~{\rm fm}^{-1}$, the bound-state is placed at the momentum $p_{n_{\rm BS}}= p_9\sim 0.145~{\rm fm}^{-1}$. Moreover, the matrix-elements of the SRG-evolved potential corresponding to momenta $p_n < p_{n_{\rm BS}}$ jump to positive values approaching the $^3S_1$ channel scattering length $a_{^3S_1} = 5.4~{\rm fm}$. The difference between the SRG evolutions of the toy model potential in the $^3S_1$ channel using the Wilson and the Wegner generators can be seen very clearly from the density plots displayed in Fig.~\ref{fig:6}.

%%%%   3S1 Wilson and Wegner

%%%%   3S1 Density Wilson & Wegner

%%%

\subsection{SRG regimes}

\begin{figure}[h]
\begin{center}
\includegraphics[width=7.5cm]{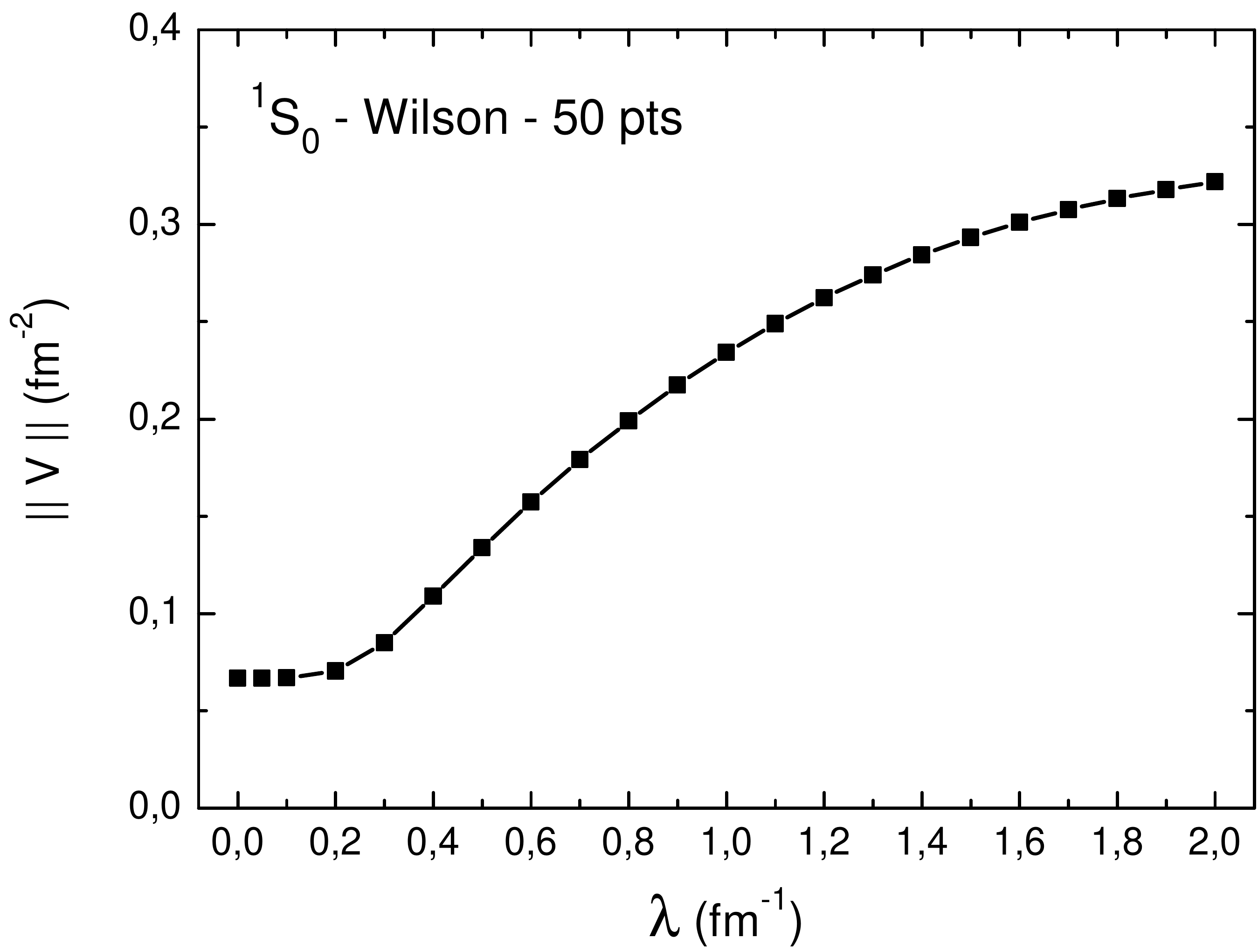}\hspace{0.5cm}
\includegraphics[width=7.5cm]{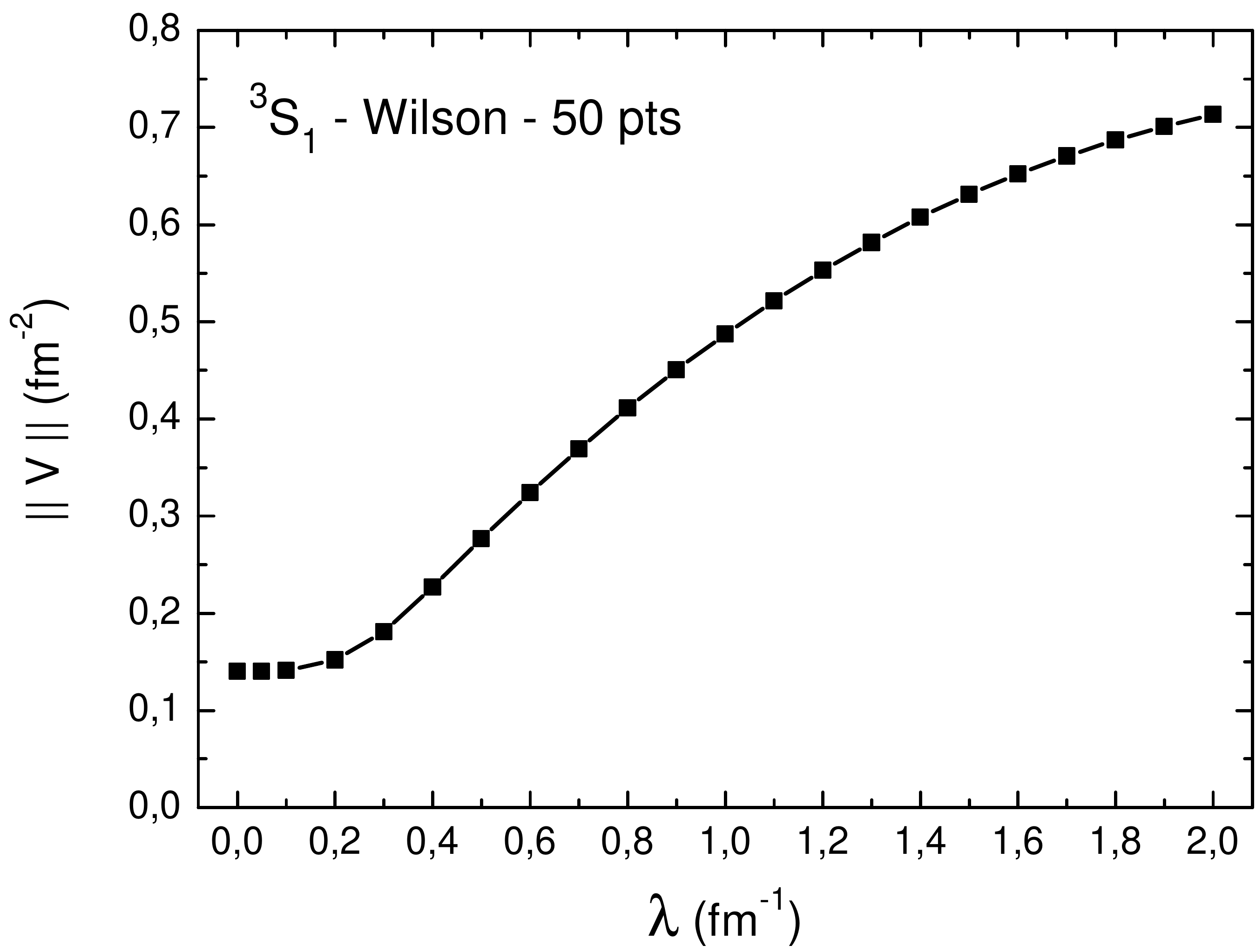} \\ \vspace{0.7cm}
\includegraphics[width=7.5cm]{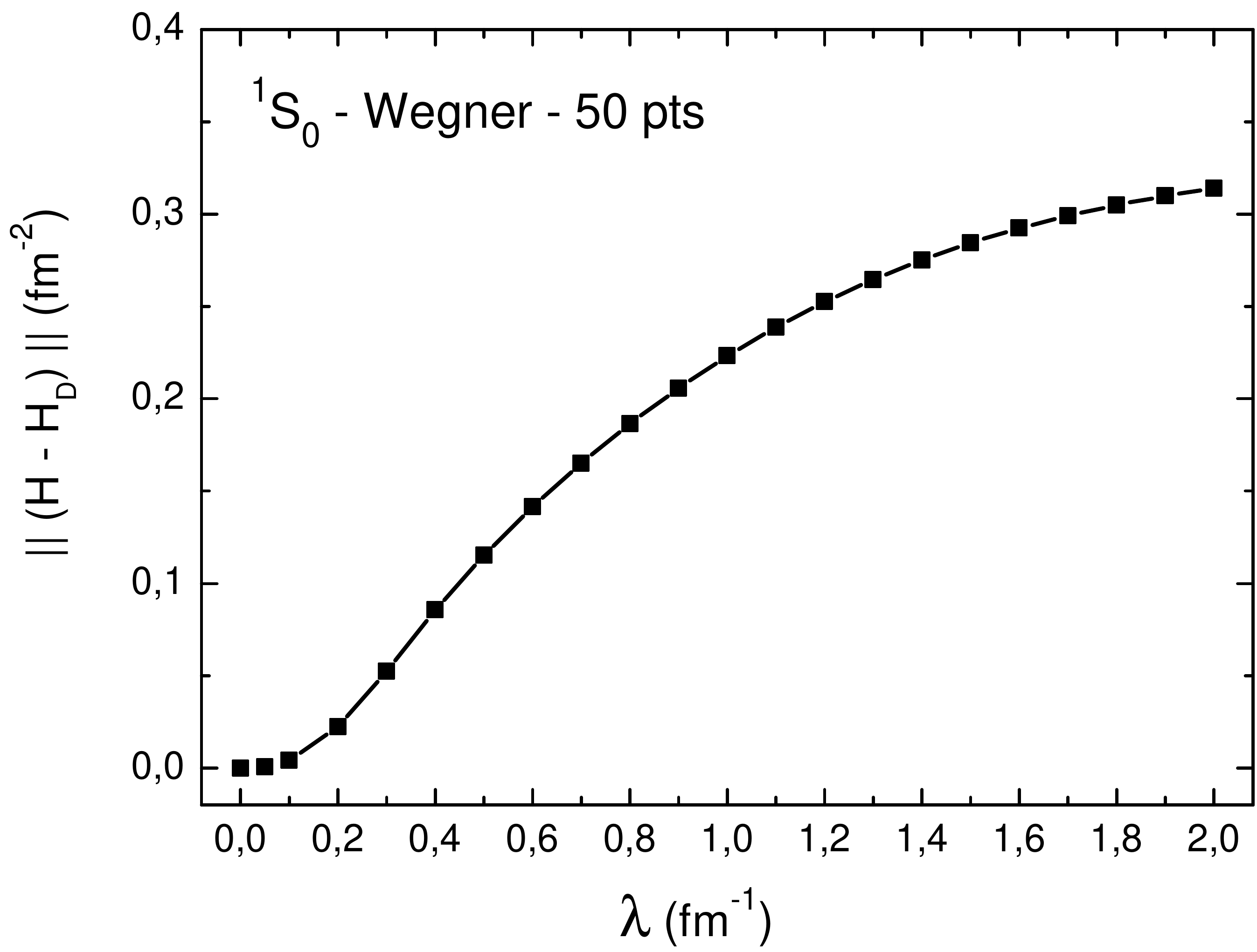}\hspace{0.5cm}
\includegraphics[width=7.5cm]{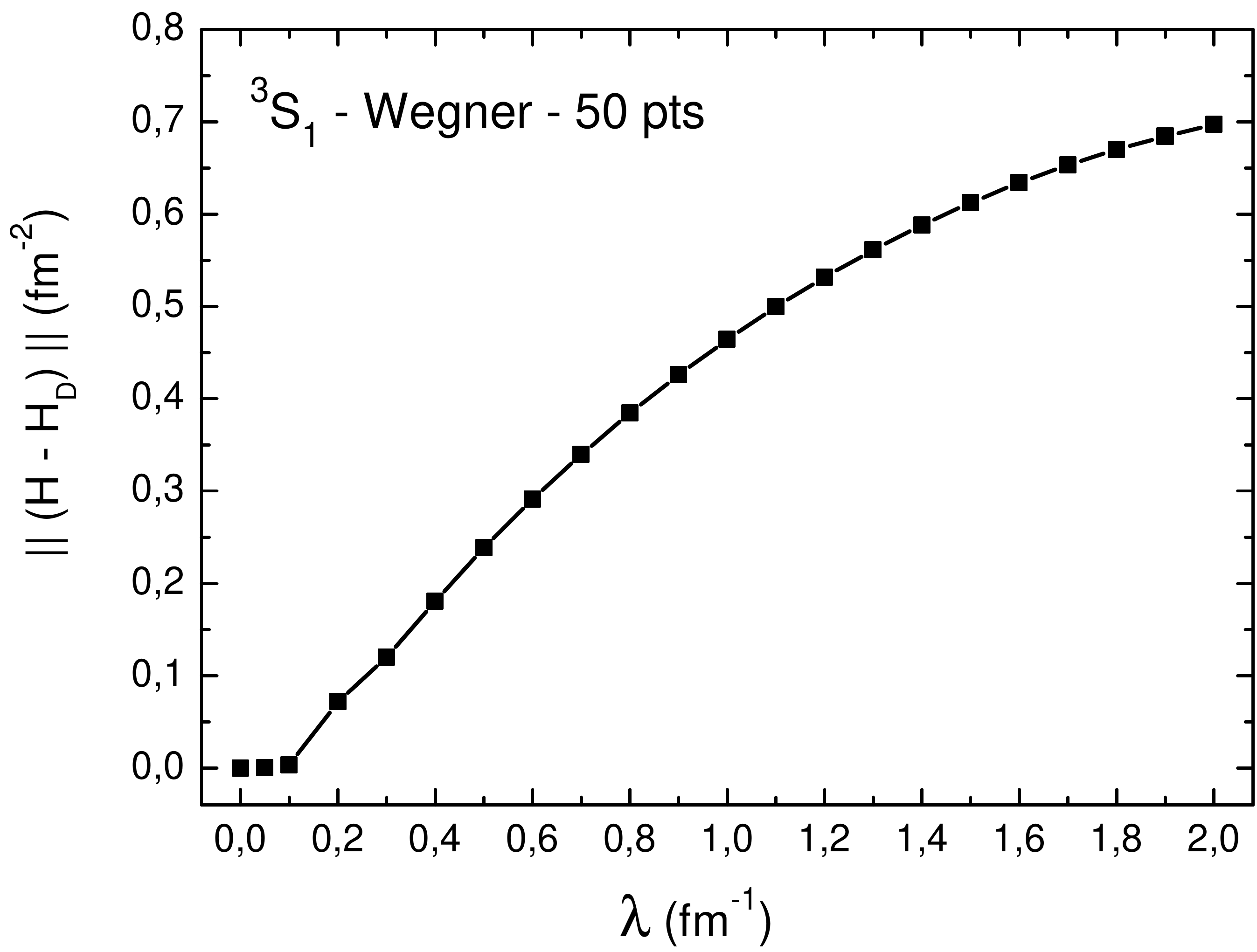}
\end{center}
\caption{Operator norms for the SRG-evolved toy model separable gaussian potential in the $^1S_0$ and the $^3S_1$ channels ($N=50$ and $\Lambda=2~{\rm fm}^{-1}$) for the Wilson and the Wegner generators as a function of the SRG cutoff $\lambda$. The infrared regime takes place below $\lambda \sim 0.1 ~ {\rm fm}^{-1}$.}
\label{fig:7}
\end{figure}

In most studies the convergence pattern of the SRG evolution towards the infrared fixed-point is determined by the stability
of single matrix-elements of the evolved potential. This corresponds to a pointwise notion of convergence. However, as we have discussed above, the proper and monotonous decreasing quantities are the operator norms.  It is interesting to illustrate to what extent the Frobenius norm and the departure from the infrared limit behave as a function of the SRG cutoff $\lambda$. The explicit expression for the 
Frobenius norm is given by
\begin{eqnarray}
|| V_\lambda ||      &=& \sqrt{ ~{\rm Tr} \, V^2_\lambda (p,p^\prime) ~ } \; ,  
\end{eqnarray}
where
\begin{eqnarray}
V^2_\lambda (p,p^\prime) &=& \frac{2}{\pi} \int dq \, q^2 ~ V_\lambda (p,q) ~  V_\lambda (q,p^\prime) \; ,  \\
&\simeq& \frac{2}{\pi} \sum_{n=1}^N  w_n  ~ q^2_n ~ V_\lambda (p,q_n) ~  V_\lambda (q_n,p^\prime) \; .  \nonumber 
\label{Fnorm}
\end{eqnarray}
This is shown in Fig.~\ref{fig:7}. As one can observe, the stationary condition is reached at about $\lambda = 0.1~{\rm fm}^{-1}$. 
This already provides the relevant scale below which the infrared regime sets in for the finite momentum grid. 
Actually, the operator norms suggest a new density plot for the quantity
\begin{eqnarray}
\frac{2}{\pi} ~ p p' \sqrt{\Delta p \Delta p'} ~ V_{\lambda}(p,p') \rightarrow  \frac{2}{\pi} ~ p_n p_m \sqrt{w_n w_m} ~ V_{\lambda}(p_n,p_m)  \; ,
\label{wdens}
\end{eqnarray}
\noindent
instead of the standard one involving only $V_{\lambda}(p,p') \rightarrow V_{\lambda}(p_n,p_m)$.  These are the normalized SRG-evolved potentials on the grid, according to our previous discussion.

\begin{figure}[h]
\begin{center}
\includegraphics[width=14cm]{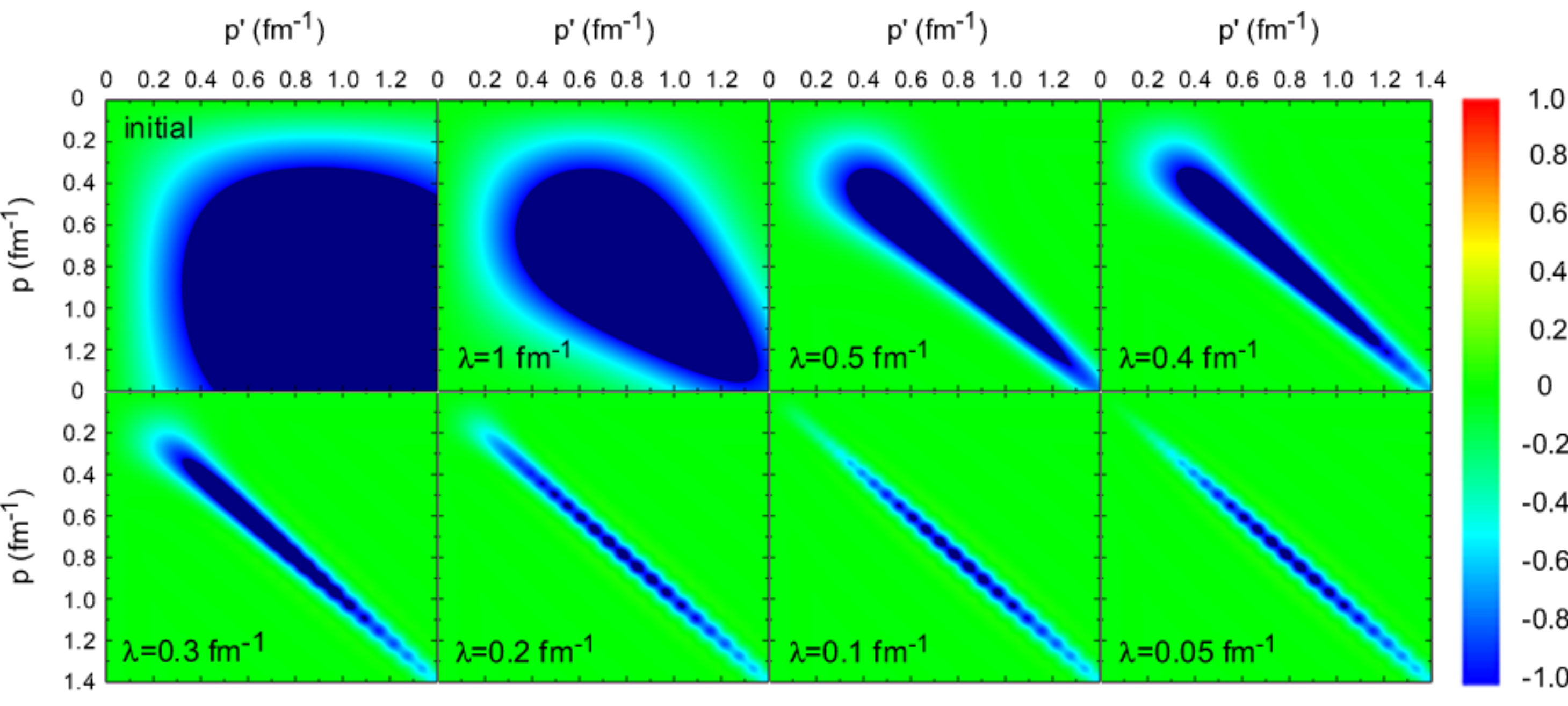}\\ \vspace{0.6cm}
\includegraphics[width=14cm]{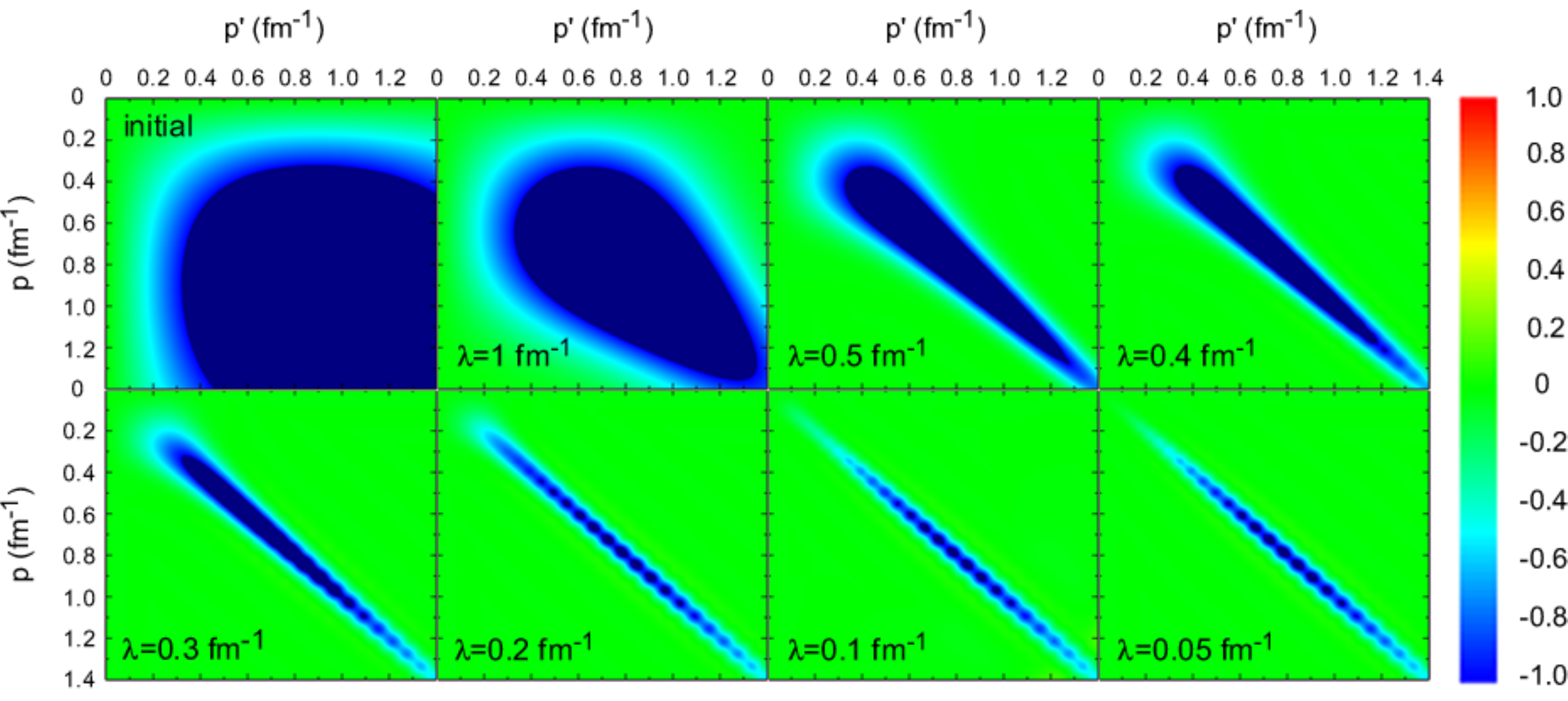}
\end{center}
\caption{Weighted density plots for the SRG evolution of the toy model separable gaussian potential in the $^1S_0$ channel ($N=50$ and $\Lambda=2~{\rm fm}^{-1}$) using the Wilson (top panel) and the Wegner (bottom panel) generators.}
\label{fig:8}
\end{figure}

\begin{figure}[h]
\begin{center}
\includegraphics[width=14cm]{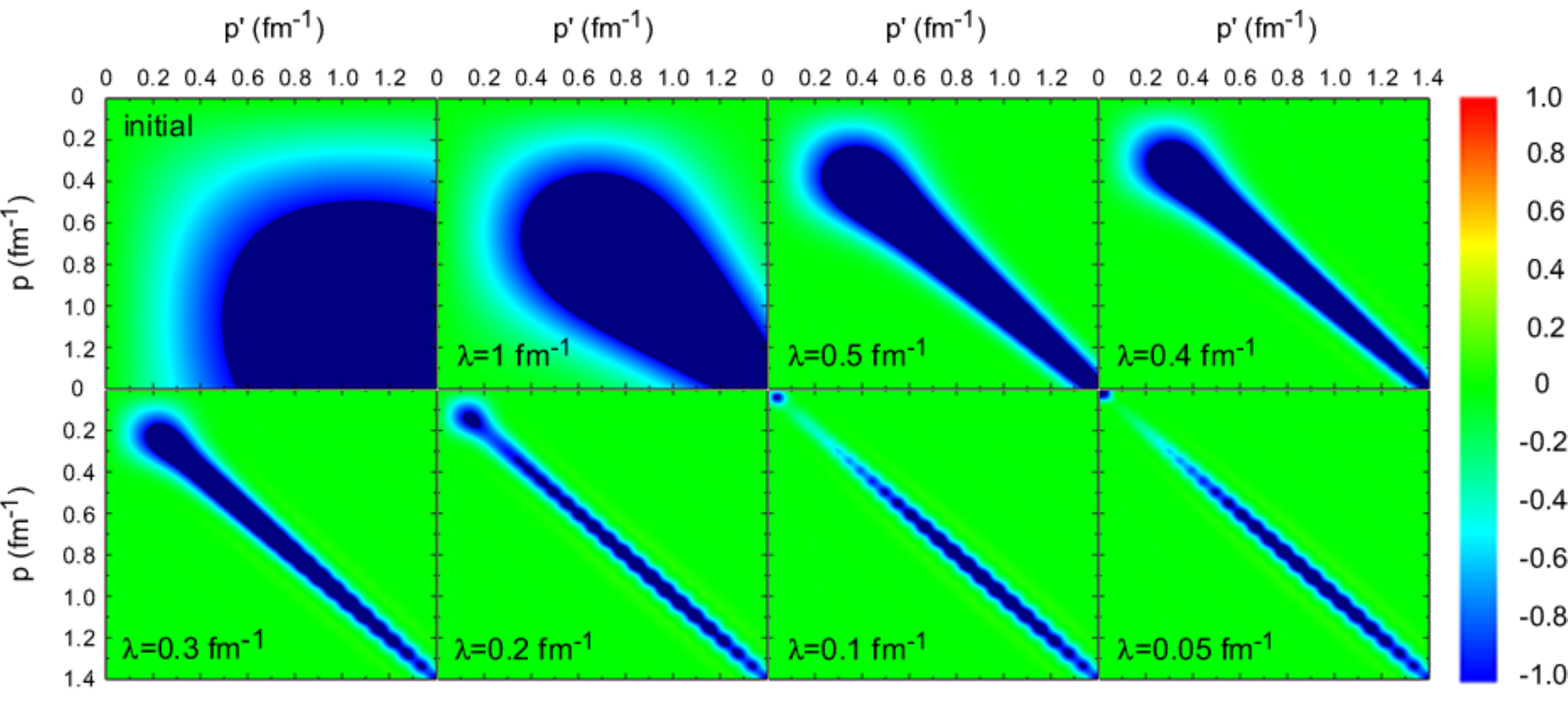}\\ \vspace{0.6cm}
\includegraphics[width=14cm]{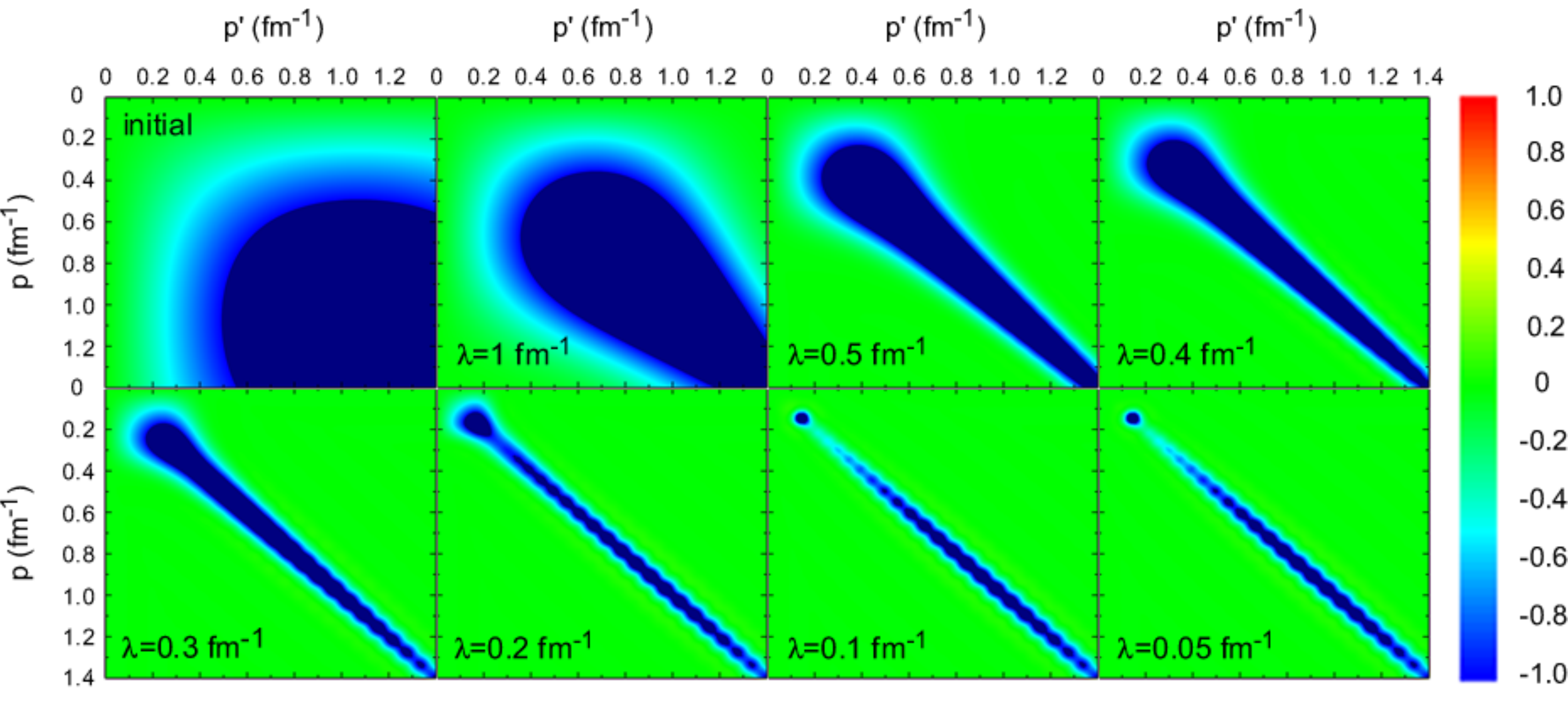}
\end{center}
\caption{Weighted density plots for the SRG evolution of the toy model separable gaussian potential in the $^3S_1$ channel ($N=50$ and $\Lambda=2~{\rm fm}^{-1}$) using the Wilson (top panel) and the Wegner (bottom panel) generators.}
\label{fig:9}
\end{figure}

The {\it weighted} density plots corresponding to Eq.~(\ref{wdens}) for the SRG evolution of the toy model potential in the $^1S_0$ and the $^3S_1$ channels are depicted respectively in Figs.~\ref{fig:8} and ~\ref{fig:9}. Note the enhancement of the diagonal region and the suppression of low-energy states as the SRG cutoff $\lambda$ decreases. This is particularly noticeable around the Deuteron bound-state in the Wegner generator case and around zero momentum in the Wilson generator case. We believe these pictures reflect more faithfully the SRG evolution pattern according to the metric induced by the operator norm relevant for the
convergence of the equations.

%%%%%%  Frobenius
%%%%%%  norm-1S0-weg.pdf norm-1S0-wil.pdf  norm-3S1-weg.pdf  norm-3S1-wil.pdf

%%%% Weighted density plots
%%   wei-1S0-wil.pdf  wei-3S1-weg.pdf  wei-3S1-wil.pdf

%%%

\subsection{SRG evolution in the infrared limit and the scattering problem on a finite momentum grid}

In the previous sections we have discussed both the SRG flow equations in the infrared limit and the scattering problem on a finite momentum grid independently. In this section we discuss in detail the relation between these two apparently disconnected topics through a surprising result involving Levinson's theorem which we have already described in a previous work \cite{Arriola:2014aia}. As mentioned above the SRG flow equations are isospectral, i.e. they maintain the spectrum of the hamiltonian invariant along the SRG trajectory. Nevertheless, the definition of the spectrum in the continuum requires some care since strictly speaking there is no finite energy-shift but the states become dense. This is illustrated by the finite box quantization formula Eq.~(\ref{eq:p-box}) together with Eq.~(\ref{eq:trace}). Therefore, if we regularize the continuum by a finite momentum grid there appears an inevitable energy-shift due to the interaction and all states in the Hilbert space become normalizable. However, once the finite momentum grid is introduced there is no scattering process and some of the wave operator properties usually assumed for continuum states, such as the intertwining property of the Moller operator, do not hold~\cite{muga1989stationary}. Actually, in the case of a hamiltonian allowing for bound-states the energy-shift does not disappear in the continuum. The question is what is the net effect of the energy-shift for positive energy and would-be scattering states when the continuum is approached as a limiting procedure starting from the finite momentum grid.

\subsubsection{SRG induced ordering of the spectrum and the energy-shift approach}

\begin{figure}[h]
\begin{center}
\includegraphics[width=7.5cm]{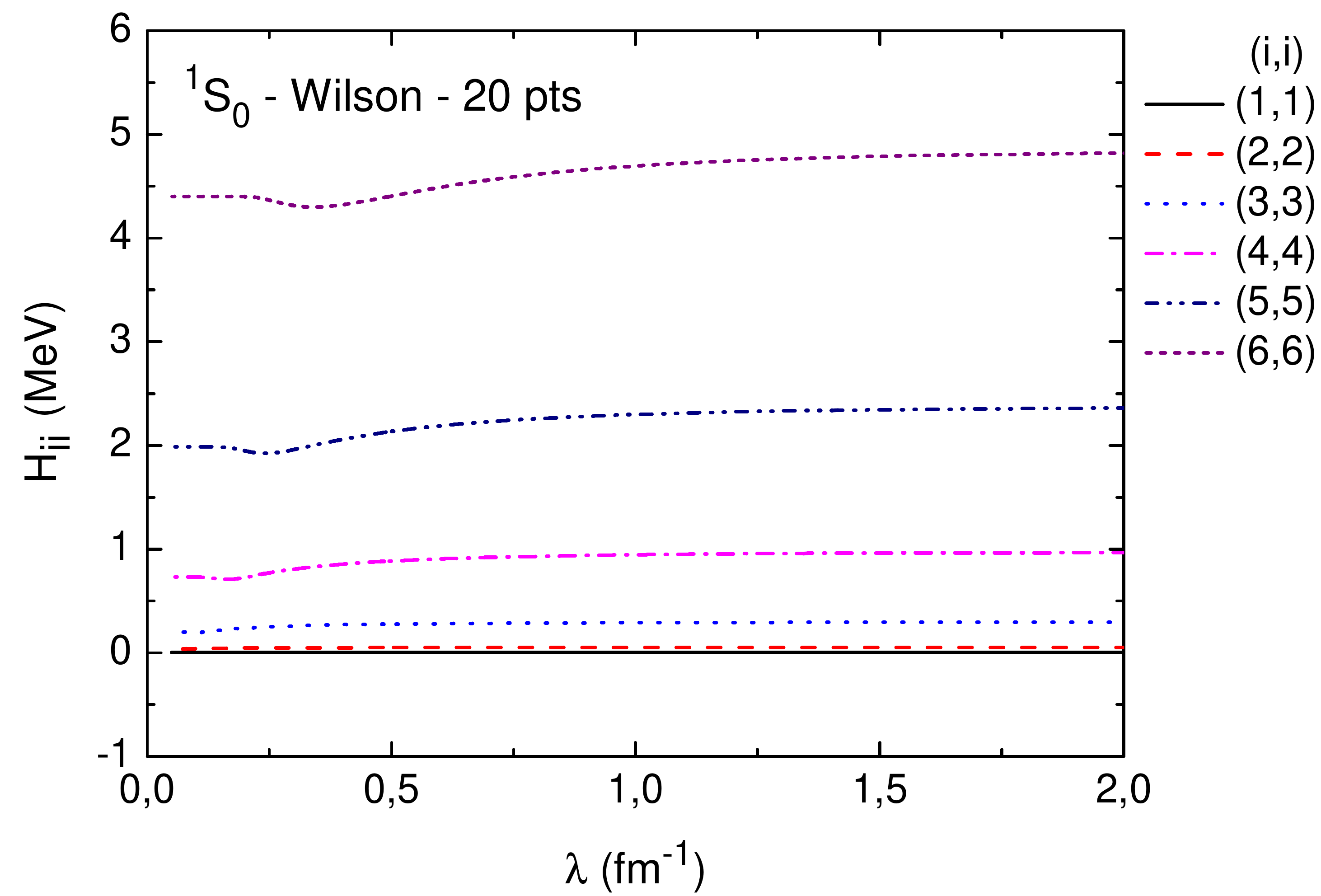}\hspace{0.5cm}
\includegraphics[width=7.5cm]{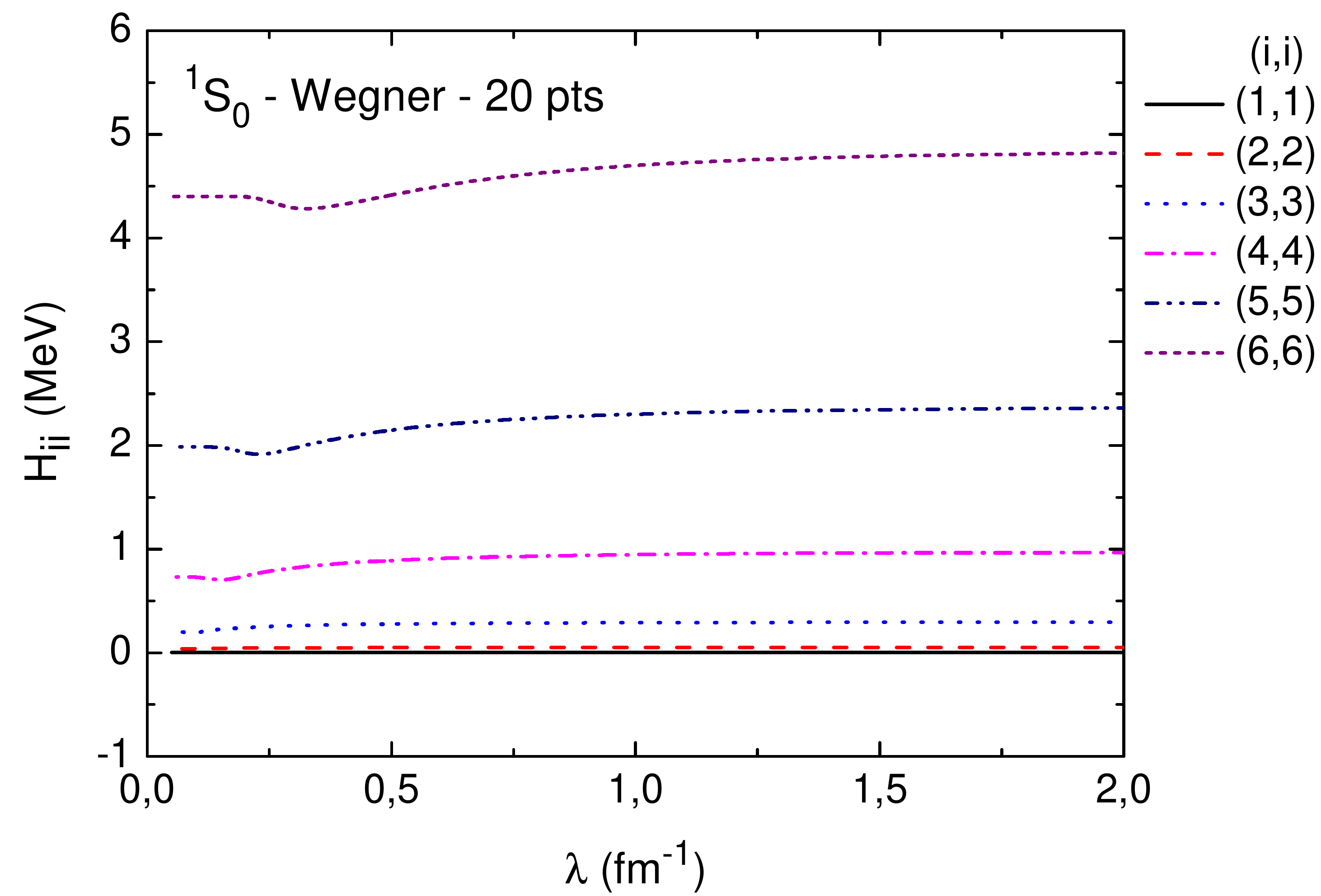} \\ \vspace{0.6cm}
\includegraphics[width=7.5cm]{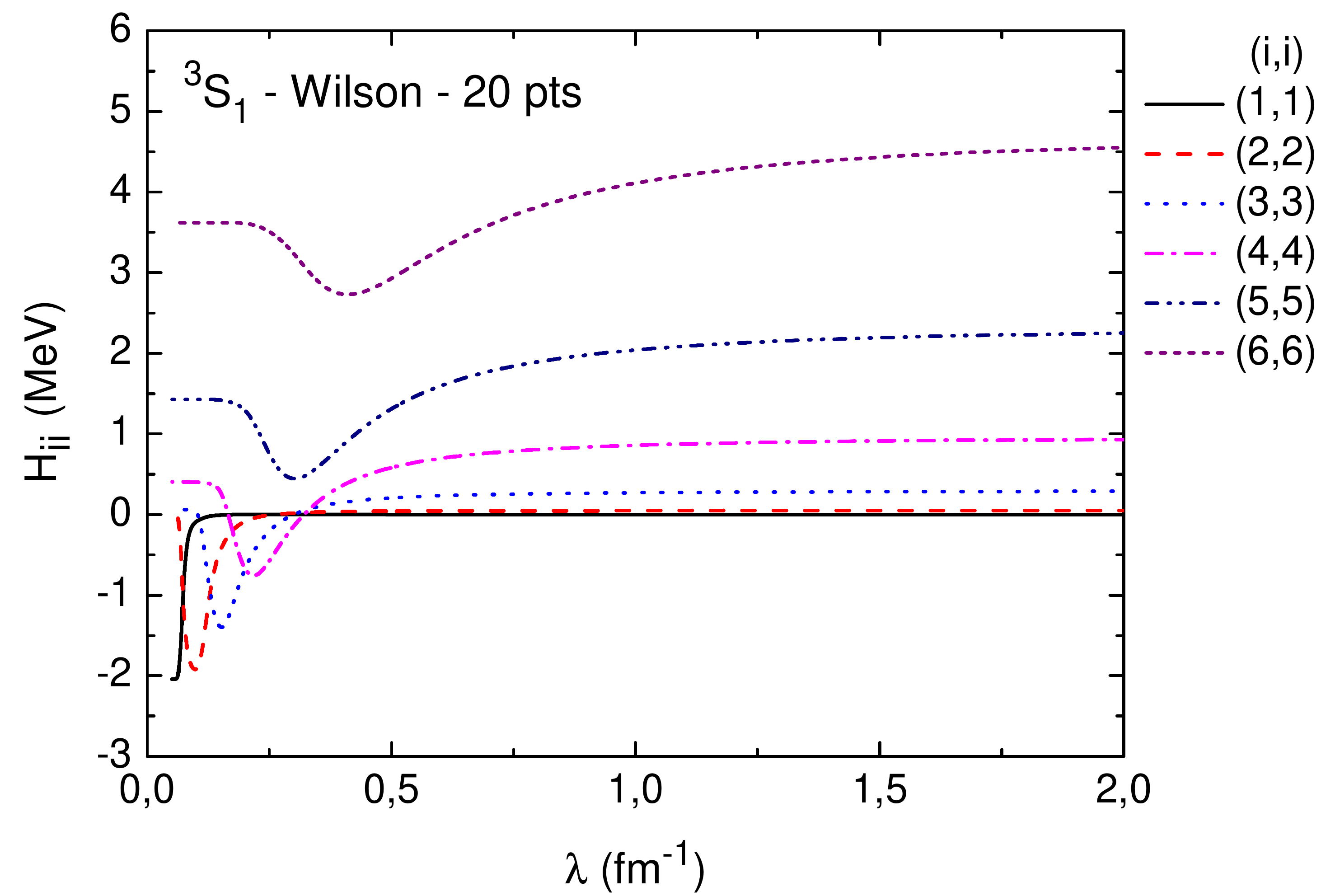}\hspace{0.5cm}
\includegraphics[width=7.5cm]{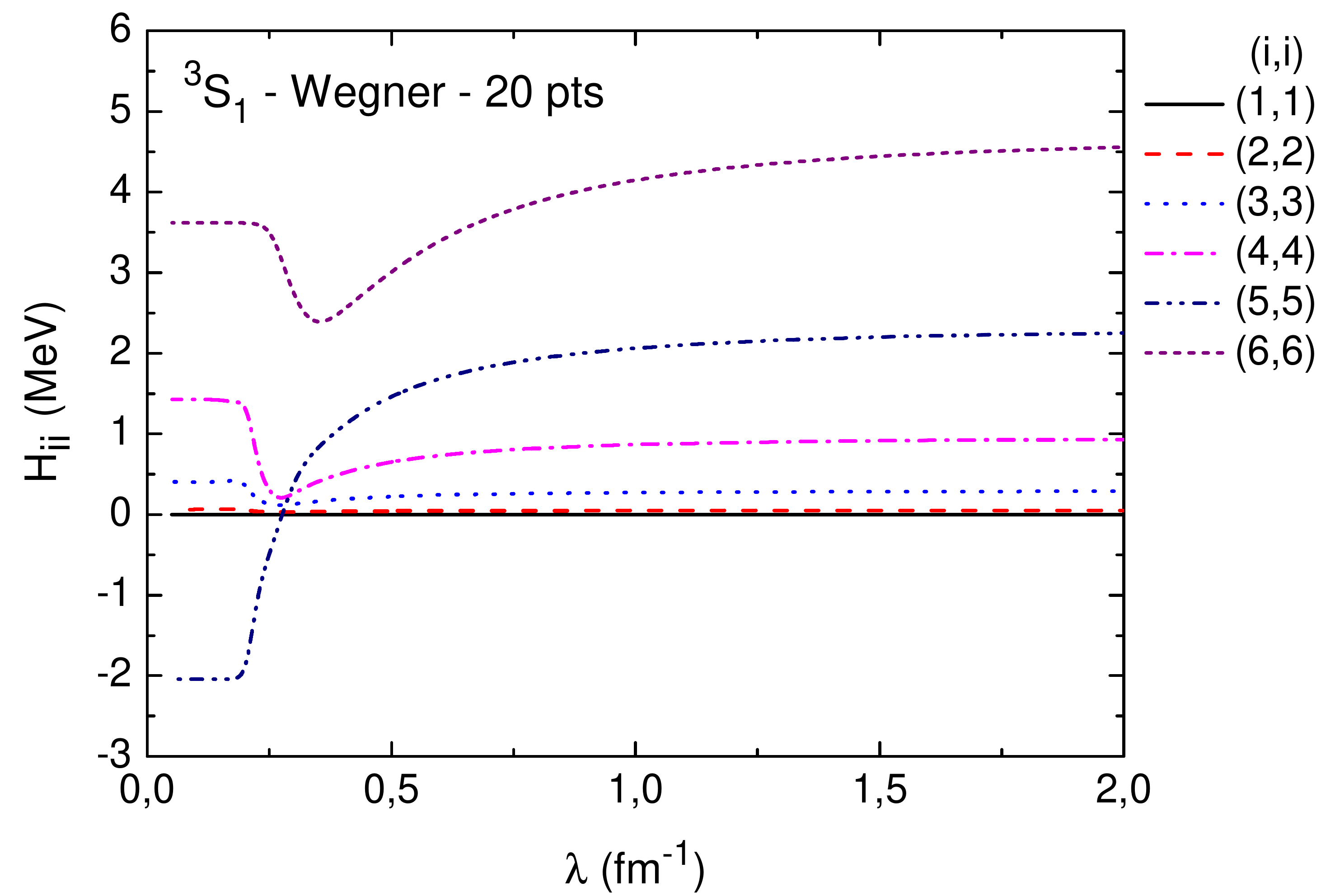}
\end{center}
\caption{SRG evolution of the lowest diagonal matrix-elements of the toy-model hamiltonian $H_{\lambda}(p_n,p_n)$ ($n=1,...,6$) in the $^1S_0$ channel and the $^3S_1$ channel ($N=20$ and $\Lambda=2~{\rm fm}^{-1}$) using the Wilson and the Wegner generators.}
\label{fig:10}
\end{figure}

The final ordering of the spectrum induced by the SRG evolution in the infrared limit $\lambda \to 0$ can be depicted through the flow of the diagonal matrix-elements of the hamiltonian along the SRG trajectory. In Fig.~\ref{fig:10} we show the SRG evolution of the lowest diagonal matrix-elements of the toy model hamiltonian $H_{\lambda}(p_n,p_n)$ ($n=1,...,6$) in the $^1S_0$ and $^3S_1$ channels, for a momentum grid with $N=20$ points and $\Lambda=2~{\rm fm}^{-1}$. As one can observe, for the $^1S_0$ channel there is no crossing amongst the diagonal matrix-elements of the hamiltonian evolved with both generators as the SRG cutoff $\lambda$ approaches the infrared limit, indicating that the initial ascending order is maintained all along the SRG trajectory. For the $^3S_1$ channel, on the other hand, there are crossings with both generators as the SRG cutoff $\lambda$ approaches the critical momentum scale $\Lambda_c \sim 0.3 ~{\rm fm}^{-1}$. In the Wilson generator case, the initial ascending order is asymptotically restored in the infrared limit $\lambda\to 0$ ~\cite{Timoteo:2011tt,Arriola:2013gya} with the lowest momentum diagonal matrix-element $H_{\lambda}(p_1,p_1)$ flowing into the Deuteron bound-state. In the Wegner generator case a re-ordering occurs, such that the diagonal matrix-element $H_{\lambda}(p_{n_{\rm {BS}}},p_{n_{\rm {BS}}})$ flows into the Deuteron bound-state in the limit $\lambda\to 0$. For this calculation with $N=20$ grid points and $\Lambda=2~{\rm fm}^{-1}$ the momentum at which the bound-state is placed on the diagonal of the hamiltonian corresponds to $p_{n_{\rm BS}}\to p_5 \sim 0.254~{\rm fm}^{-1}$. Note that the diagonal matrix-element $H_{\lambda}(p_{n_{\rm {BS}}},p_{n_{\rm {BS}}})$ flowing into the Deuteron bound-state is the one that starts to decrease rapidly towards negative values when the SRG cutoff $\lambda$ approaches $\Lambda_c$, indicating the break-up of the kinetic energy dominance,
i.e.
\begin{eqnarray}
p_{n_{\rm {BS}}}^2 < \frac{2}{\pi}~w(p_{n_{\rm {BS}}})~p_{n_{\rm {BS}}}^2~|V_{\lambda < \Lambda_c}(p_{n_{\rm {BS}}},p_{n_{\rm {BS}}})|  \; .
\end{eqnarray}
\noindent

%%%% 1S0 & 3S1 H crossing

\begin{figure}[t]
\begin{center}
\includegraphics[width=7.5cm]{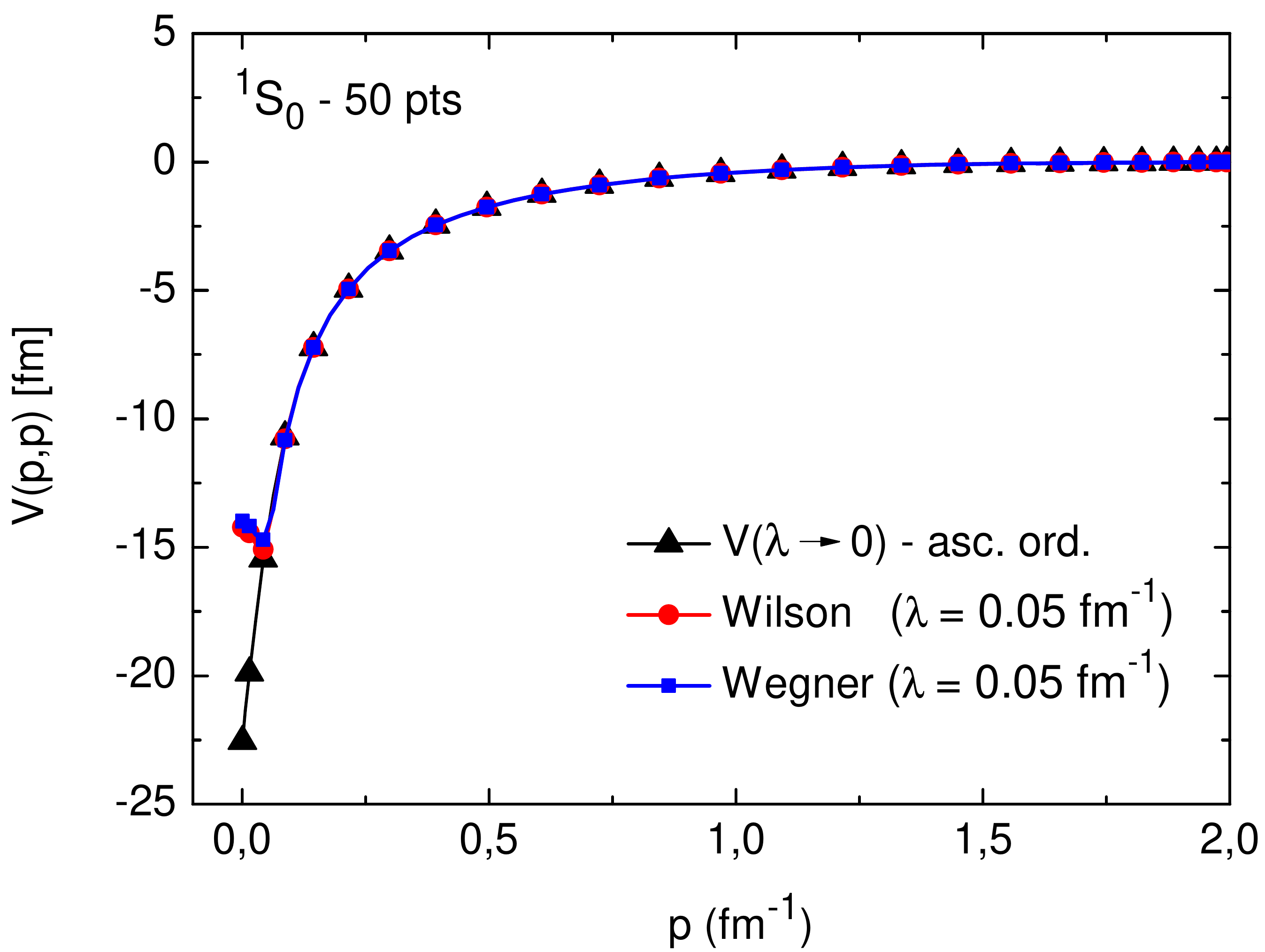}\hspace{0.5cm}
\includegraphics[width=7.5cm]{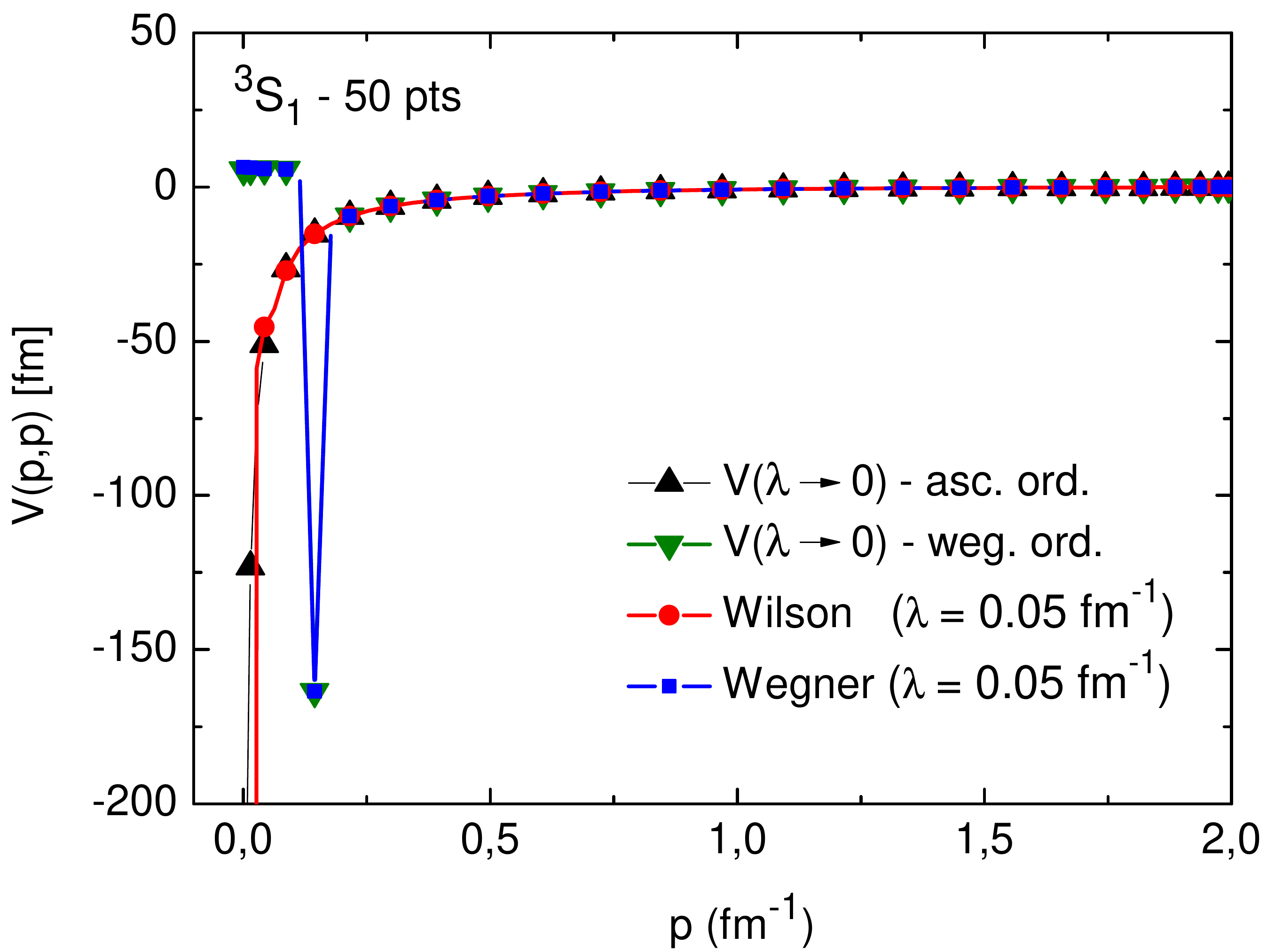}
\end{center}
\caption{Diagonal matrix-elements of the toy model separable gaussian potential in the $^1S_0$ channel and the $^3S_1$ channel ($N=50$ and $\Lambda=2~{\rm fm}^{-1}$) evolved with the Wilson and the Wegner generators up to the SRG cutoff $\lambda=0.05~{\rm fm^{-1}}$. For comparison, we also show the eigenpotentials $V_{n}^{^1S_0}(\lambda \rightarrow 0)$ and $V_{n}^{^3S_1}(\lambda \rightarrow 0)$ computed with the eigenvalues arranged according to the corresponding SRG induced orderings.}
\label{fig:11}
\end{figure}

In Fig.~\ref{fig:11} we show the diagonal matrix-elements of the toy model potential in the $^1S_0$ and the $^3S_1$
channels evolved with both the Wilson and the Wegner generators up to the SRG cutoff $\lambda = 0.05~\rm{fm}^{-1}$, compared to the corresponding eigenpotentials $V_{n}^{^1S_0}(\lambda \rightarrow 0)$ and $V_{n}^{^3S_1}(\lambda \rightarrow 0)$ evaluated from the energy-shifts through Eq.~(\ref{eigenpot}). The eigenpotential $V_{n}^{^1S_0}(\lambda \rightarrow 0)$ is computed with the eigenvalues arranged in ascending order, which corresponds to the order induced by the SRG evolution in the $^1S_0$ channel with both generators. The eigenpotential $V_{n}^{^3S_1}(\lambda \rightarrow 0)$ is computed both with the eigenvalues arranged in ascending order, which corresponds to the order induced by the SRG evolution in the $^3S_1$ channel with the Wilson generator, and in the order induced by the SRG evolution with the Wegner generator, which is obtained by placing the Deuteron bound-state at $p_{n_{\rm BS}}$. As expected by construction, the potentials evolved up to the SRG cutoff $\lambda = 0.05~\rm{fm}^{-1}$ nearly match the corresponding eigenpotentials. This result clearly shows that the SRG evolved potential indeed converges in the infrared limit $\lambda \to 0$ to the eigenpotential $V_{n}(\lambda \rightarrow 0)$ computed with the eigenvalues arranged according to the proper SRG induced ordering. Note that for the $^1S_0$ channel the eigenpotential at the lowest momentum $p_1$ on the grid approaches the scattering length, i.e. $V_{1}^{^1S_0}(\lambda \rightarrow 0)\to a_{^1S_0} = -23.7~{\rm fm}$, such that the correct behavior at low-momentum in the on-shell limit is obtained. For the $^3S_1$ channel, on the other hand, the correct behavior, with $V_{1}^{^3S_1}(\lambda \rightarrow 0)\to a_{^3S_1} = 5.4~{\rm fm}$, is obtained only when the eigenpotential is computed with the eigenvalues arranged in the order induced by the SRG evolution with the Wegner generator.

%%%%   1S0 Wilson & Wegner

\subsubsection{Ordering prescription and Levinson's theorem}

\begin{figure}[t]
\begin{center}
\includegraphics[width=7.5cm]{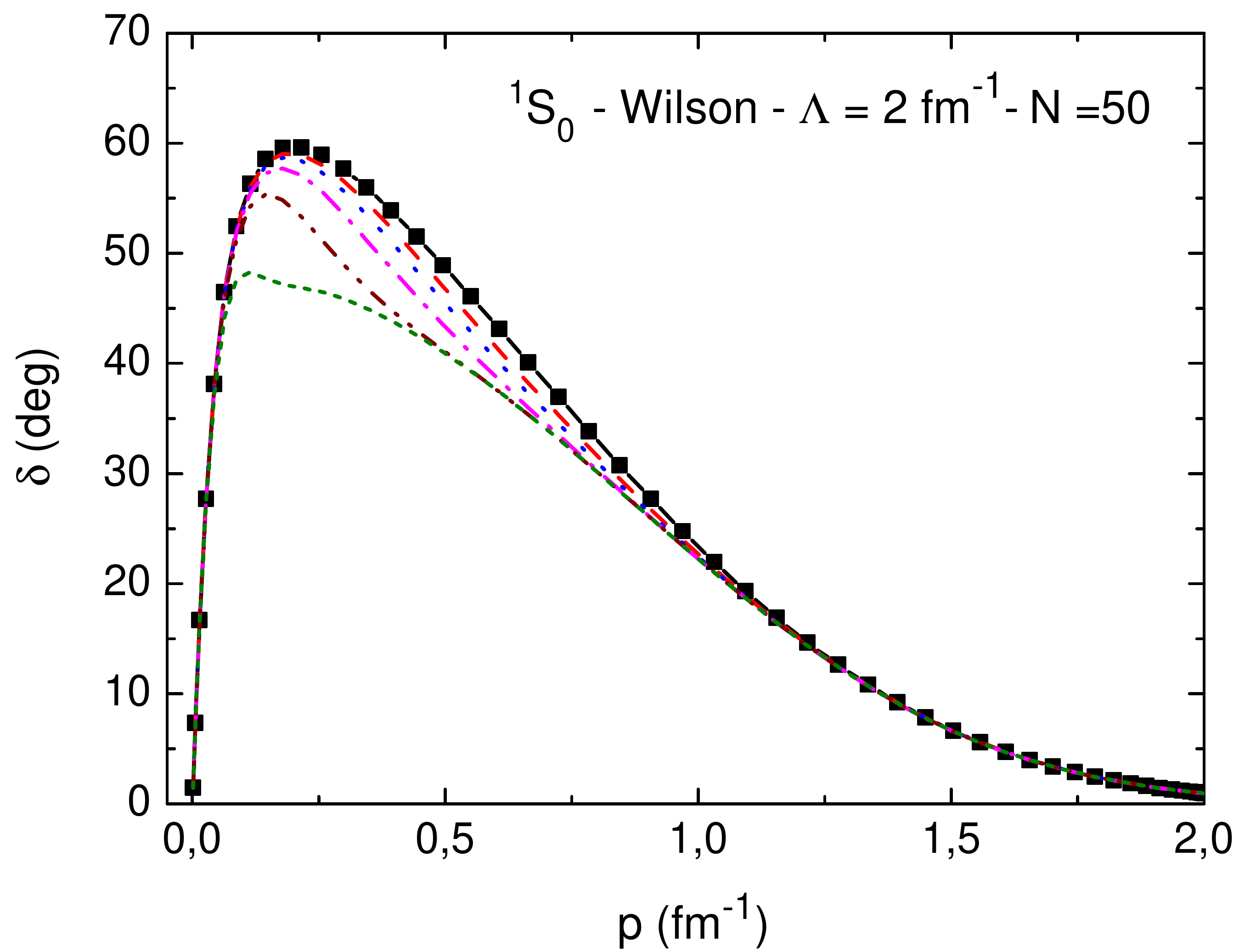}\hspace{0.5cm}
\includegraphics[width=7.5cm]{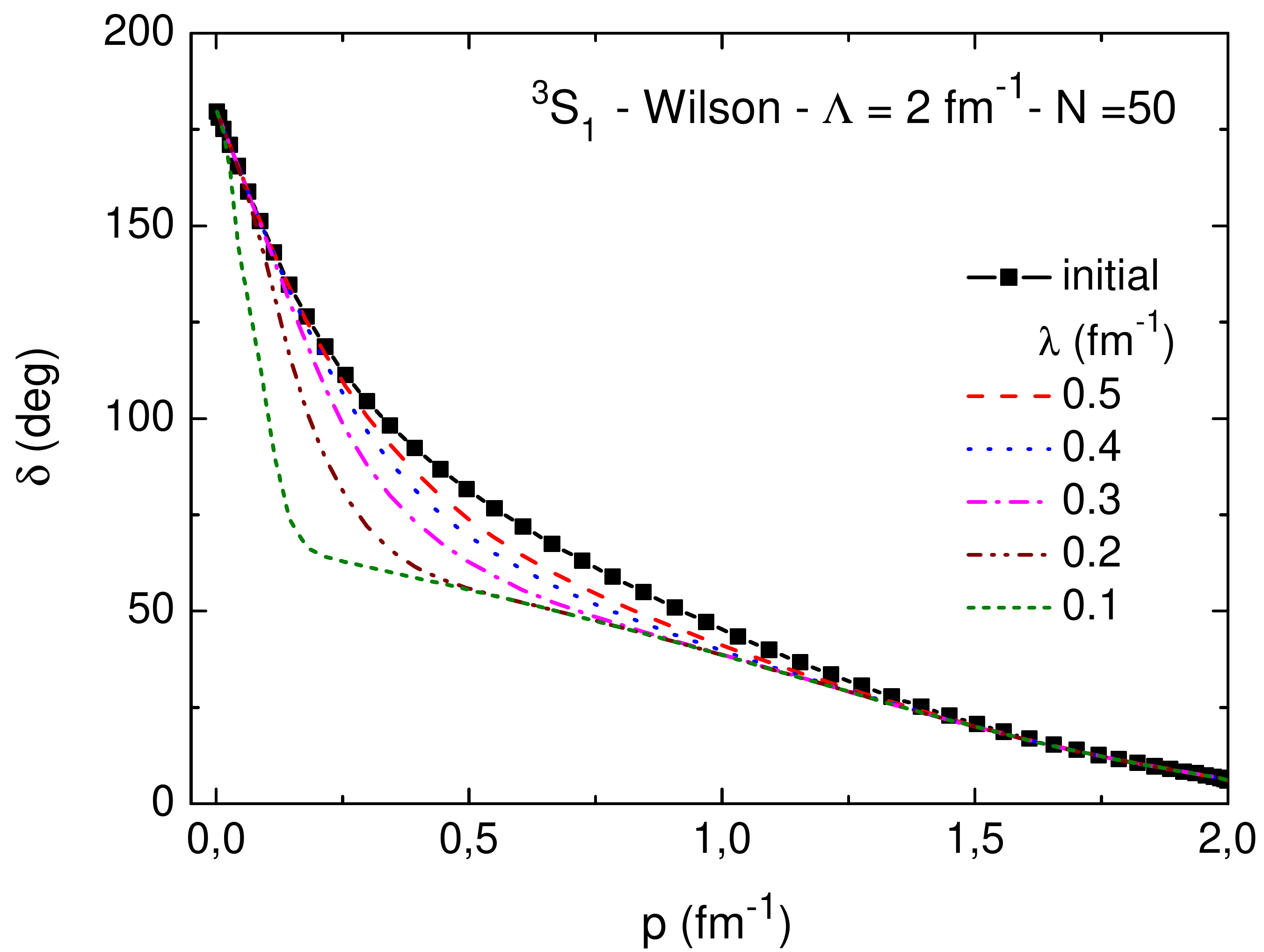}\\ \vspace{0.6cm}
\includegraphics[width=7.5cm]{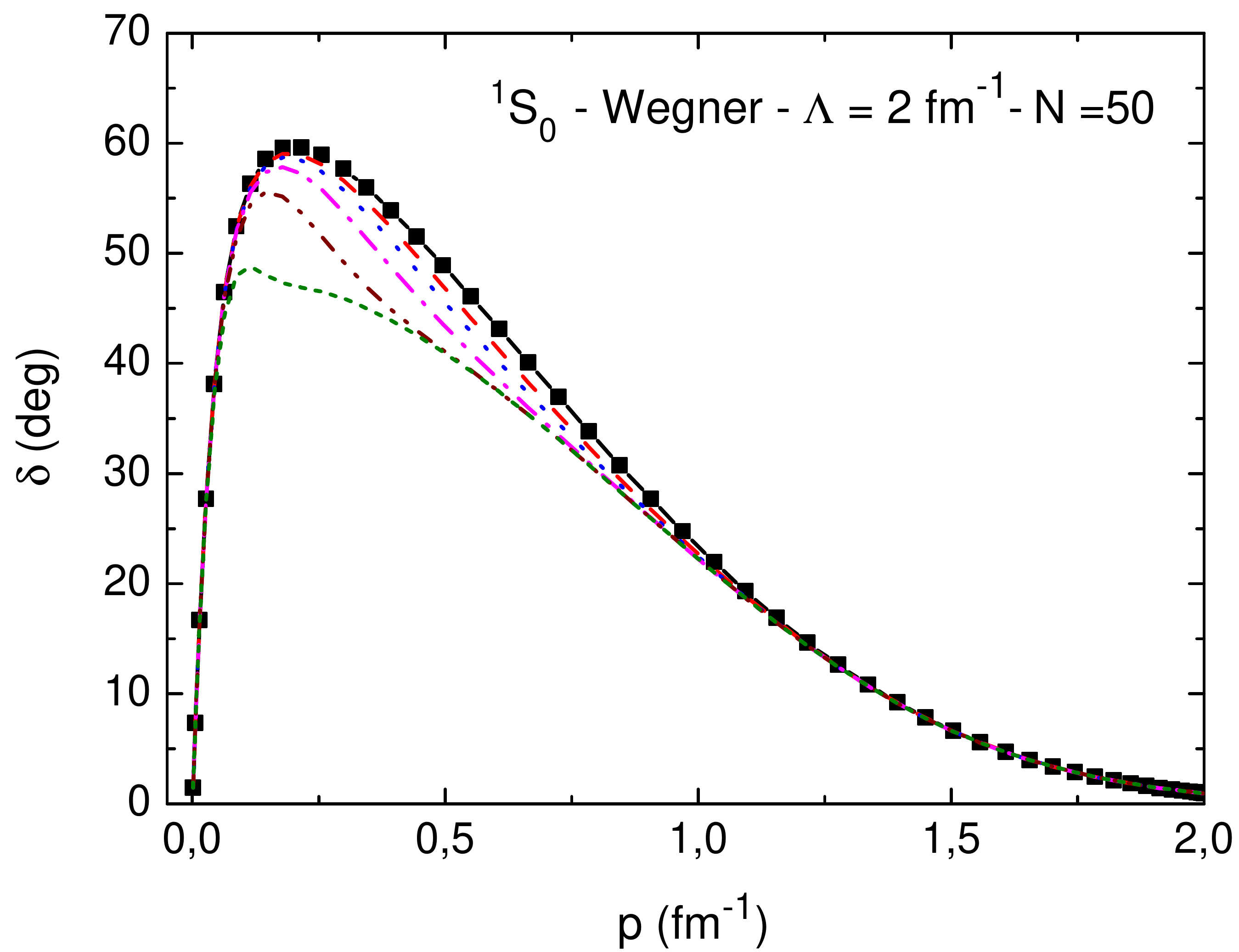}\hspace{0.5cm}
\includegraphics[width=7.5cm]{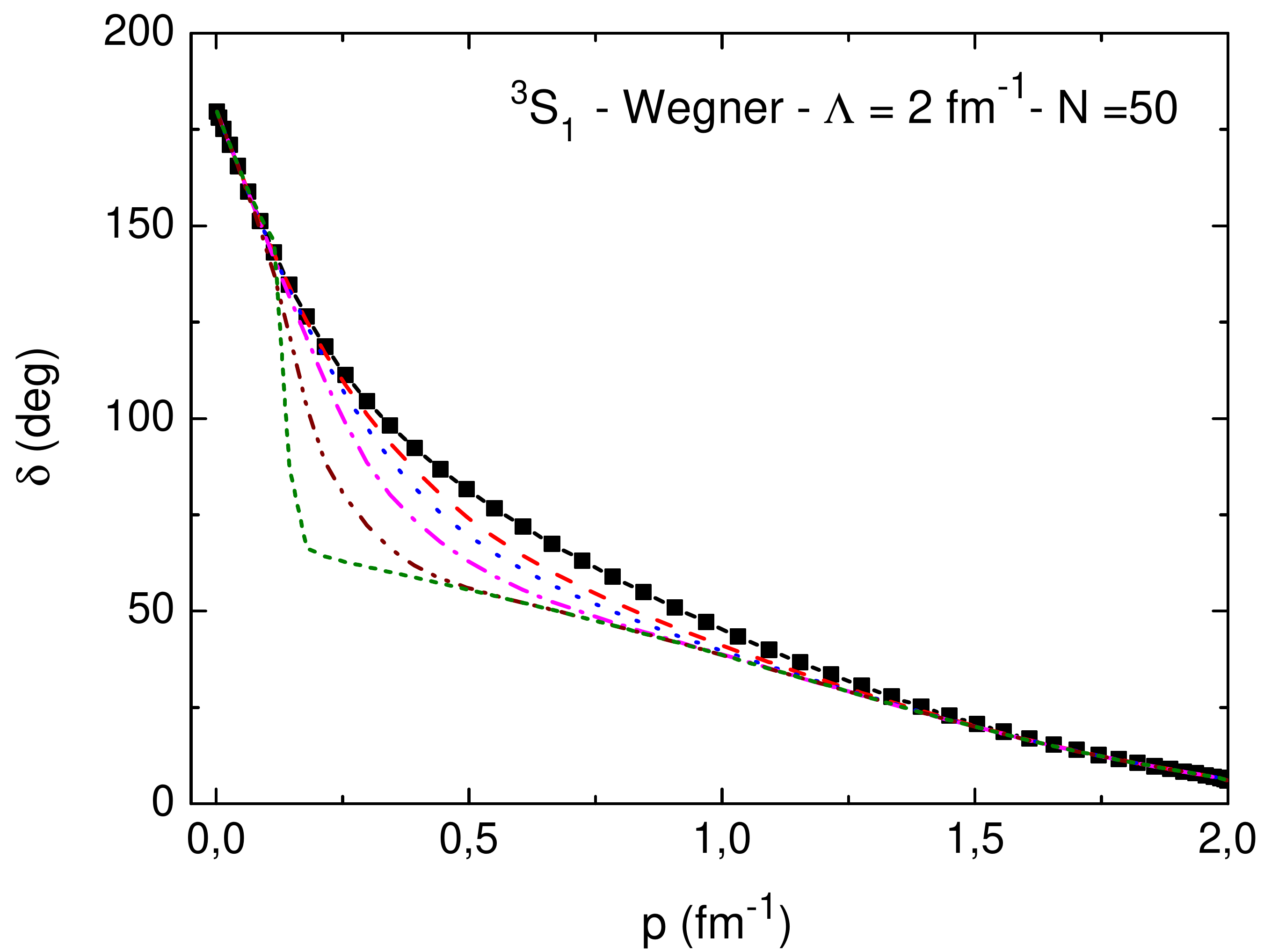}
\end{center}
\caption{Phase-shifts $\delta_\lambda^{\rm LS}(p_n)$ evaluated from the solution of the LS equation on a finite momentum grid  {\it at the grid points} ($\Lambda=2~{\rm fm}^{-1}$ and $N=50$) for the toy model separable gaussian potential in the $^1S_0$ and $^3S_1$ channels evolved through the SRG transformation with the Wilson and the Wegner generators for several values of the SRG cutoff $\lambda$.}
\label{fig:ps-LSgrid}
\end{figure}

In Fig.~\ref{fig:ps-LSgrid} we show the phase-shifts $\delta_\lambda^{\rm LS}(p_n)$ evaluated from the solution of the LS equation on a finite momentum grid {\it at the grid points} ($N=50$ and $\Lambda=2~{\rm fm}^{-1}$) for the toy model potential in the $^1S_0$ and $^3S_1$ channels evolved through the SRG transformation with the Wilson and the Wegner generators for several values of the SRG cutoff $\lambda$. As one can observe, Levinson's theorem~\cite{Ma:2006zzc} is fulfilled on the finite momentum grid for both channels, i.e.
\begin{equation}
\delta_\lambda^{\rm LS}(p_1)-\delta_\lambda^{\rm LS}(p_N)= N_B ~ \pi \; ,
\end{equation}
\noindent
with $N_B=0$ for the $^1S_0$ channel and $N_B=1$ for the $^3S_1$ channel. However, one can also clearly see that the LS phase-shifts on the finite grid are not independent of the SRG cutoff $\lambda$. As pointed out in section \ref{phaseineq}, while the lack of phase-equivalence disappears in the continuum limit, i.e. for $N \to \infty$, the isospectral definition of the phase-shifts on the finite momentum grid based on the energy-shift formula Eq.~(\ref{ES-formula}), which we refer to as the {\it eigenphases} $\delta^{\rm ES}(p_n)$, preserves phase-equivalence along the SRG trajectory for {\it any} number of grid points $N$. Furthermore, as we will show below, a proper prescription to order the spectrum of $N$ discrete eigenvalues $P_{n}^2$ of the diagonalized hamiltonian when using the energy-shift approach is required to obtain eigenphases which also fulfill Levinson's theorem, particularly when bound-states are allowed by the interaction as in the case of the toy model potential in the $^3S_1$ channel.

One possible ordering prescription is that derived by Kukulin et al.~\cite{kukulin2009discrete,Rubtsova:2010zz} for a hamiltonian allowing for $N_B$ bound-states. According to this prescription, the eigenvalues $P_{n}^2$ of the hamiltonian are arranged in ascending order and then left-shifted $N_B$ positions with respect to the eigenvalues $p_{n}^2$ of the kinetic energy operator $T$, which yields
\begin{eqnarray}
\delta_{\rm Kuk}^{\rm ES}(p_n) = - \pi \frac{P_{n+N_B}^2-p_n^2}{2 w_n p_n} \; ,
\label{deltakuk}
\end{eqnarray}
\noindent
with $n=1, \cdots , ~ N-N_B$. Note that Kukulin's prescription implies that the $N_B$ negative bound-state eigenvalues are just removed, such as to avoid discontinuities in the calculation of the eigenphases.

\begin{figure}[t]
\begin{center}
\includegraphics[width=7.5cm]{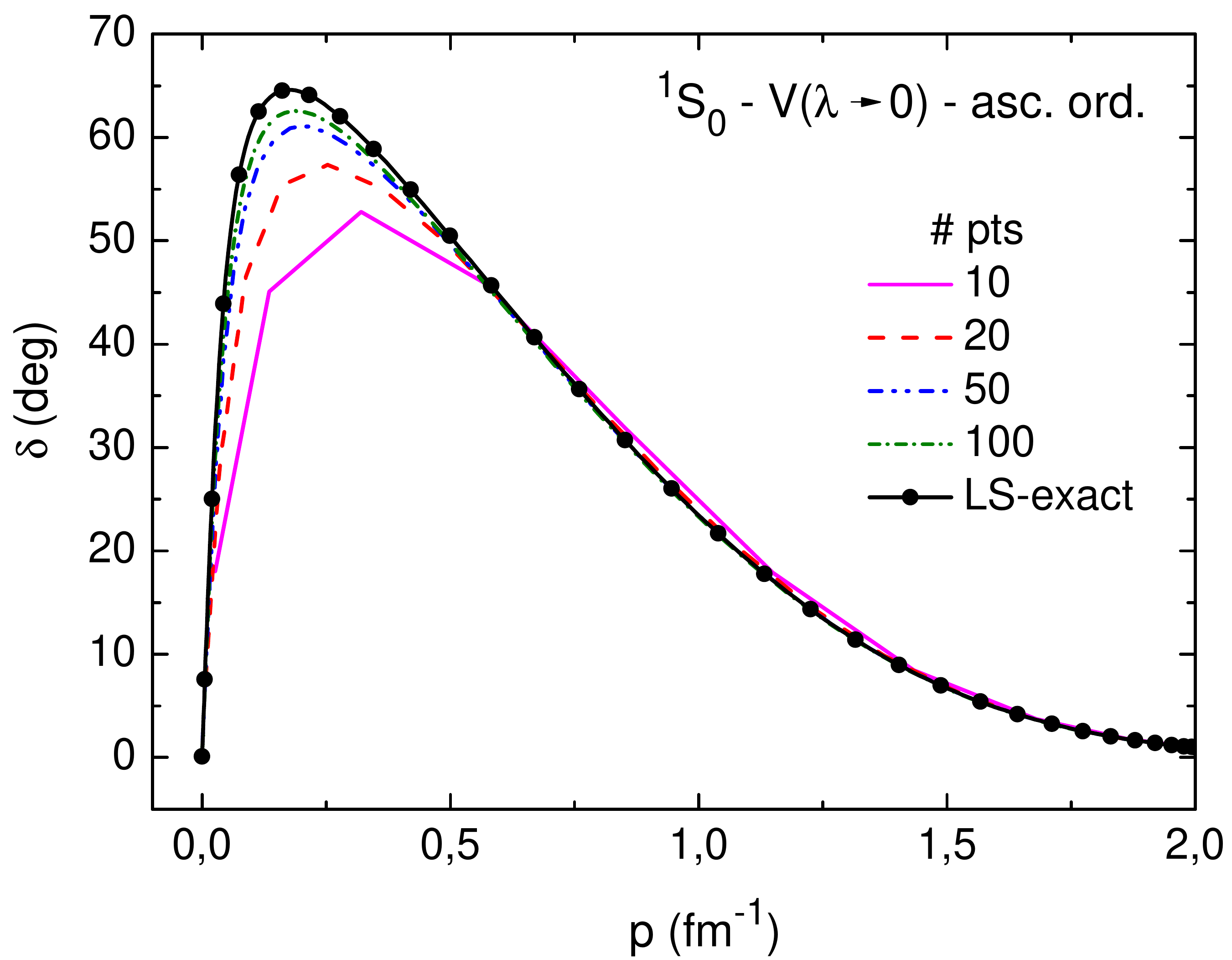}
\end{center}
\caption{Eigenphases $\delta^{\rm ES}(p_n)$ for the toy-model separable gaussian potential in the $^1S_0$ channel on a finite momentum grid ($\Lambda=2~{\rm fm}^{-1}$ and different number of grid points $N$) evaluated from the energy-shift formula with the eigenvalues arranged in ascending order. For comparison, we also show the exact phase-shifts obtained from the solution of the standard LS equation in the continuum limit ($N \to \infty$).}
\label{fig:13}
\end{figure}

In the case of the $^1S_0$ channel there is no need to shift the eigenvalues $P_{n}^2$, since there are no bound-states ($N_B=0$). Thus, we can evaluate the eigenphases $\delta^{\rm ES}(p_n)$ from the energy-shift formula just by arranging the eigenvalues in ascending order, which corresponds to the order induced by the SRG evolution in the $^1S_0$ channel with both the Wilson and the Wegner generators. The resulting eigenphases for the toy-model potential on a finite momentum grid ($\Lambda=2~{\rm fm}^{-1}$ and different number of grid points $N$) are shown in Fig.~\ref{fig:13}, compared to the exact phase-shifts obtained from the solution of the standard LS equation in the continuum limit ($N \to \infty$). As one can see, the ascending order prescription allows to obtain eigenphases which display the correct behavior in the entire range of momenta, thus complying to Levinson's theorem, i.e. $\delta^{\rm ES}(p_1)- \delta^{\rm ES}(p_N) \sim 0$.

%%%% 1S0 eigenphases ascending order

In the case of the $^3S_1$ channel, which allows for the Deuteron bound-state ($N_B=1$), we have considered three different prescriptions to evaluate the eigenphases $\delta^{\rm ES}(p_n)$ from the energy-shift formula:
\\ \\
\noindent
(i) eigenvalues $P_{n}^2$ arranged according to Kukulin's prescription, i.e. in ascending order and shifted one position to the left such as to remove the negative Deuteron bound-state eigenvalue.
\\ \\
\noindent
(ii) eigenvalues $P_{n}^2$ arranged according to the order induced by the SRG evolution in the $^3S_1$ channel with the Wilson generator, i.e. in ascending order, which corresponds to Kukulin's prescription with no left-shifting of the eigenvalues and yields
\begin{eqnarray}
\delta_{\rm Wil}^{\rm ES}(p_n) = - \pi \frac{P_{n}^2-p_n^2}{2 w_n p_n} \; ,
\label{deltawil}
\end{eqnarray}
\noindent
Note that in this prescription the negative Deuteron bound-state eigenvalue is placed on the diagonal of the hamiltonian at the lowest momentum on the grid $p_1$.
\\ \\
\noindent
(iii) eigenvalues $P_{n}^2$ arranged according to the order induced by
the SRG evolution in the $^3S_1$ channel with the Wegner generator,
i.e. with the negative Deuteron bound-state eigenvalue placed at the
momentum $p_{n_{\rm BS}}$ on the diagonal of the hamiltonian and the
remaining positive eigenvalues in ascending order. The discontinuity
in the eigenphases at the momentum $p_{n_{\rm BS}}$ can be avoided
just by removing the Deuteron bound-state, as in Kukulin's
prescription, or by interpolating between the neighboring values,
which yields
\begin{eqnarray}
\delta_{\rm Weg}^{\rm ES}(p_n) =
\begin{cases}
- \pi \frac{P_{n+1}^2-p_n^2}{2 w_n p_n} \quad {\rm if}\; \quad n < n_{\rm BS}  \\ \\
- \pi \frac{\bar P_{n_{\rm BS}}^2-p_n^2}{2 w_n p_n} \; \quad {\rm if} \quad n=n_{\rm BS} \\ \\
- \pi \frac{P_n^2-p_n^2}{2 w_n p_n} \qquad {\rm if}\; \quad n > n_{\rm BS}
\end{cases}
\label{deltaweg}
\end{eqnarray}
\noindent
where $\bar P_{n_{\rm BS}}^2=(P_{n_{\rm BS}+1}^2+P_{n_{\rm BS}-1}^2)/2$. Note that in this prescription only the eigenvalues $P_{n}^2$ corresponding to momenta $p_n < p_{n_{\rm BS}}$ are shifted one position to the left.
\\

\begin{figure}[t]
\begin{center}
\includegraphics[width=7.5cm]{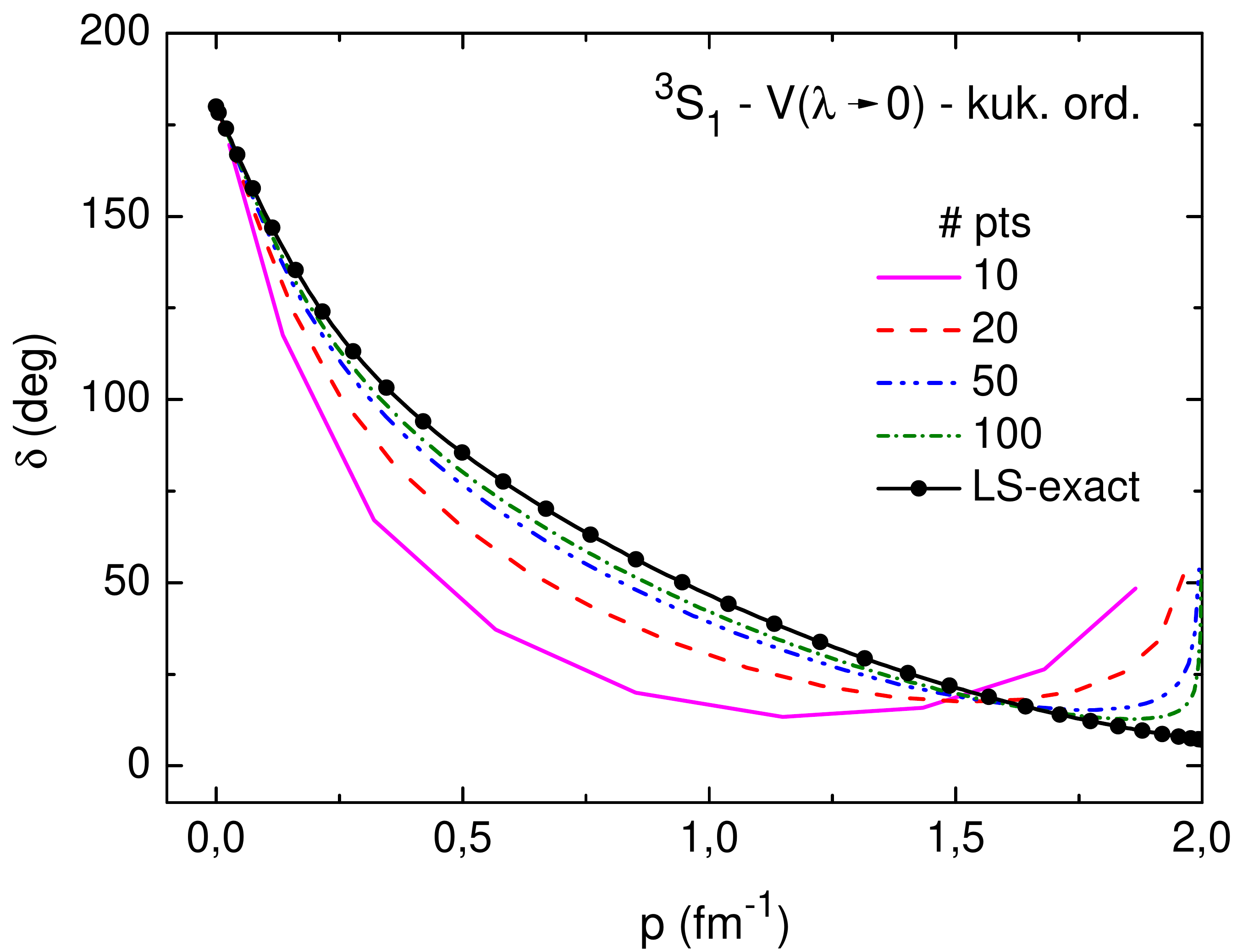}\hspace{0.5cm}
\includegraphics[width=7.5cm]{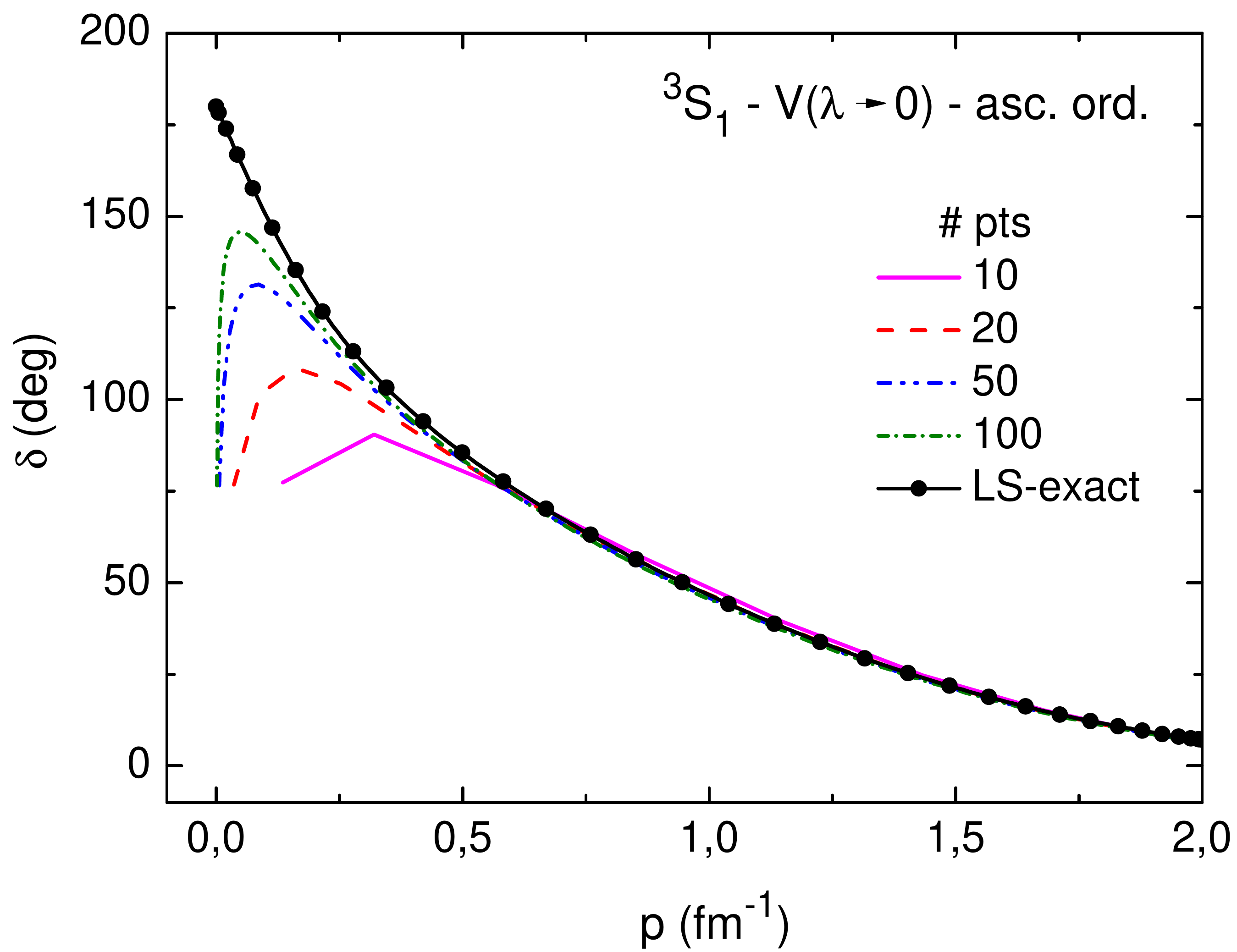}\\ \vspace{0.6cm}
\includegraphics[width=7.5cm]{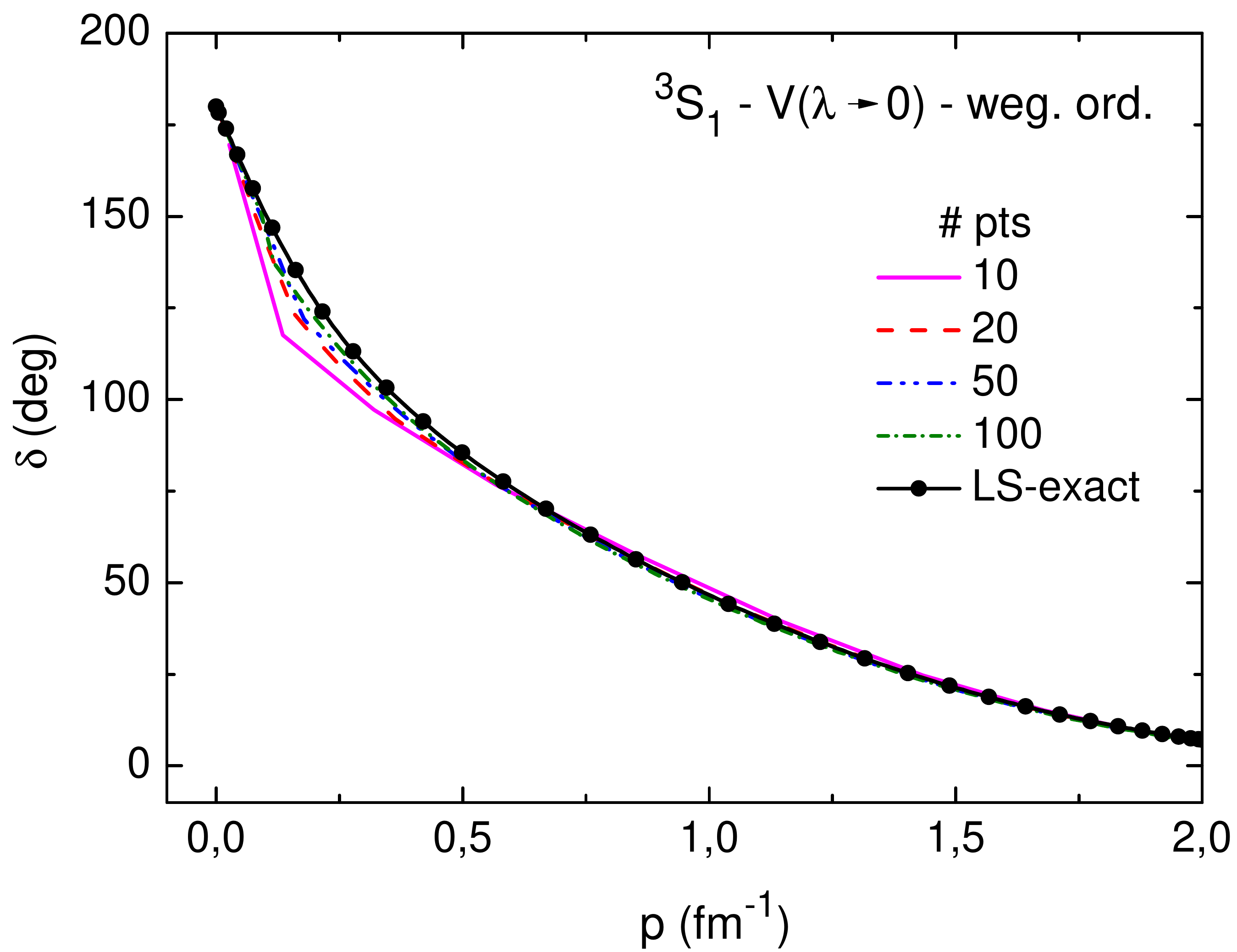}\hspace{0.5cm}
\includegraphics[width=7.5cm]{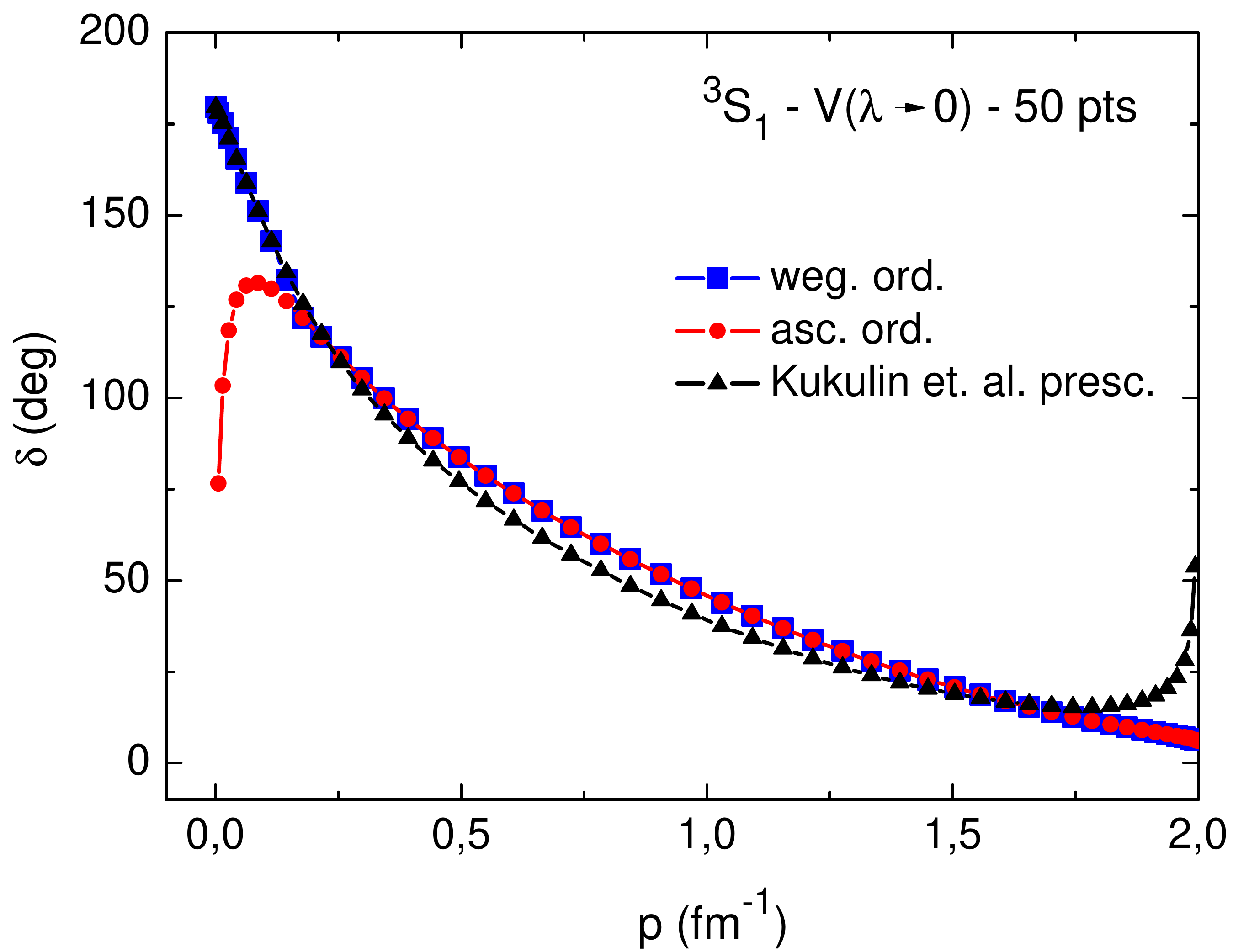}
\end{center}
\caption{Eigenphases $\delta^{\rm ES}(p_n)$ for the toy model separable gaussian potential in the $^3S_1$ channel on a finite momentum grid ($\Lambda=2~{\rm fm}^{-1}$ and different number of grid points $N$) evaluated from the energy-shift formula with the eigenvalues arranged according to Kukulin's prescription (i), the order induced by the SRG evolution with the Wilson generator (ii), i.e. in ascending order, and the order induced by the SRG evolution with the Wegner generator (iii). For comparison, we also show the exact phase-shifts obtained from the solution of the standard LS equation in the continuum limit ($N \to \infty$). In the bottom-right panel we compare the results obtained with prescriptions (i), (ii) and (iii) for $N=50$ grid points.}
\label{fig:14}
\end{figure}

In Fig.~\ref{fig:14} we show the eigenphases for the toy-model potential on a finite momentum grid ($\Lambda=2~{\rm fm}^{-1}$ and different number of grid points $N$) evaluated from the energy-shift formula with the eigenvalues $P_{n}^2$ of the hamiltonian  arranged according to the three prescriptions outlined above, compared to the exact phase-shifts obtained from the solution of the standard LS equation in the continuum limit ($N \to \infty$). As one can see, the eigenphases obtained with prescription (i) and prescription (ii), which are shown respectively in the top-left and the top-right panels, clearly violate Levinson's theorem. Prescription (i) yields eigenphases which have a proper low-momenta behavior but become distorted at high-momenta due to the left-shifting of the eigenvalues. Prescription (ii), on the other hand, yields eigenphases which have a proper high-momenta behavior but are distorted at low-momenta due to the presence of the Deuteron bound-state~\footnote{The point corresponding to the lowest momentum $p_1$ on the grid, at which the Deuteron bound-state eigenvalue is placed on the diagonal of the hamiltonian, is out of scale and so is omitted from the plot.}. One should note that both in the case of prescription (i) and prescription (ii) the distortion of the eigenphases is generated by a mismatch between the eigenvalues $P_{n}^2$ of the hamiltonian and the eigenvalues $p_{n}^2$ of the kinetic energy operator $T$, respectively at high-momenta and low-momenta. As shown in the bottom-left panel, prescription (iii) yields eigenphases which have a proper behavior both at low-momenta and high-momenta, since the left-shifting is applied {\it only} to the eigenvalues corresponding to momenta $p_n < p_{n_{\rm BS}}$. As mentioned before, the position $p_{n_{\rm BS}}$ of the Deuteron bound-state eigenvalue induced by the SRG evolution with the Wegner generator depends both on the number of grid points $N$ and the cutoff $\Lambda$. In the bottom-right panel, we compare the results obtained with the three prescriptions for $N=50$ grid points. As one can see, the eigenphases obtained with prescription (iii) indeed match those obtained with prescription (i) and prescription (ii) respectively for momenta below and above $p_{n_{\rm BS}}$. Thus, we find that the final ordering of the spectrum of discrete eigenvalues $P_{n}^2$ of the toy-model hamiltonian induced by the SRG evolution with the Wegner generator in the infrared limit ($\lambda \to 0$) remarkably provides a prescription which allows to obtain isospectral phase-shifts from the energy-shift formula that comply to Levinson's theorem in the presence of the Deuteron bound-state.

%%%% 3S1 eigenphases ascending, wegner and Kukulin order

\begin{figure}[t]
\begin{center}
\includegraphics[width=7.5cm]{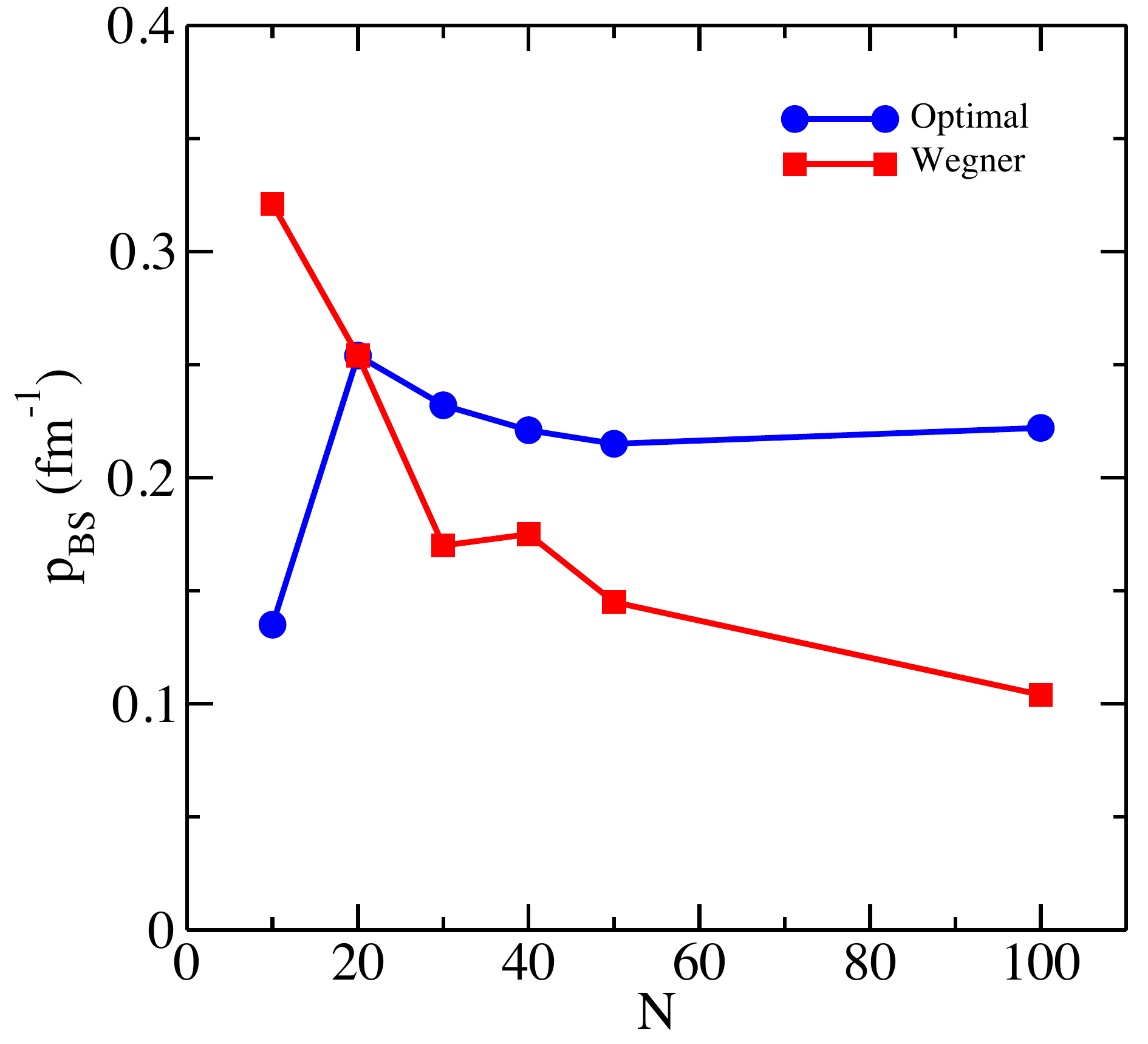}
\end{center}
\caption{Comparison between the position of the Deuteron bound-state which minimizes the RMS errors in the eigenphases $\delta^{\rm ES}(p_n)$ with respect to the exact phase-shifts obtained from the solution of the standard LS equation in the continuum limit ($N \to \infty$) and the position induced by the SRG evolution with the Wegner generator in the infrared limit ($\lambda \to 0$), for different number of grid points $N$.}
\label{fig:15}
\end{figure}

The effectiveness of ordering prescription (iii) can be traced to the decoupling of the Deuteron bound-state from the low-momentum scales when the toy-model potential in the $^3S_1$ channel is evolved through the SRG transformation using the Wegner generator. The key point is that by placing the Deuteron bound-state eigenvalue at the position $p_{n_{\rm BS}}$ induced by the SRG evolution with the Wegner generator in the infrared limit ($\lambda \to 0$) we obtain eigenphases from the energy-shift formula which display the correct behavior in the entire range of momenta (within the expected uncertainties of the finite momentum grid). One should note, however, that such an ordering does not necessarily correspond to the optimal one. We have evaluated the eigenphases $\delta^{\rm ES}(p_n)$ for the toy model potential in the $^3S_1$ channel on a finite momentum grid ($\Lambda=2~{\rm fm}^{-1}$ and different number of grid points $N$) using prescription (iii) but varying the position of the Deuteron bound-state eigenvalue $p_{n_{\rm BS}}$, and compared the results to the exact phase-shifts obtained from the solution of the standard LS equation in the continuum limit ($N \to \infty$). By computing the corresponding root mean square (RMS) errors as a function of $p_{n_{\rm BS}}$, we find that for each value of the number of grid points $N$ there is a well-defined minimum. As one can see in Fig.~\ref{fig:15}, the position $p_{n_{\rm BS}}^{\rm opt}$ minimizing the RMS errors, which seems to converge in the continuum limit to the characteristic Deuteron momentum scale $\gamma = 0.23~{\rm fm}^{-1}$, is different from that induced by the SRG evolution with Wegner generator in the infrared limit.

As pointed out before, the final ordering of the spectrum induced by the SRG evolution in the infrared limit may depend on the choice of the SRG generator when bound-states are allowed by the interaction. In this way, it is possible that a specific generator exists for which the position of the Deuteron bound-state induced by the SRG evolution corresponds to $p_{n_{\rm BS}}^{\rm opt}$, thus providing the optimal ordering prescription for the energy-shift approach.

%% Bound State Position

%%%

\section{Consequences of the fixed points for nuclear binding: Triton and Helium nuclei}
\label{sec:nucl}

The ambiguities arising from off-shell uncertainties has been the main
source of problems in Nuclear Physics calculations. There are a
variety of $NN$ potentials in the literature providing a high-quality
description of scattering data and Deuteron properties. They produce,
however, different results when comparing to nuclear matter and
nuclear structure calculations. So far, we have dealt with the
two-body problem. Of course, much of the motivation to undertake the
SRG evolution concerns the applications to nuclear structure. Our
purpose here is to illustrate the different implications of the SRG
using the Wegner and the Wilson generators along the lines discussed
above using our simple toy model potential. Most calculations in
Nuclear Physics use a mean field reference state, which is often a
harmonic oscillator (HO) shell-model state, upon which correlations
are built, namely $|\Psi \rangle = U | \phi \rangle $, where $U$ is
some unitary transformation.  The SRG interpretation is that
correlations are shifted from the reference state to the interaction
via the unitary transformation $U$.

Within the shell-model we will restrict to the cases with $A=3$ and $A=4$, which correspond respectively to the Triton and the $\alpha$-particle, as only $S$-wave interactions are required. In order to simplify matters as much as
possible we will describe the states as $t \equiv (1s)^3$ and $ \alpha
\equiv (1s)^4$ respectively,
\begin{eqnarray}
\psi_{t, \uparrow} (p_1,p_2,p_3)  &\equiv& \left[ \prod_{i=1}^3\varphi_{1s} (p_i) \right]
 {\cal A} (p \uparrow , n \uparrow
, n \downarrow ) \; , \\
\psi_{t, \downarrow} (p_1,p_2,p_3)  &\equiv& \left[ \prod_{i=1}^3\varphi_{1s} (p_i) \right]
 {\cal A} (p \downarrow , n \uparrow
, n \downarrow ) \; , \\
\psi_{\alpha} (p_1,p_2,p_3,p_4)  &\equiv& \left[ \prod_{i=1}^4\varphi_{1s} (p_i) \right]
{\cal A} (p \uparrow , n \uparrow , p \downarrow , n \downarrow) \; ,
\end{eqnarray}
\noindent
where ${\cal A}$ is a normalized antisymmmetrizer and $\varphi_{1s}$ is
the ${\rm HO}-1s$ wave-function with a $b$ parameter, given by
\begin{eqnarray}
\sqrt{\frac{2}{\pi}} \varphi_{1s}(p) = \frac{2 \sqrt{b^3} e^{-\frac{1}{2} b^2 p^2}}{\sqrt[4]{\pi }} \; .
\end{eqnarray}
\noindent
The energy calculation for the $A=3$ and $A=4$ systems at rest, $\sum_{i=1}^A p_i=0$,
is straightforward and reads
\begin{eqnarray}
\{ E_t(\lambda), E_\alpha(\lambda)\}=\{-B_t,-B_\alpha\} = \min_b \left[
  (A-1) \left\langle \frac{p^2}{2M} \right\rangle_{1s} + \frac{A(A-1)}{2} \frac12
  \langle V_{^1S_0,\lambda} + V_{^3 S_1,\lambda} \rangle_{{\rm rel},1s} \right]
\Big|_{A=3,4} \; .
\label{eq:b34}
\end{eqnarray}
\noindent
The single particle kinetic energy reads
\begin{eqnarray}
\left\langle \frac{p^2}{2M} \right\rangle_{1s} = \frac2{\pi}
  \int_0^\infty p^2 \, dp \, \left[\varphi_{1s}(p)\right]^2
  \frac{p^2}{2M} \; ,
\end{eqnarray}
\noindent
and the potential matrix-element is defined as
\begin{eqnarray}
\langle \phi_{\rm rel} | V_{\lambda} | \phi_{\rm rel} \rangle =
 \frac2{\pi} \int_0^\infty dp \,  [p^2 \phi_{\rm rel} (p) ]
\frac2{\pi}\int_0^\infty dp'\, [p'^2 \phi_{\rm rel} (p') ] ~ V_{\lambda}(p',p) \;,
\end{eqnarray}
\noindent
where the {\it relative} wave function is given by (note the $\sqrt{2}$
factor),
\begin{equation}
\phi_{\rm rel}(p)= \varphi_{\rm 1s} \left(p,b/\sqrt{2}\right)\; .
\end{equation}
\noindent
Eq.~(\ref{eq:b34}) can be interpreted in terms of the number of
pairs in the $^1S_0$ and $^3S_1$ states being $n_{^1S_0,~t}=n_{^3S_1,~t}=3/2$ for the Triton and $n_{^1S_0,~\alpha}=n_{^3S_1,~\alpha} = 6/2 $ for the $\alpha$-particle.

Within this shell-model calculation scheme we may also define a variational Deuteron energy,
\begin{equation}
E_d(\lambda)= \min_b ~ \langle p^2/M + V_{^3S_1,\lambda} \rangle_{{\rm rel},1s} \; ,
\end{equation}
\noindent
which unlike the exact one will depend on the SRG cutoff $\lambda$. For $\lambda \to \infty$
the Deuteron is unbound by $0.2 ~ {\rm MeV}$, and the binding-energy obtained for the
Triton is $B_t^{\rm Var} = 5.9661 ~ {\rm MeV}$ to be compared with the
exact Faddeev equation result $B_t = 6.65543 ~ {\rm MeV}$, whereas the
$\alpha$-particle yields $B_\alpha^{\rm Var} = 32.1054 ~ {\rm MeV}$.

%% Binding energies

\begin{figure}[h]
\begin{center}
\includegraphics[width=7cm]{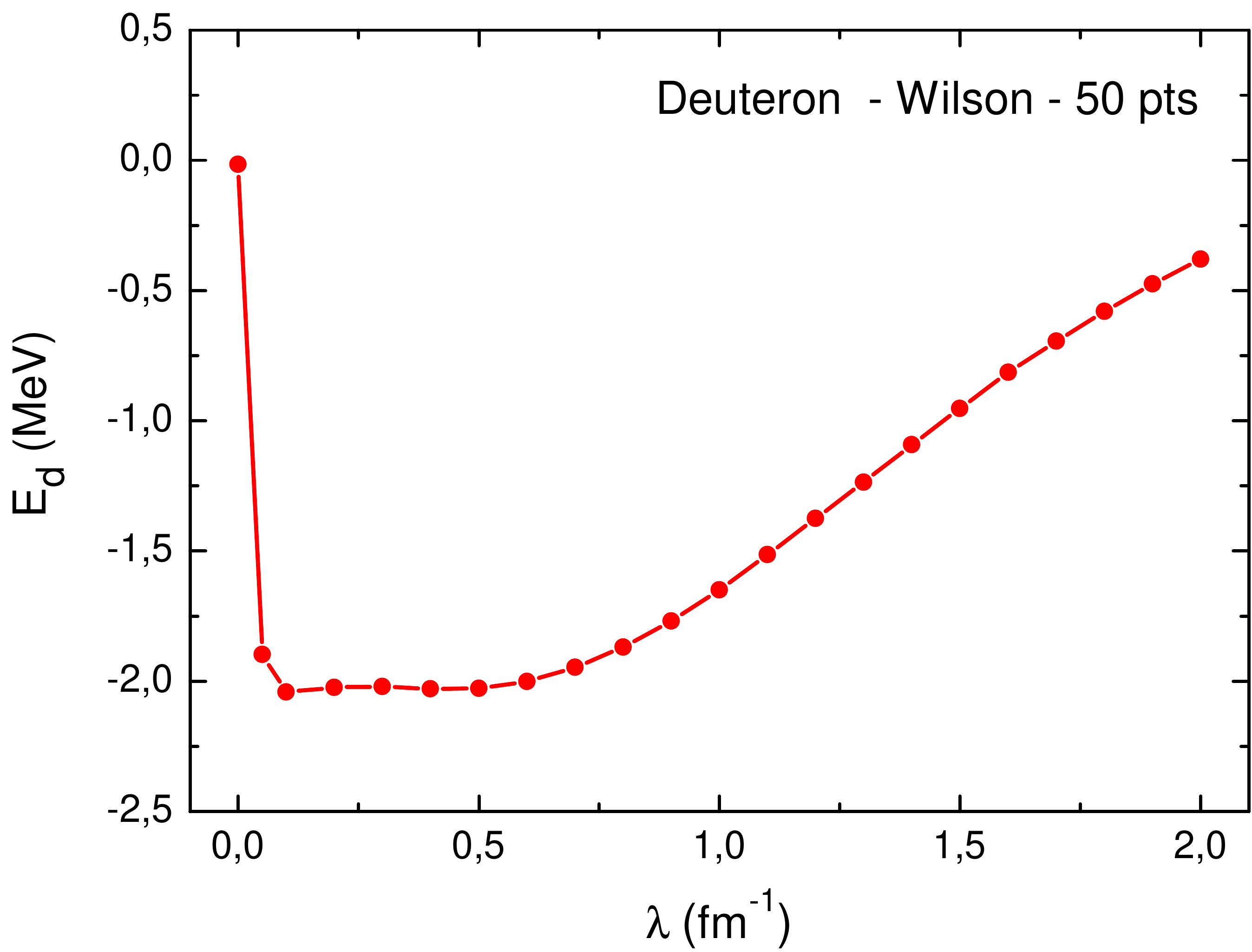}\hspace{0.5cm}
\includegraphics[width=7cm]{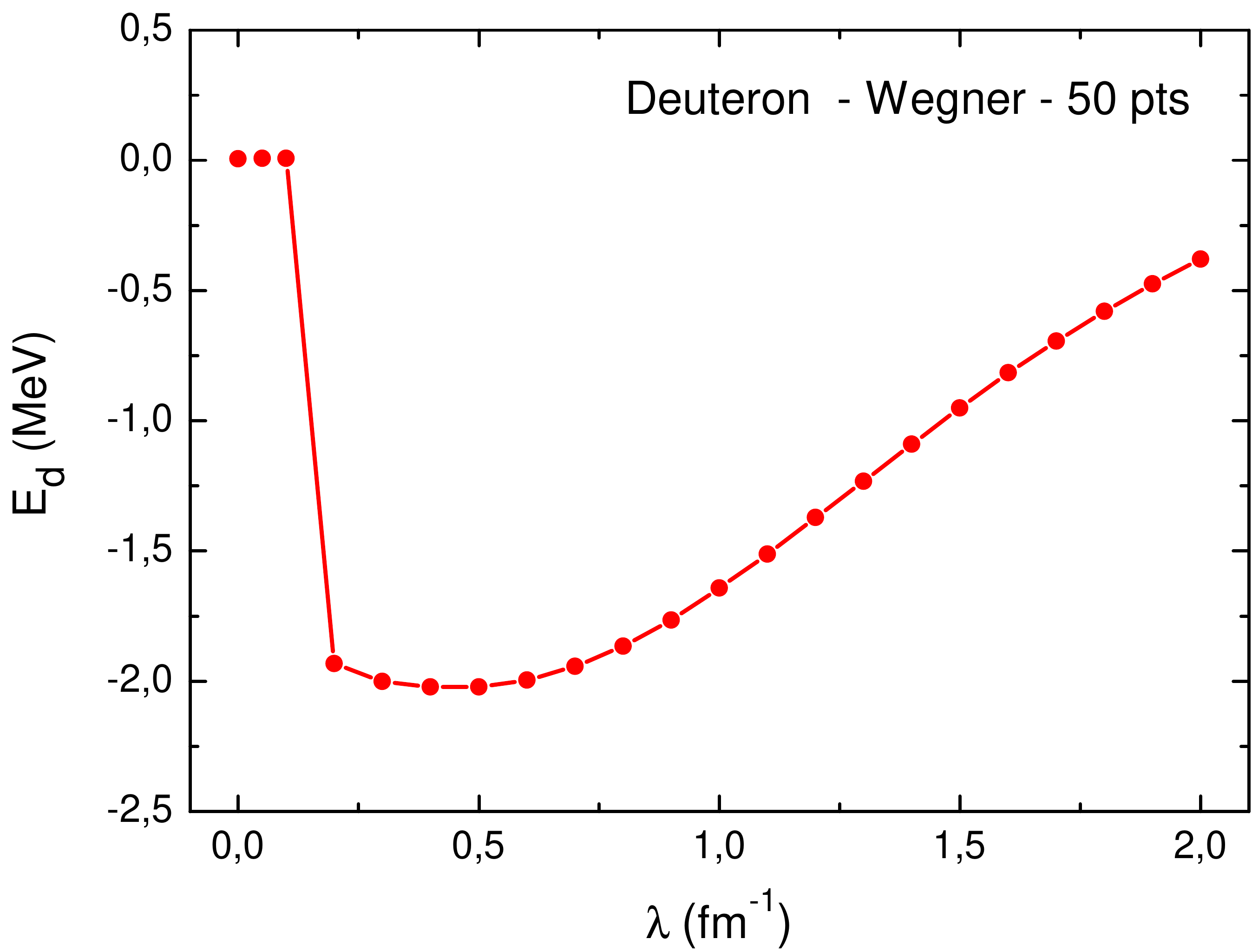}\\\vspace{0.6cm}
\includegraphics[width=7cm]{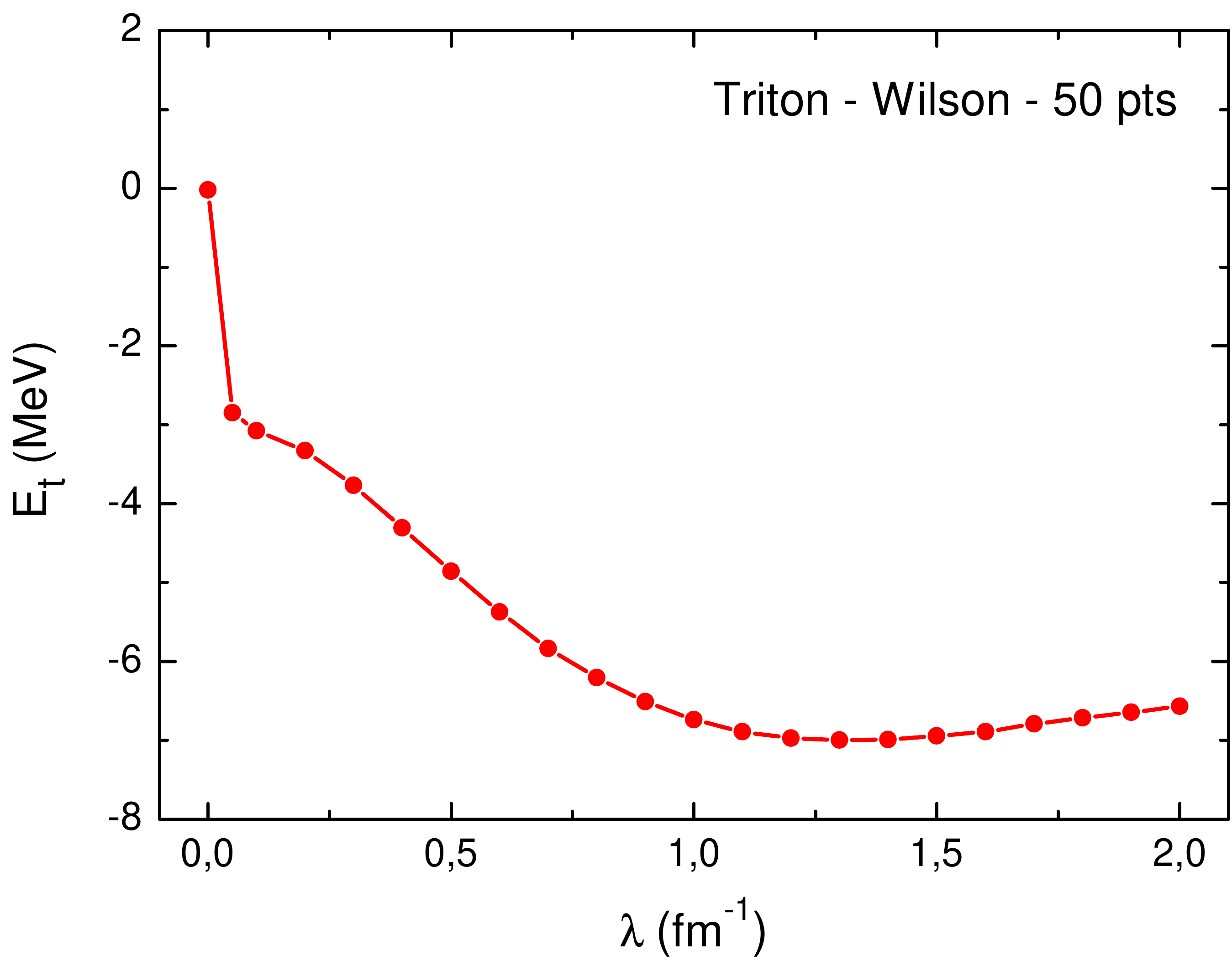}\hspace{0.5cm}
\includegraphics[width=7cm]{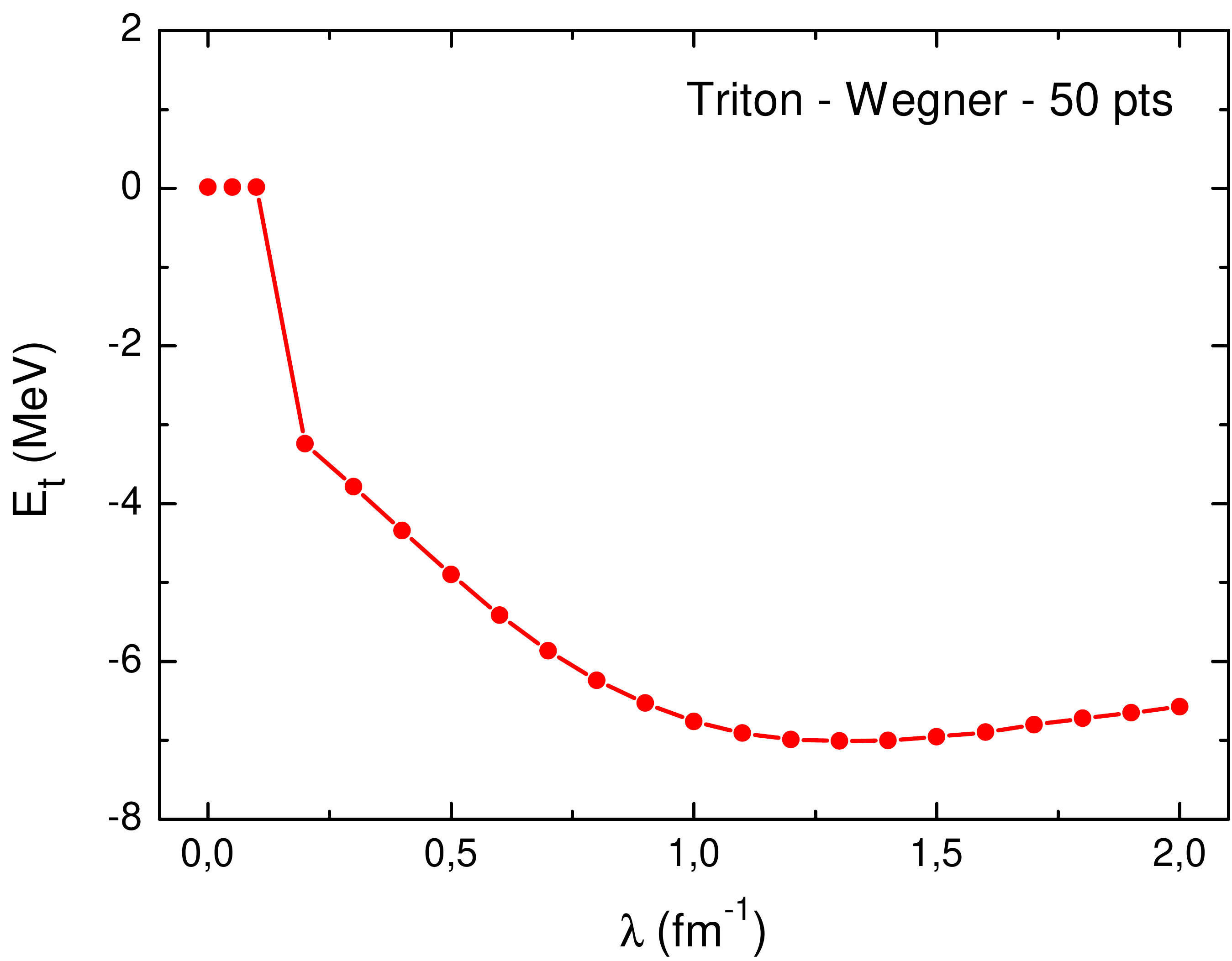}\\\vspace{0.6cm}
\includegraphics[width=7cm]{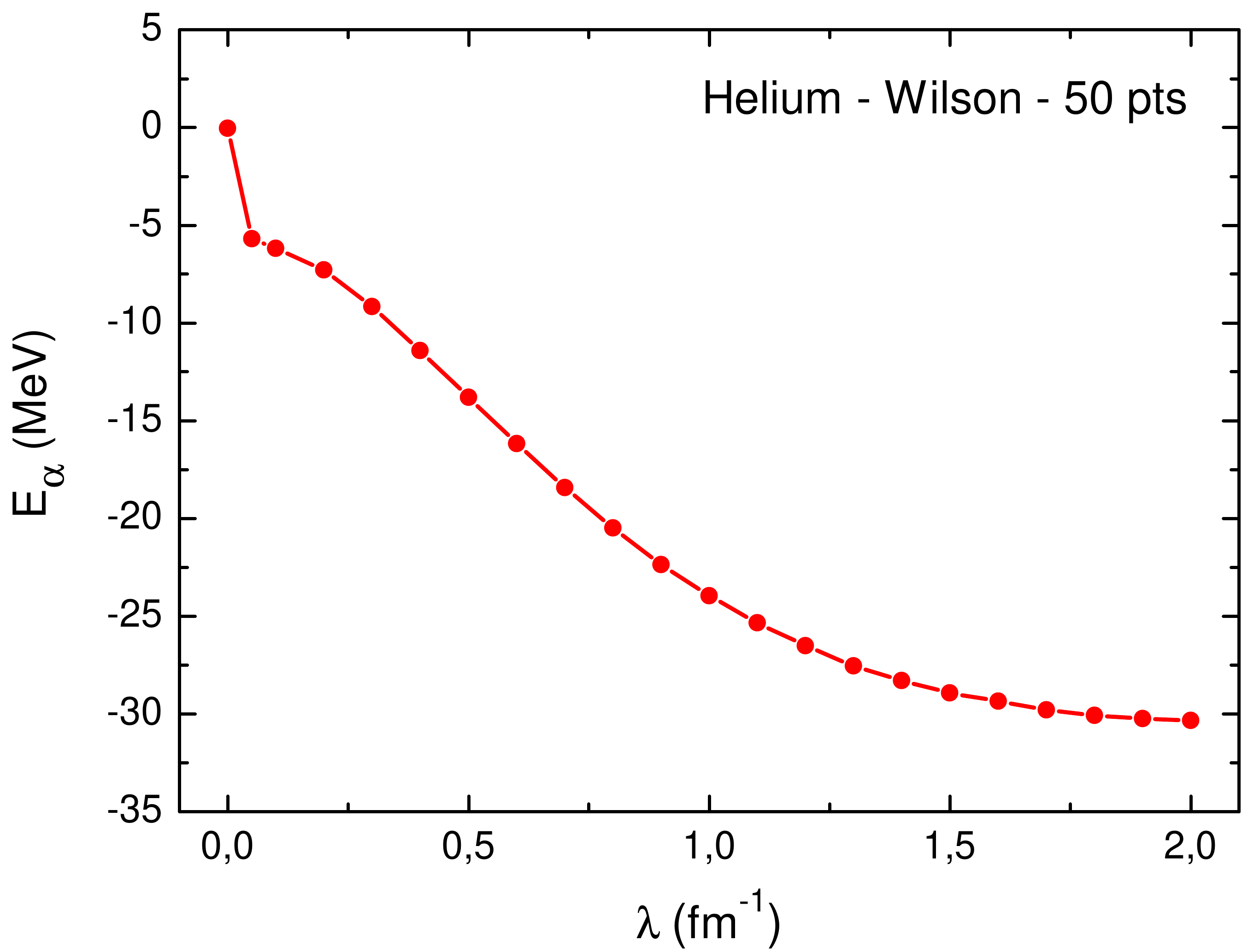}\hspace{0.5cm}
\includegraphics[width=7cm]{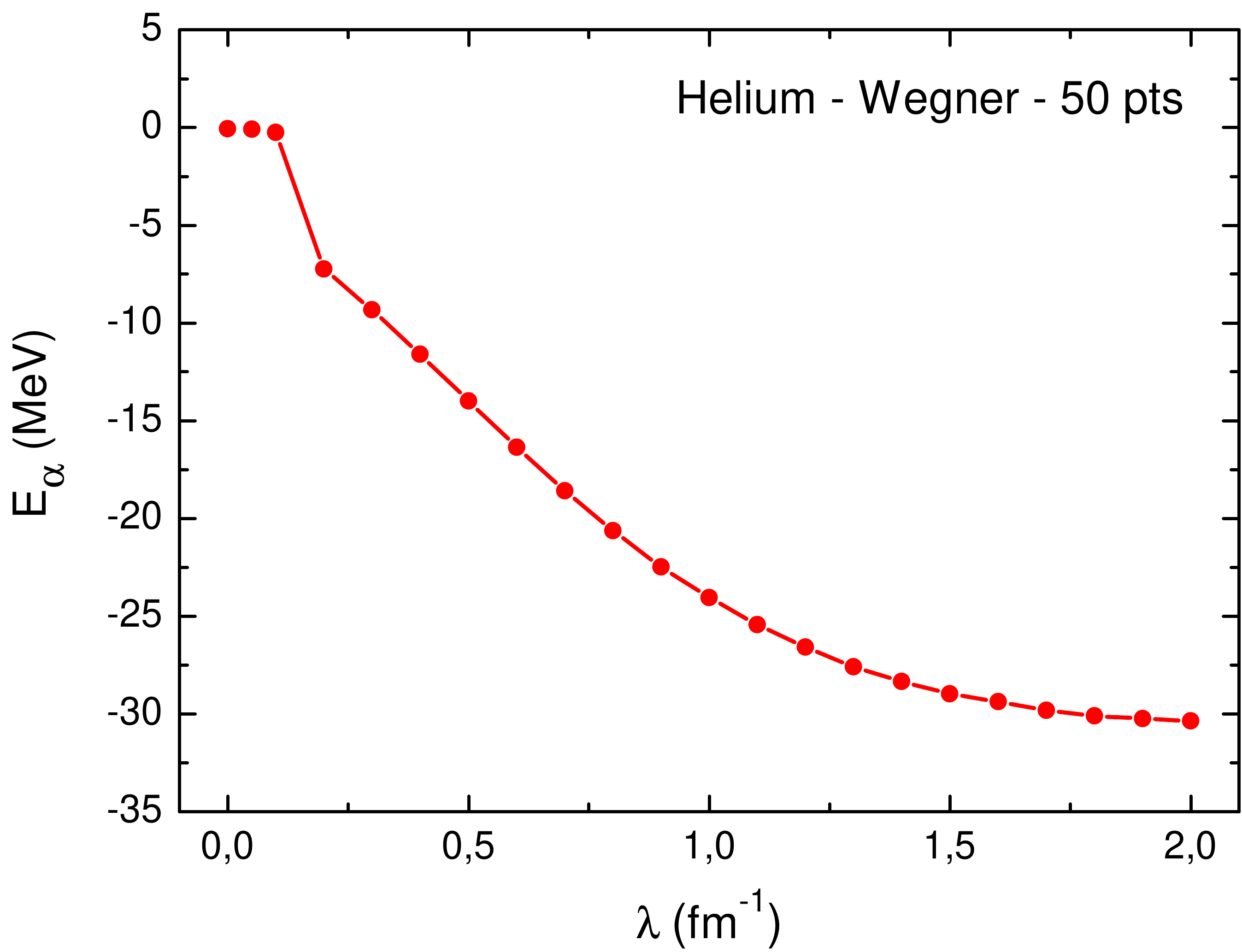}
\end{center}
\caption{Shell-model variational energies for Deuteron, Triton and Helium obtained from the toy-model separable gaussian potentials ($\Lambda=2~{\rm fm}^{-1}$ and $N=50$) evolved through the SRG transformation using the Wilson and the Wegner generators as a function of the SRG cutoff $\lambda$.  }
\label{fig:binding}
\end{figure}

%% Tjon-lines

\begin{figure}[h]
\begin{center}
\includegraphics[width=7.5cm]{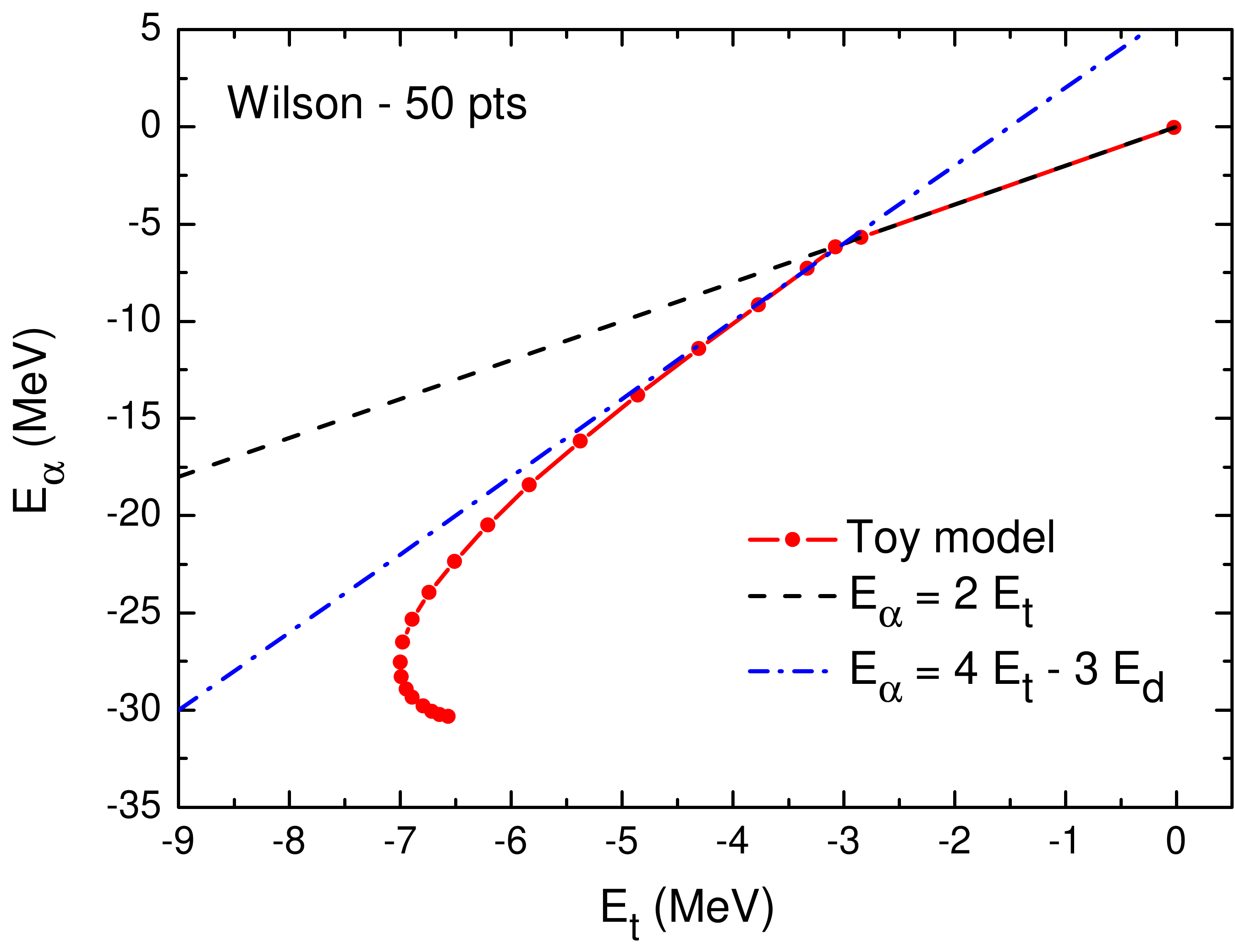}\hspace{0.5cm}
\includegraphics[width=7.5cm]{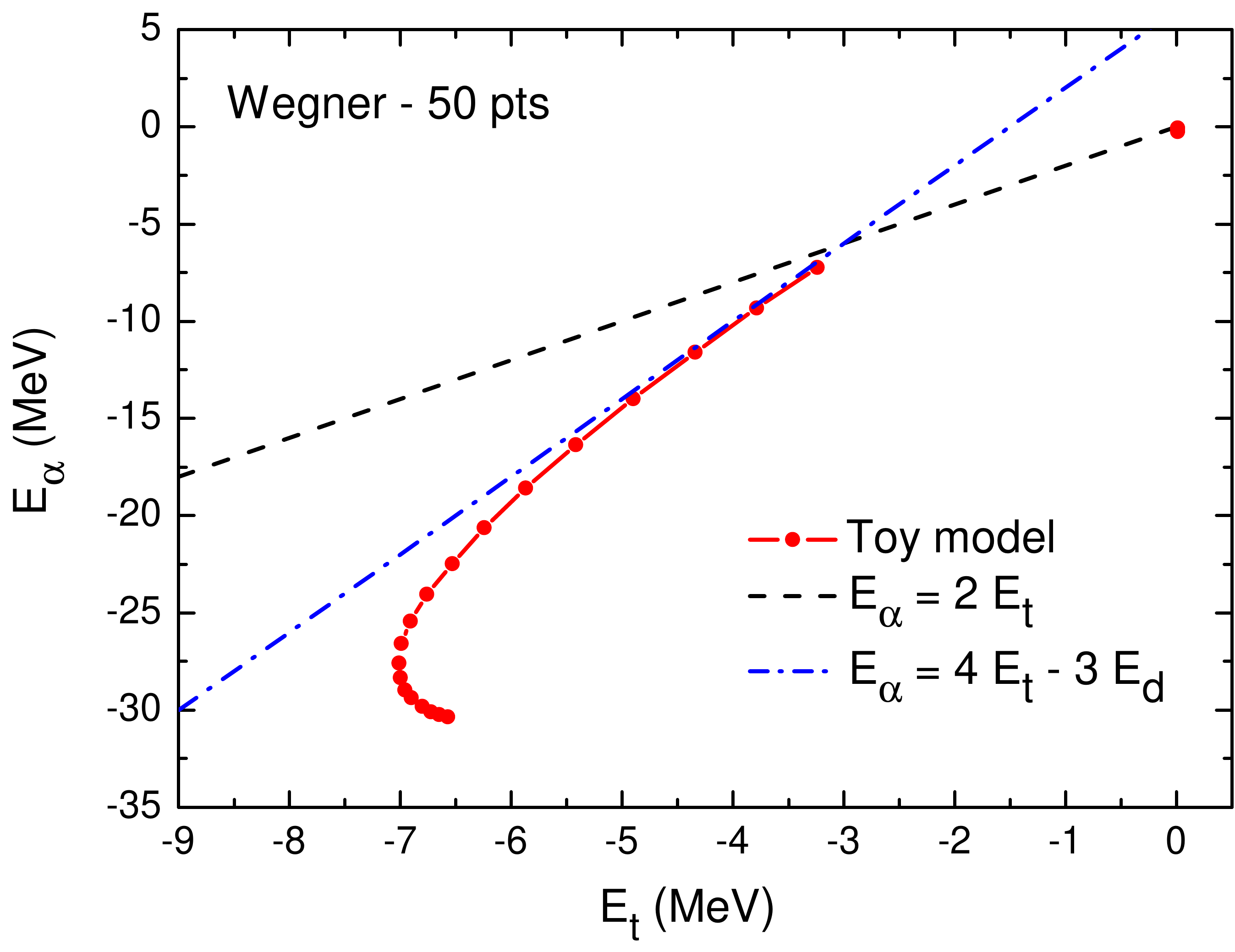}
\end{center}
\caption{Tjon-lines obtained from the toy-model separable gaussian potentials ($\Lambda=2~{\rm fm}^{-1}$ and $N=50$) evolved through the SRG transformation using the Wilson and the Wegner generators. We compare with the asymptotic lines
$E_\alpha = 4 E_t - 3 E_d$ and $E_\alpha = 2 E_t$. }
\label{fig:Tjon}
\end{figure}

It is interesting to analyze the SRG evolution from $\lambda \to \infty$ down to $\lambda \to 0$,
disregarding here explicit three-body or four-body forces (see Ref.~\cite{Arriola:2013gya} for
some further details on this issue). In Fig.~\ref{fig:binding} we see
the effect of the SRG evolution on a simple variational calculation for
the cases with $A=2,3,4$. Here and for illustration purposes we take a grid with $\Lambda=2~{\rm fm}^{-1}$ and $N=50$ gauss points. A very similar behavior in the region $\lambda \gtrsim 1 {\rm fm}^{-1}$ is found for both
the Wilson and the Wegner generators. However, as we can see, in the region of small $\lambda$ the
behavior for the Wilson and the Wegner generators is quite different and,
near $\lambda \sim 0.2~ {\rm fm}^{-1}$, a clear jump in the energies is observed. To
understand how this surprising result comes about let us consider the
calculation of the energy in further detail for the finite momentum
grid which is used for the SRG evolution. The kinetic and potential
energy contributions become
\begin{equation}
\left\langle \frac{p^2}{2M} \right\rangle_{1s} \to  \frac{2}{\pi} \sum_n w_n p_n^2 \frac{p_n^2}{M}  \phi_{N,\rm rel} (p_n)^2 \;
\end{equation}
\noindent
and
\begin{equation}
\langle \phi_{\rm rel} |  V_{\lambda} | \phi_{\rm rel} \rangle \to
 \left(\frac2{\pi} \right)^2 \sum_{n,m} w_n w_m
[p_n^2 \phi_{\rm rel,N} (p_n)]
[p_m^2 \phi_{\rm rel,N} (p_m)] V_{nm}(\lambda) \; .
\end{equation}
\noindent
The ${\rm HO}-1s$ states need to be renormalized on the grid as follows,
\begin{eqnarray}
\varphi_{1s,N} (p_n) = Z_N \varphi_{1s} (p_n) \; ,  \qquad Z_N^2 = \frac{2}{\pi}
\sum_n w_n p_n^2 \left[ \varphi_{1s}(p_n) \right]^2 \; .
\end{eqnarray}
\noindent
If one takes the infrared limit $\lambda \to 0$, the potential becomes diagonal
and one gets
\begin{eqnarray}
\langle \phi_{\rm rel} |  V_{\lambda \to 0} | \phi_{\rm rel} \rangle \to
\left(\frac2{\pi} \right)^2 \sum_{n} w_n^2 p_n^4 \phi_{\rm rel} (p_n)^2 V_{n}(\lambda \to 0)\; .
\end{eqnarray}
\noindent
We see the weight squared factor $w_n^2$ and {\it naively} in the $\Lambda$-fixed and $N \to \infty $ continuum limit, we obtain
\begin{eqnarray}
V(N \to \infty)=\langle \phi_{\rm rel} | V_{\lambda \to 0} | \phi_{\rm rel} \rangle_{N \to \infty} \to
\left(\frac2{\pi} \right) \frac{\Lambda}{N} \int_0^\Lambda p^4 \phi_{\rm rel} (p)^2\left[\frac{-\delta(p)}{p} \right]\to 0 \; ,
\end{eqnarray}
\noindent
regardless on the HO parameter b. Thus, the kinetic energy gets
minimal for $b \to \infty$ and hence the minimum takes place at zero
energy. While this argument agrees with the behavior for the Wegner generator observed
in Fig.~\ref{fig:binding} it fails to reproduce the trend found for
the Wilson generator.  The flaw in the argument can be seen more clearly in the Deuteron case, where
\begin{eqnarray}
\lim_{\lambda \to 0} E_d (\lambda)=  \frac{2}{\pi} \sum_n w_n p_n^2 \frac{P_n^2}{M}  \phi_{\rm rel,N} (p_n)^2 \; .
\end{eqnarray}
\noindent
Note that with the exception of the Deuteron state $P_d^2 =
-\gamma_d^2$ all others fulfill $P_n^2 >0$. The difference, however,
is in the location of the bound-state after the SRG evolution. The bound-state is located at the lowest point on the grid for
the Wilson generator and at some intermediate point for the Wegner
generator,
\begin{eqnarray}
P_d^2 = P_1^2 \quad ({\rm Wilson~ generator}) \qquad {\rm and} \qquad P_d^2 = P_{n_B}^2 \quad ({\rm
  Wegner~ generator}) \; .
\end{eqnarray}
\noindent
However, the trial wave function is peaked at the origin,
which on the finite momentum grid corresponds to the lowest momentum
$p_1$. Using the fact that all $P_n^2 >0$ with the exception of the
bound-state we obtain
\begin{eqnarray}
\lim_{\lambda \to 0} E_d (\lambda) &\ge& \frac{2}{\pi} w_1 p_1^2
\frac{P_1^2}{M} \phi_{\rm rel,N} (p_1)^2 = -\frac{\gamma_d^2}{M}
\frac{w_1 p_1^2 e^{-p_1^2 b^2}}{\sum_n w_n p_n^2 e^{-p_n^2 b^2}} \to
-\frac{\gamma_d^2}{M}\qquad {\rm (Wilson~ generator)} \; , \\
\lim_{\lambda \to 0} E_d (\lambda) &\ge& \frac{2}{\pi} w_{n_B} p_{n_B}^2 \frac{P_{n_B}^2}{M}
\phi_{\rm rel,N} (p_{n_B})^2 = -\frac{\gamma_d^2}{M} \frac{w_{n_B} p_{n_B}^2
  e^{-p_{n_B}^2 b^2}}{\sum_n w_n p_n^2 e^{-p_n^2 b^2}}\to 0 \qquad {\rm (Wegner~ generator)} \; .
\end{eqnarray}
\noindent
Of course, this argument holds in the infrared limit $\lambda \to 0$. A
  qualifying remark here becomes necessary: a distinction between the
  infrared limit $\lambda \to 0^+$ and the value $\lambda=0$ should be made as
  we are dealing with variational wave functions. For finite
$\lambda$, as long as the bound-state is not either shifted or
anchored at some fixed place, the behavior for both generators will not
essentially differ.  The dramatic evolution of the potential can be more
clearly seen by inspecting the sequence previously displayed. We
see that the jump in the Deuteron energy in the case of the Wegner generator is
directly related to the crossing of the hamiltonian diagonal matrix-element corresponding to the bound-state. A similar argument for the variational shell-model calculations for the cases with $A=3$
and $A=4$ yields the results
\begin{eqnarray}
&&\lim_{\lambda \to 0 } E_d(\lambda) = -B_d \, ,\qquad
\lim_{\lambda \to 0 } E_t(\lambda) = -\frac32 B_d \, , \qquad
  \lim_{\lambda \to 0 } E_\alpha (\lambda) = - 3 B_d \, , \qquad {\rm (Wilson~ generator)} \; , \\
&&\lim_{\lambda \to 0 } E_d(\lambda) = 0 \, ,\qquad \quad
\lim_{\lambda \to 0 } E_t(\lambda) = 0 \, , \qquad \qquad
  \lim_{\lambda \to 0 } E_\alpha (\lambda) = 0\, , \qquad \qquad {\rm (Wegner~ generator)} \; ,
\end{eqnarray}
\noindent
which is checked by the numerical calculations as shown in
Fig.~\ref{fig:binding}. Note that the
only contribution in the infrared limit $\lambda \to 0$ stems in this case from the two-body
bound-state. The continuum contributions in both the $^1S_0$ and
$^3S_1$ channels vanish.

For completeness, let us mention that as we evolve along the SRG
trajectory we also find linear correlations in two regimes
\begin{eqnarray}
\Delta B_\alpha / \Delta B_t \sim 2 ~ ~ (\lambda \to 0)
\qquad {\rm and} \qquad
\Delta B_\alpha / \Delta B_t \sim 4 ~ ~ (\lambda \sim 1) \; .
\end{eqnarray}
\noindent
The corresponding Tjon-lines are depicted in Fig.~\ref{fig:Tjon}.  The
gap in the Wegner generator case can be clearly identified. We compare
with the asymptotic lines $E_\alpha = 4 E_t - 3 E_d$ and $E_\alpha = 2
E_t$ which were identified in our previous work~\cite{Arriola:2013gya}
invoking three-body forces. The extrapolation of this formula gives
$B_\alpha = 4 \times 8.482 - 3 \times 2.225 = 27.53~{\rm MeV}$, while
the experimental value is $B_\alpha^{\rm exp} = 28.296 ~ {\rm
  MeV})$. As we see, the sharp intersection between the two lines
happens exactly at the SRG cutoff $\lambda$ where the jumps in the
binding energies are observed in Fig.~\ref{fig:binding}, which
corresponds to the critical values $E_t=3 E_d/2$ and $E_\alpha=3
E_d$. The phenomenon we observe resembles closely an avoided crossing
pattern, a phenomenon familiar from molecular physics in the
Born-Oppenheimer approximation~\cite{landau2013quantum}, where the
adiabatic parameter is the Triton binding-energy, $E_t$. As we see,
the variational calculation takes the minimal energy from the two
possible branches in the Wilson generator case since the variational
$(1s)^4$ state for the $\alpha$-particle contains $dd$ states with
opposite spin polarization at rest. Hence, the critical point
corresponds to the break up process $\alpha \to dd$.  As we see from
Fig.~\ref{fig:Tjon}, this break-up component is absent in the Wegner
generator case and one jumps to the free $4N$ state at rest at
$\lambda \to 0$.

Our results generate a Tjon-line, with a slope 4 for the SRG cutoff
below $\lambda \sim 1 {\rm fm}^{-1}$. We remind that this is the
regime which so far remained unaccessible for realistic interactions,
basically due to numerical difficulties triggered by the long momentum
tail of the potentials. Of course, while the generic features found here are expected
to become universal as we approach the infrared, the accuracy of our
results regarding the specific implications for nuclear binding should
be tested, particularly since our solution is just a variational
approximation. A thorough analysis of both the $A=3,4,16,40$ as well
as neutron and nuclear matter extending preliminary results for
realistic potentials~\cite{Arriola:2013nja} along the present lines
will be presented elsewhere.

%%%

%%% Needs to be revised

%%%

\section{Summary, conclusions and outlook}
\label{sec:summary}

The SRG flow equations provide an example of an isospectral flow,
i.e. a transformation which preserves the eigenvalues of the
hamiltonian but rotates the eigenfunctions in such a way that
transition matrix elements are increasingly and exponentially
suppressed in the corresponding energy differences. This approach has
offered new opportunities in Nuclear Physics as it allows the
possibility of describing phase-equivalent interactions which
conveniently become more diagonal. The practical implementation of the
SRG approach usually requires a discretization in momentum space.

Our analysis has therefore been performed on a finite momentum grid
which makes the problem computationally manageable but also leads to
specific features related to the formulation of the scattering
problem. In particular two unitarily equivalent hamiltonians do not
provide the same phase-shifts obtained by the solution of the LS
equation on the grid. Therefore, we have adopted the energy-shift
definition of the discretized Hamiltonian suggested many years ago by
Lifschits and extensively developped by Kukulin {\it et al.} in
more recent years within the context of the few-body problem. We have
found suitable formulas correctly incorporating Levinson's theorem. As
a by-product we have also deduced momentum grid based generalized trace
identities or finite energy sum rules put forward by Jaffe {\it et
  al.} on the basis of analyticity.

The energy-shift definition of the phase-shift is invariant along the
SRG trajectory but for the case of different generators such as those
of Wegner and Wilson the behaviour when approaching the infrared limit
turns out to be significantly different when bound-states are
present. For $N$-dimensional hamiltonians the SRG flow has always one
unique stable fixed point. In the case of the Wilson generator there
are $N!$ fixed points for non-degenerate hamiltonians but only one is
exponentially stable. In the case of the Wegner generator all $N!$
fixed points are stable. We remind that the SRG evolution with the different generators,
while preserving the two-body spectrum, yields asymptotically to infrared fixed-points which only correspond to re-orderings of the eigenvalues and hence contain the same physics of the initial bare two-body Hamiltonian.

With a simple toy model for the nuclear force, we explored the infrared
limit of the SRG evolution with both the Wilson and the Wegner generators.
While the fixed point is the same for the two generators
when no bound-state is supported by the interaction, we have two
distinct fixed points when a bound-state is allowed. The evolution
with the Wegner generator provides the correct behavior of the
interaction in the infrared limit $\lambda \to 0$ and defines an ordering of the
eigenvalues which can be used to compute the phase-shifts complying to
Levinson's theorem without solving the scattering equation. We provide
a consistent prescription to shift the eigenvalues in order to obtain
the nuclear force in the infrared limit $\lambda \to 0$ and also to compute the
phase-shifts from the eigenvalues with the correct behavior in both low-
and high-momentum regions.

Even though the ordering of states induced by the SRG evolution with the
Wegner generator make the phase-shift at low-momentum compliant to
Levinson's theorem, there is an optimal ordering which gives the smoothest phase-shifts around the bound-state scale.  The optimal ordering corresponds to a shift of the eigenvalues below the bound-state scale
while keeping the ascending order above it. The optimal value for the
position of the deuteron bound-state, which minimizes the RMS errors, seems to approach the characteristic deuteron momentum scale $\gamma = 0.23~{\rm fm}^{-1}$. Of course, it must be verified through explicit calculations if this a general result, which holds for any weakly or strongly coupled bound-state.

With the SRG evolution carried out towards $\lambda \to 0$, the difference
between the $^1S_0$ and $^3S_1$ channels due to the Deuteron bound-state
can be seen very clearly. The infrared limit of the SRG evolution with
the Wilson generator actually corresponds to an ascending order for the
eigenvalues of the interaction, while the Wegner generator preserves the
ascending order only above the bound-state scale.

We have also analyzed the $A=3$ and $A=4$ systems within a simple
harmonic oscillator scheme where the main differences between
different generators may be clearly appreciated as the critical scale
is approached. There appear two branches $E_\alpha= 4 E_t - 3 E_d$ and
$E_\alpha= 2 E_t $ and the system takes the lowest energy at any rate
among all possible states contained in the trial wave function. Our
results generate a Tjon-line which matches the expression $E_\alpha= 4
E_t - 3 E_d$ found by invoking three-body forces in our previous work
for the SRG cutoff below $\lambda \sim 1 {\rm fm}^{-1}$ (the factor 4
corresponds to 4 triplets)~\cite{Arriola:2013gya}. We remind that this
is the regime which so far remained unaccessible for realistic
interactions. The accuracy of the Tjon formula is rather satisfactory
and a complementary investigation going beyond the simple variational
ansatz and including specifically many-body forces would be most
illuminating. The application of the on-shell interactions to
few-nucleon systems, light nuclei and nuclear many-body problems
(neutron and nuclear matter) for realistic interactions is beyond the
scope of this work but is certainly an interesting study we will
pursue in forthcoming works on the light of the present findings.

\section*{Acknowledgements}

E.R.A. would like to thank the Spanish Mineco (Grant FIS2014-59386-P) and
Junta de Andalucia (grant FQM225).  S.S. and V.S.T. are supported by
FAPESP (grant 2014/04975-9). V.S.T. also thanks FAEPEX (grant
1165/2014) and CNPq (grant 310980/2012-7) for financial support.

\end{document}